\documentclass[preprint]{aastex}

\newcommand{\teff}{$T_{\rm eff}$}
\newcommand{\rxj}{RXJ0529.4$+$0041}
\newcommand{\ntt}{NTT~045251$+$3016}

\slugcomment{Accepted by ApJ}

\shorttitle{Dynamical Masses of PMS Stars}
\shortauthors{Stassun et al.}

\received{2003 November 4}
\begin{document}

\title{Dynamical Mass Constraints on Low-Mass Pre--Main-Sequence Stellar
Evolutionary Tracks: An Eclipsing Binary in Orion with a 1.0 M$_\odot$
Primary and an 0.7 M$_\odot$ Secondary\footnote{Based on data
collected with the Hobby-Eberly Telescope and the WIYN Telescope.}}

\author{Keivan G.\ Stassun\altaffilmark{2,3,4},
Robert D.\ Mathieu\altaffilmark{4},
Luiz Paulo Vaz\altaffilmark{5},
Nicholas Stroud\altaffilmark{4},
Frederick Vrba\altaffilmark{6}}

\altaffiltext{2}{Visiting Astronomer, Cerro Tololo Inter-American Observatory
and Kitt Peak National Observatory.
CTIO and KPNO are operated by AURA, Inc.\ under contract to the National Science
Foundation.}
\altaffiltext{3}{Hubble Fellow}
\altaffiltext{4}{Department of Astronomy, University of Wisconsin---Madison,
Madison, WI 53706, and Department of Physics \& Astronomy, Vanderbilt
University, Nashville, TN 37235; keivan.stassun@vanderbilt.edu}
\altaffiltext{5}{Departamento de F\'{\i}sica, Universidade Federal de Minas
Gerais, C.P. 702, 30.123-970, Belo Horizonte, MG, Brazil}
\altaffiltext{6}{US Naval Observatory, Flagstaff Station, Flagstaff, AZ}

\begin{abstract}
We report the discovery of a double-lined, spectroscopic, eclipsing binary
in the Orion star-forming region. We analyze the system spectroscopically
and photometrically to empirically determine precise, distance-independent 
masses, radii, effective temperatures, and luminosities for both components.
The measured masses for
the primary and secondary, accurate to $\sim 1$\%, are 
$1.01\;{\rm M}_\odot$ and 
$0.73\;{\rm M}_\odot$, respectively; thus the primary is a definitive
pre--main-sequence solar analog, and the secondary is the 
lowest-mass star yet discovered among pre--main-sequence eclipsing binary 
systems. We use these fundamental measurements to test the predictions of
pre--main-sequence stellar evolutionary tracks. None of the models
we examined correctly predict the masses of the two components
simultaneously, and we implicate differences between the theoretical and 
empirical effective temperature scales for this failing. All of the
models predict the observed slope of the mass-radius relationship
reasonably well, though the observations tend to favor models with low
convection efficiencies. Indeed,
considering our newly determined mass measurements together with other dynamical
mass measurements of pre--main-sequence stars in the literature, as
well as measurements of Li abundances in these stars, we show that 
the data strongly favor evolutionary models with inefficient convection in the
stellar interior, even though such models cannot reproduce the properties 
of the present-day Sun. 
\end{abstract}


\section{Introduction\label{intro}}
Pre--main-sequence stellar evolutionary models are fundamental to
our paradigm of star formation and early stellar evolution.
These models are the means by which masses and ages are assigned to 
young stars and brown dwarfs, and
are therefore central 
to our understanding of physical processes that depend, directly
or indirectly, on knowledge of stellar masses and ages.
Indeed, pre--main-sequence (PMS) stellar evolutionary models
touch upon the most basic questions in star
formation research: the nature and origin of the initial mass function; 
the timescale for circumstellar disk evolution and planet formation;
and the initial distribution of stellar angular momentum, its evolution in time,
and the dependence of that evolution on stellar mass, accretion, and
other mass-dependent properties.
Thus, the absolute calibration of PMS evolutionary models with respect 
to stellar mass is of foundational importance to star formation research.

Empirical mass
determinations exist in the literature for 15 individual PMS stars 
\citep{popper,andersen,casey,simon,covino,steffen,aksco}, with only 6 of these in
the mass domain of 1 M$_\odot$ or below. Moreover, many of these 
measurements, while of high precision, potentially suffer from systematic
errors (primarily distance uncertainties) that limit their accuracy.
Indeed, there is no PMS star with $M < 1\;{\rm M}_\odot$ whose mass has
been measured with an accuracy of better than a few percent.

Consequently, the absolute mass calibration of theoretical PMS evolutionary
tracks remains weakly constrained by observations. The dearth of accurate
measurements below 1 M$_\odot$ is particularly salient because low-mass
stars dominate studies of young stellar populations, and it is in this
mass domain that key, outstanding questions remain with respect to, e.g.,
the nature of the initial mass function \citep{hill97,muench} and the evolution 
of stellar angular momentum \citep{stass99,herbst,rebull01}.

Eclipsing binary stars provide a uniquely powerful means of determining
absolute stellar masses because the mass determinations are distance-independent. 
Eclipsing binaries also yield direct measurement of the stellar
radii, permitting the determination of stellar luminosities that are
also distance independent.
To date, only four eclipsing binaries with PMS components
have been reported in the literature \citep{andersen,popper,casey,covino}.

Here we report the discovery of a previously unknown, double-lined,
spectroscopic, eclipsing binary system in the Orion star-forming region,
\objectname{V1174~Ori}. We combine time-series photometry with high-resolution
spectroscopy to determine all of the fundamental physical properties
of the system and of its stellar components. We then use these measurements
to test the predictions of a variety of PMS stellar evolutionary models.

We begin by describing the discovery of \objectname{V1174~Ori} and its identification
as a PMS eclipsing binary in \S \ref{discovery}. In \S \ref{data}, we
present our photometric and spectroscopic observations and the 
methods used in their
reduction. We analyze these observations in \S \ref{analysis} to
determine the system ephemeris and to derive a precise double-lined orbit 
solution, providing absolute stellar masses. We conduct a thorough 
spectroscopic analysis that
includes determination of the stellar rotational velocities, 
measurement of the primary-to-secondary flux ratio, spectral
classification and determination of effective
temperatures, measurement of Li abundances, and 
measurement of the equivalent width of H$\alpha$ emission. We also
present a full analysis of our multi-band time-series photometry using 
synthetic light-curve fitting to determine precise, absolute stellar
dimensions. The results of our analysis are presented in \S \ref{results},
including an independent determination of the distance to \objectname{V1174~Ori}.

In \S \ref{tracks} we use our empirically determined stellar masses, radii,
effective temperatures, and luminosities to test a variety of PMS
stellar evolutionary tracks possessing a variety of characteristics,
including different atmosphere models and treatment of convection.
We assess the performance of the theoretical models both against the 
components of \objectname{V1174~Ori} alone as well as against the 
ensemble of all PMS stars 
with empirical mass determinations in the literature. We consider 
how the tracks compare quantitatively to the data in both high- and
low-mass regimes, in both binary and single systems, and in both the
H-R diagram and mass-radius observational planes. Moreover, we examine the 
general behavior of the tracks to suggest, in qualitative terms,
how their performance against the data might be improved.

In addition, we
assay the ensemble of empirical mass determinations independent of the
models to explore the degree of internal consistency in the data
themselves, and we show how Li abundances can be used in some cases to 
resolve apparent contradictions. Finally, we discuss more general
insights that can be gleaned from the Li abundances, arguing that the
data strongly favor models with inefficient convection in the stellar
interior during the PMS stage of evolution, even though such models 
cannot reproduce the
properties of the present-day Sun. We present a summary of our 
conclusions in \S \ref{summary}.

\section{Discovery\label{discovery}}
Here we describe the discovery of \objectname{V1174~Ori} as
a PMS eclipsing binary system. Its position is 
$\alpha = 05$:34:27.85, $\delta = -05$:41:37.8 (J2000.0). 

\subsection{Data from the Literature\label{background}}
\objectname{V1174~Ori} has been included in a number of previous surveys of stars in 
the vicinity of the Orion Nebula Cluster (ONC), dating back to the
catalog of \citet{parenago} (star \# 1478). The General Catalog of Variable Stars,
whose designation we use in this paper,
reports a range of 1.1 mag (in the photographic system) between 
its high and low states. 

More recently, the ROSAT observatory included \objectname{V1174~Ori} (designation
ROS-ORI~166) in a deep PSPC pointing centered on the Trapezium region.
The analysis of these data by \citet{geier95} assigns a net of 
75 X-ray photons to \objectname{V1174~Ori}, for a count rate of 17.02 cts/ks. 
This corresponds to an X-ray luminosity of 
$L_X \approx 1 \times 10^{30}$ erg/s in the ROSAT bandpass (0.07 to 2.4
keV), assuming the ONC distance to \objectname{V1174~Ori} of 470 pc (e.g.\
\citet{genzel81}). \objectname{V1174~Ori} has not been included in any of the {\it Chandra}
observations of the ONC currently in the public archive, but this
X-ray luminosity is fairly typical among the PMS stars that have
been observed by {\it Chandra\/} (e.g.\ \citet{feigelson02}).

\objectname{V1174~Ori} was observed in the $K$-band survey of the Orion A cloud by
\citet{alidepoy95} (\# 2176 in their catalog), but their reported 
$K$-band magnitude of 9.00 is
at the bright limit of their survey, making this measurement suspect.
More recently, \citet{carpenter01} presented $JHK$ light curves of
\objectname{V1174~Ori} (star number 4008 in their catalog), as part of their 
analysis of time-series photometry of the Orion star-forming region,
conducted as an extension of the 2MASS project\footnote{See
\url{http://www.ipac.caltech.edu/2mass} for information about 2MASS}.
Their light curves show rms
scatters of 0.08--0.09 mag. In addition to low-level variability, 
their light curves display evidence for a possible eclipse event in
the form of a single observation (seen in all three filters) that 
is 0.3--0.4 mag dimmer than the mean brightness level.

Finally, \objectname{V1174~Ori} was included in the study of \citet{rebull00} 
(\# 1-665 in their catalog), whose $UVI$ photometry places
\objectname{V1174~Ori} on the locus of Orion PMS stars in the $(U-V)$ vs.\ $(V-I)$
color-magnitude diagram. They report a spectral type for 
\objectname{V1174~Ori} of M0 on the basis of low-resolution spectroscopy, but
derive extinctions of $A_V = -0.50$ and $A_I = -0.31$, indicating
potential problems with the spectral classification or photometry.
While \objectname{V1174~Ori} was included in the region covered by the time-series
study of \citet{rebull01}, they do not report it as a
periodic variable. \objectname{V1174~Ori} was in eclipse during at least 7 of 
the 30 nights covered by their observations (assuming the ephemeris
we adopt in \S \ref{analysis}), but with only one
observation per night their light curves were presumably too 
sparse to detect periodic behavior (if they detected the eclipses
at all).

\subsection{Identification as a PMS Double-Lined Eclipsing Binary\label{identaseb}}
We first identified \objectname{V1174~Ori} as a candidate eclipsing binary 
from an $I$-band light
curve obtained by us in 1994 December as part of a photometric study of
stellar rotation in the ONC (Stassun et al.\ 1999; hereinafter 
\citet{stass99}). The discovery light curve
was by itself very compelling, showing 
eclipse-like variability with a period of $\approx 2.63$ days.

\begin{figure}[ht]
\epsscale{1.0}
\plotone{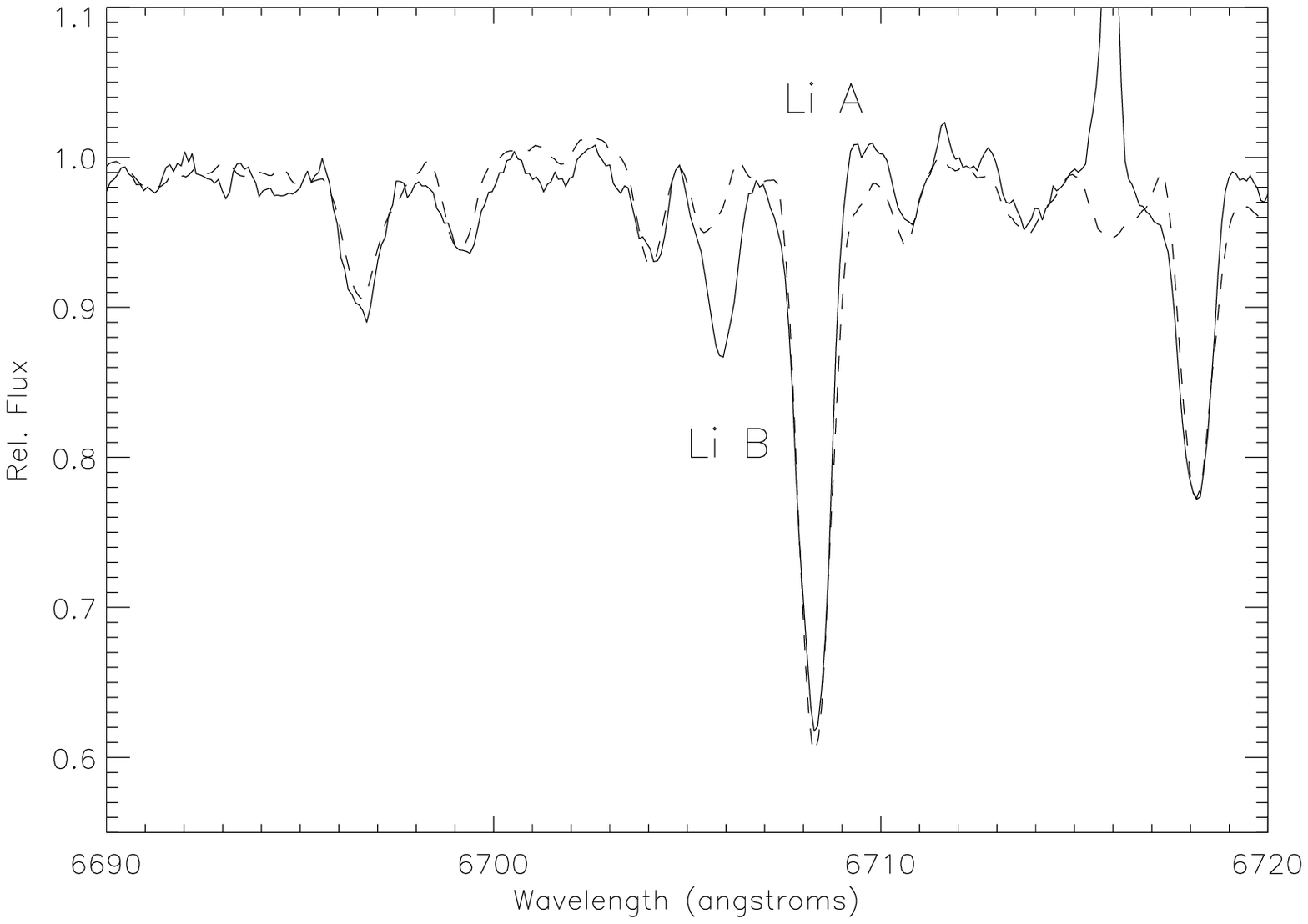}
\caption{Keck HIRES spectrum of V1174~Ori, in the vicinity of the Li $\lambda$6708
line, is shown as a solid curve. The dashed curve is a spectrum of the V1174~Ori
primary, obtained by averaging together all of the HET HRS spectra. The Li line
from the secondary (labeled Li B) is clearly present, in this case blueward of 
the primary's Li line (labeled Li A) and partially blended with an Fe line from
the primary. The emission feature near $\lambda$6716 is nebular [\ion{S}{2}].
\label{li-hires-fig}}
\end{figure}

\objectname{V1174~Ori} was also included in several pointings during our spectroscopic 
survey of the region with the WIYN\footnote{The WIYN Observatory is a joint 
facility of the University of Wisconsin-Madison, Indiana University, 
Yale University, and the National Optical Astronomy Observatories.}
multi-object spectrograph, which 
showed \objectname{V1174~Ori} to have strong Li absorption (an indicator of stellar
youth) and revealed it to be a 
single-lined spectroscopic binary. We confirmed the
spectroscopic presence of the secondary in a spectrum kindly obtained
by G.\ Basri with the Keck HIRES spectrograph, which also clearly revealed
the presence of Li absorption due to the secondary (Fig.\ \ref{li-hires-fig}).
This spectrum also allowed us to determine that the primary was of roughly
mid-K spectral type.

\section{Observations and Data Reduction\label{data}}
Having identified \objectname{V1174~Ori} as a PMS eclipsing binary
and double-lined spectroscopic
binary, we obtained multiple high-S/N, high-resolution spectra
with which to determine a precise, double-lined orbit solution. We
also obtained high-precision, high-cadence light curves at multiple
wavelengths in order to (a) firmly establish the system ephemeris,
and to (b) determine, via light-curve synthesis and modeling, such 
parameters as the orbital 
inclination, component radii, and ratio of component effective 
temperatures. The one remaining piece of information needed to fully 
establish all system parameters is the effective temperature of one
of the components, a datum also provided by our high-resolution
spectroscopy.

Here we present the photometric and spectroscopic data we obtained, and
describe the procedures we employed in their reduction.

\subsection{Photometry\label{photometry}}
\begin{figure}[ht]
\plotone{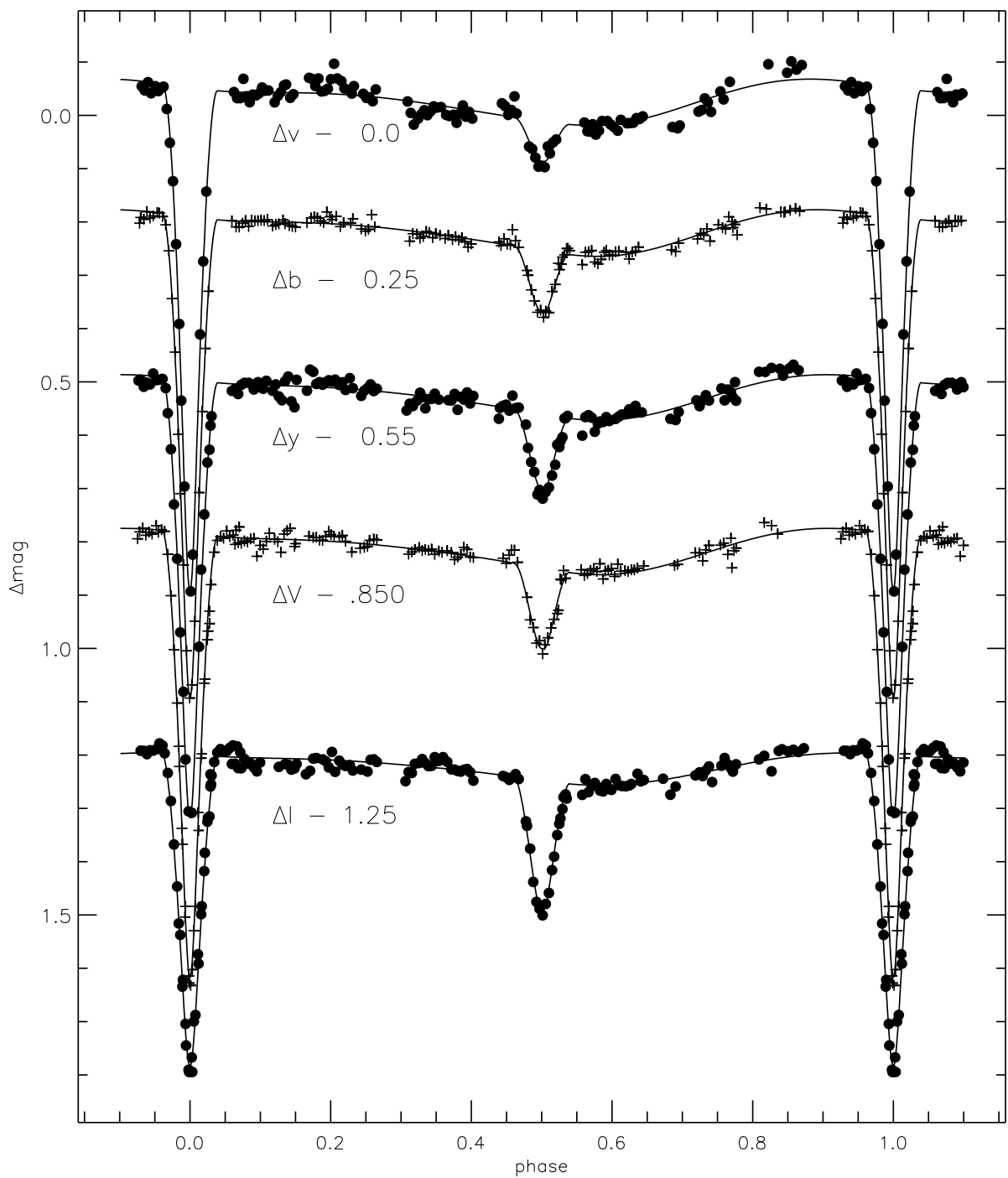}
\caption{The $vbyVI$ light curves observed in Jan 2001 together with
the theoretical light curves synthesized with a Wilson-Devinney based
algorithm (see Tables \ref{wd5} and
\ref{spotsall}). For clarity, the light curves were shifted by the
amount indicated in the figure,
but the relative scale is the same for all passbands. Note the
changes in the depths of both eclipses with wavelength.
\label{IVybvlc}}
\end{figure}

We observed \objectname{V1174~Ori} photometrically with CCDs at 1m-class telescopes at 6 separate
epochs spaced over 8 years. Table \ref{phottable} summarizes the
time-series photometry we obtained. The individual measurements are available
electronically and are shown in 
Figs.\ \ref{IVybvlc}, \ref{IVBI01lc}, and \ref{JHKI94lc}.

\begin{deluxetable}{lccccrc}
\tablewidth{0pt}
\tabletypesize{\scriptsize}
\tablecolumns{7}
\tablecaption{Photometric time-series observations of V1174~Ori\label{phottable}}
\tablehead{
\colhead{UT Dates} & \colhead{N (nights)} &
   \colhead{${\rm Range~of~Julian~Dates}\atop{(2400000+)}$} &
   \colhead{Observatory} & \colhead{Filter} &
   \colhead{${\rm \#~of}\atop{\rm obs.}$} & \colhead{Ref.} }
\startdata
1994 Dec 11--27 & 16 & 49698.340--49714.491 & KPNO 0.9m, USNO 1m, Wise 1m & $I_C$ & 128 & \citet{stass99} \\
2001 Jan 20--27 & 8  & 51929.588--51936.775 & CTIO 0.9m & $I_C$ & 152 & this study \\
                &    & 51929.608--51936.768 &           & $V$   & 144 & this study \\
                &    & 51929.590--51936.776 &           & Str.\ $y$ & 158 & this study \\
                &    & 51929.594--51936.783 &           & Str.\ $b$ & 154 & this study \\
                &    & 51929.623--51936.735 &           & Str.\ $v$ & 130 & this study \\
2001 Jan 31     & 1  & 51940.591--51940.819 & USNO 1m   & $I_C$ & 20 & this study \\
                &    & 51940.589--51940.817 &           & $V$   & 20 & this study \\
2001 Nov 14--24 & 9  & 52227.746--52237.996 & WIYN 0.9m & $I_C$ & 78 & this study \\
2002 Nov 17--26 & 10 & 52595.745--52605.000 &           & $I_C$ & 110 & this study \\
                &    & 52595.802--52605.004 &           & $V$   & 96 & this study \\
                &    & 52595.806--52604.984 &           & $B$   & 96 & this study \\
2002 Dec 05--16 & 11 & 52613.810--52624.946 &           & $I_C$ & 24 & this study \\
\enddata
\end{deluxetable}

All CCD data were processed in the manner described in \citet{stass99}. 
Briefly, we reduced our CCD frames using the IRAF\footnote{IRAF is 
distributed by the National Optical Astronomy Observatories, which is operated 
by the Association of Universities for Research in Astronomy, Inc.,
under cooperative agreement with the National Science Foundation.}
{\tt CCDRED} or {\tt XCCDRED} (for CTIO quad-readout data) packages, and performed
aperture photometry on all stars in the field of view using the IRAF APPHOT 
package. We then applied an
algorithm similar to that of \citet{honeycutt} to our raw light curves to remove 
non-cosmic frame-to-frame photometric variations (due to changes in, e.g., seeing, 
sky brightness, atmospheric transparency). In all cases, the resulting photometric 
precision in the light curves of \objectname{V1174~Ori} is limited by systematics 
(flat-fielding, PSF variations, etc.) at the 
$\sim 0.01$ mag level, which we determine from inspection of 
non-variable stars in the field of comparable brightness.

In addition to these differential time-series observations, we obtained 
calibrated $BVI$ photometry of \objectname{V1174~Ori} on 1995 Feb 02 at the 
USNO 1m telescope.
These data were obtained out of eclipse, at orbital phase 0.35, based on the
ephemeris presented in \S \ref{ephemeris}.
Calibrated $UVI$ photometry of \objectname{V1174~Ori} has been reported by 
\citet{rebull00},
and \citet{carpenter01} report $JHK$ measurements. These 
measurements are summarized in Table \ref{calphot}.

\citet{rebull00} do not report
the precise times of their observations. However, the $V-I$
color reported by them is identical to that measured by us, which leads us to
believe that their observations were also obtained out of eclipse. 
\citet{carpenter01} report mean $JHK$ magnitudes and colors, averaged over all of
their observations which span 36 days (excluding the two non-contiguous 
measurements obtained in 1998 Mar and 2000 Feb). The $J$
magnitude we adopt is the mean of their non-eclipse observations.
The $J-H$ and $H-K$ colors do not show
significant variations in their observations and so we do not adjust the
values reported by those authors.

\begin{deluxetable}{crrll}
\tablewidth{0pt}
\tabletypesize{\scriptsize}
\tablecolumns{5}
\tablecaption{Calibrated photometry of V1174~Ori\label{calphot}}
\tablehead{ \colhead{Filter(s)} & \colhead{Mag./Color} & \colhead{Mag./Color} & \colhead{UT Date(s)} & 
\colhead{Ref.} \\
\colhead{} & \colhead{observed} & \colhead{calculated\tablenotemark{a}} & \colhead{} & \colhead{} }
\startdata
$V$    & $13.95 \pm 0.03$ & 13.95 & 1995 Feb 02 & 1 \\
$B-V$  & $1.25 \pm 0.02$  &  1.24 &             & 1 \\
$V-I$  & $1.62 \pm 0.02$  &  1.48 &             & 1 \\
       & $1.61 \pm 0.02$  &       & 1998 Jan    & 2 \\
$U-V$  & $2.22 \pm 0.02$  &  2.33 &             & 2 \\
$J$\tablenotemark{1}    & $11.22 \pm 0.05$ & 11.52 & 2000 Mar 04 -- Apr 08\tablenotemark{2} & 3 \\
$J-H$  & $0.65 \pm 0.02$  & 0.65 &              & 3 \\
$H-K$  & $0.19 \pm 0.03$  & 0.14 &              & 3 \\
\enddata
\tablerefs{
1: this study, 2: \citet{rebull00}, 3: \citet{carpenter01}
}
\tablenotetext{a}{Assuming main-sequence colors, bolometric corrections, and
$R_V = 3.12$. See \S \ref{results-fundamental}.}
\tablenotetext{1}{Value adjusted from that reported in \citet{carpenter01},
in order to account for eclipses.}
\tablenotetext{2}{Contiguous portion of light curve only.}
\end{deluxetable}

\subsection{Spectroscopy\label{spectroscopy}}
In order to derive a precise, double-lined orbit solution for \objectname{V1174~Ori}, we 
obtained 15 high S/N observations in queue observing mode with the High 
Resolution Spectrograph (HRS)
on the Hobby Eberly Telescope\footnote{The Hobby-Eberly Telescope 
is operated by McDonald Observatory on behalf 
of The University of Texas at Austin, the Pennsylvania State University, 
Stanford University, Ludwig-Maximilians-Universit\"{a}t M\"{u}nchen, and 
Georg-August-Universit\"{a}t G\"{o}ttingen. The observations described
here were obtained through community access made possible by NOAO.} 
(HET) on 15 
different nights between 2001 Nov 22 and 2002 Feb 23. The spectrograph 
setup we used yielded a resolving power of $R \approx 30000$, with 
wavelength coverage from 5095\AA\ to 8860\AA, and centered at 6948\AA.
The 51 echelle orders in each spectrum are imaged onto two CCDs, 
with the ``blue" chip imaging the 32 orders from 5095\AA\ to 6803\AA\ 
and the ``red" chip imaging the remainder. The spectrograph was also 
set up with a single sky fiber so that simultaneous sky spectra could
be obtained with each observation. The CCD images were binned on readout
such that a spectral resolution element corresponds to approximately 3
pixels in the extracted spectra.

The exposure times were initially $3\times 600 = 1800$ sec, and later
increased to $3\times 800 = 2400$ sec, and we combined the individual 
exposures in each set with cosmic-ray rejection. Each observation 
of \objectname{V1174~Ori} was bracketed by an observation of a ThAr lamp. In 
addition, all but one observation was accompanied by an observation of 
a bright radial-velocity standard star (typically obtained shortly 
after the \objectname{V1174~Ori} exposure) in order to closely monitor and correct 
for instrumental drifts in the radial velocities.

In addition, we obtained a single, very high S/N ($\sim 250$) 
observation of a late-type radial-velocity standard star for use as a 
radial-velocity template in our cross-correlation analysis (\S
\ref{orbit}). Once we had established the approximate spectral types
of the components of \objectname{V1174~Ori} (\S \ref{spectral}), we also obtained single 
observations of two appropriate spectral-type standard stars, of types 
K3 and K7\footnote{More careful analysis of the \objectname{V1174~Ori}
spectra after the observations were obtained indicates that the
components of \objectname{V1174~Ori} have spectral types of K4.5 and $\sim$M1.5 (see 
\S\ref{spt}), but this does not significantly affect our analysis.}. 
Table \ref{het-table} summarizes our spectroscopic observations with 
HET HRS.

\begin{deluxetable}{rlccrrl}
\tablewidth{0pt}
\tabletypesize{\scriptsize}
\tablecolumns{7}
\tablecaption{Log of HET HRS Observations\label{het-table}}
\tablehead{ \colhead{\#} & \colhead{UT Date} & \colhead{HJD\tablenotemark{1}} & 
\colhead{Object} & \colhead{Exp.\ Time (s)} & \colhead{S/N\tablenotemark{2}} 
& \colhead{Comments} }
\startdata
0 & 2001 Oct 29 & 2212.74347 & \objectname{HD 18884} & 30 & 245 & R.V.\ template, SpT = M2 III \\
1 & 2001 Nov 22 & 2235.85217 & \objectname{V1174~Ori}            & 1800 & 85 &          \\
  &          & 2235.87681 & \objectname{HD 26162} &  5 & 120 & R.V.\ standard, SpT = K2 III \\
2 & 2001 Nov 23 & 2236.84827 & \objectname{V1174~Ori}            & 1800 & 80 &          \\
  &          & 2236.89321 & \objectname{HD 26162} & 15 & 145 &         \\
3 & 2001 Dec 10 & 2253.80559 & \objectname{V1174~Ori}            & 1800 & 60 &          \\
  &          & 2253.82748 & \objectname{HD 26162} & 10 & 85 &          \\
4 & 2001 Dec 18 & 2261.78887 & \objectname{V1174~Ori}            & 1800 & 40 &          \\
  &          & 2261.80695 & \objectname{HD 26162} & 10 & 90 &          \\
5 & 2001 Dec 21 & 2264.77976 & \objectname{V1174~Ori}            & 1800 & 80 &          \\
  &          & 2264.79712 & \objectname{HD 26162} & 10 & 95 &          \\
6 & 2001 Dec 22 & 2265.76822 & \objectname{V1174~Ori}            & 1800 & 75 &          \\
  &          & 2265.83415 & \objectname{HD 26162} & 10 & 95 &          \\
7 & 2001 Dec 23 & 2266.76687 & \objectname{V1174~Ori}            & 1800 & 105 &         \\
  &          & 2266.81776 & \objectname{HD 26162} & 10 & 120 &         \\
8 & 2002 Jan 26 & 2300.67336 & \objectname{V1174~Ori}            & 2400 & 125 &         \\
9 & 2002 Jan 27 & 2301.66941 & \objectname{V1174~Ori}            & 2400 & 95 &          \\
  &          & 2301.72972 & \objectname{HD 26162} & 10 & 100 &         \\
10 & 2002 Jan 28 & 2302.66839 & \objectname{V1174~Ori}            & 2400 & 105 &         \\
  &          & 2302.72798 & \objectname{HD 26162} & 10 & 140 &         \\
11 & 2002 Feb 07 & 2312.64124 & \objectname{V1174~Ori}            & 2400 & 100 &         \\
  &          & 2312.70694 & \objectname{HD 26162} & 10 & 105 &         \\
12 & 2002 Feb 09 & 2314.63298 & \objectname{V1174~Ori}            & 2400 & 120 &         \\
  &          & 2314.69301 & \objectname{HD 26162} & 10 & 160 &         \\
13 & 2002 Feb 13 & 2318.63282 & \objectname{V1174~Ori}            & 2400 & 95 &          \\
  &          & 2318.68518 & \objectname{HD 26162} & 10 & 130 &         \\
14 & 2002 Feb 21 & 2326.61211 & \objectname{V1174~Ori}            & 2400 & 90 &          \\
  &          & 2326.62791 & \objectname{HD 26162} & 10 & 115 &         \\
15 & 2002 Feb 23 & 2328.60823 & \objectname{V1174~Ori}            & 1800 & 100 &         \\
  &          & 2328.62179 & \objectname{HD 26162} & 10 & 110 &         \\
16 & 2002 Mar 07 & 2340.93213 & \objectname{HD 110463} & 180 & 140 & SpT standard, SpT = K3 V \\
  &          & 2340.64503 & \objectname{HD 237903} & 240 & 160 & SpT standard, SpT = K7 V \\
\enddata
\tablenotetext{1}{Heliocentric Julian Date (2450000+)}
\tablenotetext{2}{Signal-to-noise is per 3-pixel resolution element, measured near 6500\AA.}
\end{deluxetable}

All spectra were processed with the IRAF {\tt CCDPROC} and {\tt ECHELLE} packages. 
The steps involved in producing the final spectra include bias subtraction,
flat-fielding, bad-column interpolation, order tracing (for both object and
sky apertures), sky subtraction, wavelength calibration, and heliocentric 
velocity correction. Extraction of the spectra from the raw data was not
straightforward, particularly with respect to order tracing which required
a high degree of manual intervention.  The orders are
highly angled on the CCD, have flat-topped profiles in the cross-dispersion
direction, and are separated from the sky orders by only a few pixels. We
thus encountered difficulties with the automated order-finding and tracing 
frequently ``skipping" between the object and sky apertures. 
However, we found that these problems could be circumvented manually.

ThAr arc spectra were extracted similarly. We found that relatively
high-order polynomials both along and across the orders were required to 
provide a good mapping between pixel, order number, and wavelength. In the end we
adopted third-order and seventh-order polynomial fits along and across the
orders, respectively. The wavelength solutions were applied to the object
spectra by taking an average of the ThAr spectra taken before and
after the object spectra, weighted by their proximity in time to the
object spectra.

Heliocentric velocity and date corrections were computed via the
IRAF {\tt RV} package.
We carefully assigned mid-exposure observation times to each combined
spectrum, taking into account the flux levels of the 3 sub-exposures 
(which occasionally differed from one another) and their time spacing
relative to one another.

As ours was the first community program to use HET HRS, we unfortunately
encountered an unforseen problem that rendered the ``red" orders
in our spectra unusable. The ``red" HRS CCD in use at the time of our
observations suffers from severe fringing, which would normally be
removed by a spectrum of a bright continuum source such as an internal
lamp. However, in the case of the HRS, the fringing pattern produced
by the internal flat-field calibration lamp is different from the pattern
present in the data. This is in part due to 
the fact that a different fiber is used for calibration purposes, which
results in the flat-field beam having a different $f/$-ratio as
compared to the science-object beam (M.\ Shetrone, priv.\ comm.). Thus
in our spectroscopic analyses we use only the orders on the ``blue" CCD,
which show no such fringing problems.

\section{Analysis\label{analysis}}

\subsection{Eclipse timings: System ephemeris\label{ephemeris}}
We use eclipse timing measurements from our photometric light curves 
(\S \ref{photometry}) to determine a precise period and ephemeris 
for \objectname{V1174~Ori}. Our $I$-band data, which span a total of 2927 days, 
include 7 primary and 5 secondary eclipses. Some of these eclipses were also
observed in other filters, providing additional (though not fully independent)
eclipse timing information. We measure the time of each
eclipse minimum by fitting a gaussian to the points in
the light curve in and around the eclipse. These eclipse timings and their
uncertainties are reported in Table \ref{eclipses-table}.

We searched the combined $I$-band lightcurves for a period via the 
Phase-Dispersion Minimization (PDM) technique \citep{stellingwerf},
and found a best period of 2.61156 days. However, visual inspection
of the combined data folded on this period revealed that the algorithm
had chosen a beat period whereby the primary eclipses from the 2001
data were overlaid on the secondary eclipses from the 1994 data. Thus
we refined the period by searching our radial-velocity measurements
(\S \ref{orbit}) for periods near this value. This yielded a best 
period of $2.63469 \pm 0.00012$ days. A final visual inspection of the light
curve folded on this period suggested a small manual refinement to
$P = 2.634727 \pm 0.000004$ days. 

\begin{figure}[ht]
\epsscale{0.8}
\plotone{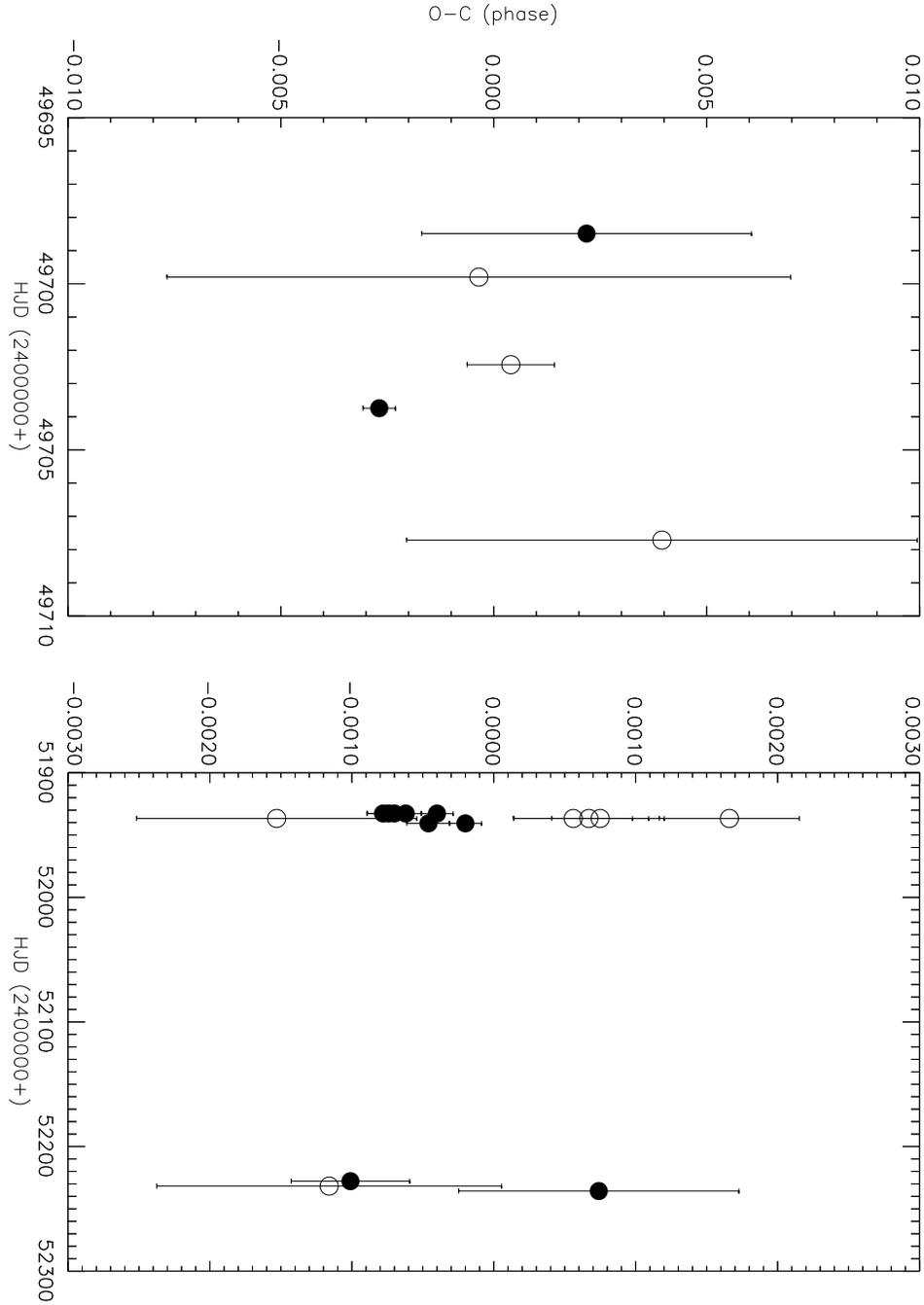}
\caption{The $O-C$ phase residuals of the eclipse timings of V1174~Ori are
shown based on a best-fit period of
2.634727 days. Filled circles represent primary eclipses, open circles represent
secondary eclipses. See Table \ref{eclipses-table}. \label{ephem-oc-fig}}
\end{figure}

In Fig.\ \ref{ephem-oc-fig} we show the $O-C$ phase residuals of the 
eclipse timings from Table \ref{eclipses-table} using this period.
To show the residuals of the secondary eclipse together with those of
the primary,
we have subtracted 0.5 phase from the secondary residuals, which
assumes a circular orbit.
These residuals have an r.m.s.\ of 0.0015 in phase (5.5 min), and show no trends 
over the 2900-day span (some 1100 cycles) of the data. While there appears
to be a weak effect in the Jan 2001 data for the primary eclipse $(O-C)$
values to 
be systematically negative and for the secondary eclipse $(O-C)$ values to 
be systematically positive, 
it should be borne in mind that the eclipse timings from the different
filters in Jan 2001 are not truly independent, as these light curves share a 
common sampling pattern.  
The eclipse timings are therefore consistent with the assumption of a 
circular orbit.

We thus adopt the following ephemeris for the remainder of our analysis:
\begin{displaymath}
{\rm Min\; I: HJD\; 2,449,703.7504(15)}\, + 2.634727(4)E{\rm ,}
\end{displaymath}
where $E$ is the epoch number, and the zero-point 
corresponds to primary eclipse (i.e.\ the deeper minimum). 

\begin{deluxetable}{lcclr}
\tablewidth{0pt}
\tabletypesize{\scriptsize}
\tablecolumns{5}
\tablecaption{Times of Eclipse Minima\label{eclipses-table}}
\tablehead{ \colhead{UT Date} & \colhead{HJD\tablenotemark{1}} &
\colhead{Filter} & \colhead{Eclipse type} & \colhead{${{O-C}\atop{\rm (phase)}}$} } 
\startdata
1994 Dec 11 & $49698.4867 \pm 0.0102$ & $I_C$ & Primary   &  $0.0022$ \\
1994 Dec 13 & $49699.7974 \pm 0.0193$ & $I_C$ & Secondary & $-0.0004$ \\
1994 Dec 15 & $49702.4341 \pm 0.0027$ & $I_C$ & Secondary &  $0.0004$ \\
1994 Dec 17 & $49703.7433 \pm 0.0010$ & $I_C$ & Primary   & $-0.0027$ \\
1994 Dec 21 & $49707.7129 \pm 0.0158$ & $I_C$ & Secondary &  $0.0040$ \\
1994 Dec 27 & $49714.2850 \pm 0.0011$ & $I_C$ & Primary   & $-0.0016$ \\
2001 Jan 23 & $51932.7275 \pm 0.0004$ & $I_C$ & Primary   & $-0.0007$ \\
            & $51932.7284 \pm 0.0003$ & $V$   &           & $-0.0004$ \\
            & $51932.7274 \pm 0.0003$ & $y$   &           & $-0.0008$ \\
            & $51932.7276 \pm 0.0002$ & $b$   &           & $-0.0007$ \\
            & $51932.7278 \pm 0.0003$ & $v$   &           & $-0.0006$ \\
            & $51932.72774 \pm 0.00020$ & mean\tablenotemark{a} &       & $-0.00065$ \\
2001 Jan 27 & $51936.6835 \pm 0.0009$ & $I_C$ & Secondary &  $0.0008$ \\
            & $51936.6833 \pm 0.0014$ & $V$   &           &  $0.0007$ \\
            & $51936.6859 \pm 0.0013$ & $y$   &           &  $0.0017$ \\
            & $51936.6830 \pm 0.0011$ & $b$   &           &  $0.0006$ \\
            & $51936.6775 \pm 0.0026$ & $v$   &           & $-0.0015$ \\
            & $51936.68264 \pm 0.00155$ & mean\tablenotemark{a}  &       & $0.00042$ \\
2001 Jan 31 & $51940.6324 \pm 0.0004$ & $I_C$ & Primary   & $-0.0005$ \\
            & $51940.6331 \pm 0.0003$ & $V$   &           & $-0.0002$ \\
            & $51940.63275 \pm 0.00049$ & mean\tablenotemark{a}  &       & $-0.00033$ \\
2001 Nov 14 & $52227.8162 \pm 0.0011$ & $I_C$ & Primary   & $-0.0010$ \\
2001 Nov 18 & $52231.7679 \pm 0.0032$ & $I_C$ & Secondary & $-0.0012$ \\
2001 Nov 22 & $52235.7250 \pm 0.0026$ & $I_C$ & Primary   &  $0.0007$ \\
\enddata
\tablenotetext{1}{Heliocentric Julian Date (2400000+)}
\tablenotetext{a}{Mean of the measurements obtained from the individual filters.}
\tablecomments{$O-C$ values are with respect to the ephemeris for primary
minimum HJD 2,449,703.7504 $+$ 2.634727$E$ (see text).}
\end{deluxetable}

\subsection{Radial velocities and orbit solution\label{speedometry}}
We use our HET HRS observations (Table \ref{het-table}) to measure 
precise radial velocities for both components of the \objectname{V1174~Ori} binary. Here
we describe our procedure for measuring radial velocities, and discuss
our assessment
of potential systematic effects that may degrade the accuracy of the
masses that we derive from them.

\subsubsection{Radial velocities\label{radvels}}
We measure radial velocities by cross-correlating
each \objectname{V1174~Ori} spectrum with a high-S/N spectrum of the radial-velocity
standard \objectname{HD 18884} (see Table \ref{het-table}), for which we
adopt the heliocentric radial velocity of $-26.1$ km s$^{-1}$ as determined 
by \citet{udry99}, whose long-term monitoring of this star shows it to be
non-variable to a precision level of 0.3 km/s 
over nearly 300 observations.  We chose this late-type (M2) template
in the hopes of maximizing the cross-correlation peak of the faint, 
late-type secondary star.

\begin{figure}[ht]
\epsscale{0.8}
\plotone{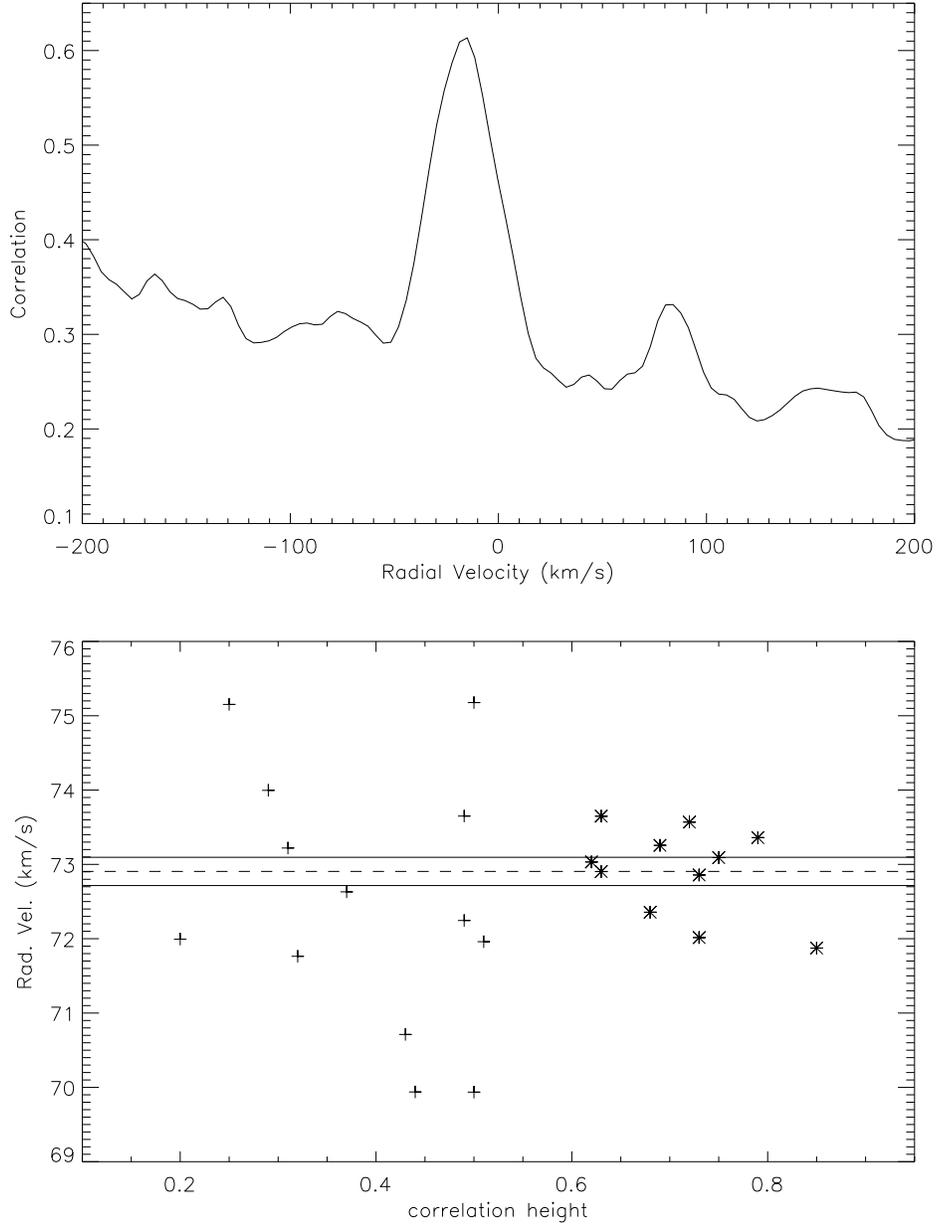}
\caption{\footnotesize Example of our procedure for measuring radial velocities from
cross-correlation.
(a) Shown is the cross-correlation function from a single spectral order
in the spectrum of UT 2002 Jan 28, in which we detect both the primary and
the secondary peak. In a typical spectrum, most of the orders yield
a velocity measurement for the primary, whereas only a few orders produce 
an unambiguous secondary peak, such as the one seen in this example.
(b) Shown as a function of correlation peak height are the radial velocities 
from the 24 orders in the spectrum of UT 2001 Dec 18 that give radial 
velocities (km/s) for the primary star. We use only velocities with correlation heights
greater than 0.6 (asterisks), which we found empirically yield more reliable
velocities. 
The horizontal lines indicate the mean velocity
of the primary (72.91 km/s) and the standard deviation of the mean (0.19 km/s);
see Table \ref{vel-table}.
\label{xcorl-fig}}
\end{figure}

We cross-correlate each of the 32 spectral orders in a given spectrum
separately, and measure the centroid of the primary and secondary star peaks 
(see Fig.\ \ref{xcorl-fig}a).
For each spectrum we thus in principle have 32 separate (but 
not independent) radial-velocity determinations for each of the
primary and secondary. In practice, however, some orders consistently produce
poor radial velocities, typically due to the presence of very 
broad lines (e.g.\ the Na D lines) which result in cross-correlation peaks
that do not lend themselves well to centroiding. In addition, we were 
unable to securely identify the peak corresponding to the
secondary star in most orders, due to other low-level structure in the
cross-correlation function. To be conservative, we thus typically 
only accepted a few very good orders, orders where the spectral features 
of the secondary are particularly strong. 
Finally, we found that it was necessary to avoid spectral regions 
with strong nebular emission, which often subtracted poorly. 
By carefully selecting the spectral regions to be used in each order
for each spectrum, we were able to maximize the number of orders yielding
radial velocities.

For the secondary star,
we use all orders that unambiguously show a secondary peak to define 
a mean radial velocity in each spectrum.
With the primary star, for which most of the spectral orders
yield a velocity measurement, we can be even more discriminating to
maximize our precision.  As shown in Fig.\ \ref{xcorl-fig}b, 
the scatter in the primary radial velocities produced by the
different spectral orders is a function of correlation peak height;
empirically, we find that the scatter is smallest among 
spectral orders that yield correlation peak heights above a certain 
threshhold value, typically around 0.6 (as in the example of 
Fig.\ \ref{xcorl-fig}b). 
We thus determine primary radial velocities from each spectrum by taking
the mean of only those spectral orders with correlation peak heights
greater than this threshold. 
Finally, we further refined the primary velocities
by iteratively excluding orders that deviated by greater than 2$\sigma$
from the mean, which typically resulted in the rejection of at most 
one order per spectrum.

In Table \ref{vel-table} we report the primary and secondary radial
velocities that we derive in this manner. In addition to the mean
velocities adopted for each star, we give the number of spectral
orders participating in defining the mean. 
In addition, we report the $O-C$
velocity residuals compared to our best-fit orbit solution (see 
below) for each of the primary and secondary velocities. 
The uncertainty reported with each velocity is simply the
standard deviation of the mean. 
These uncertainties, which average around 0.3 km/s for the primary
and 1.4 km/s for the secondary, thus represent an estimate of our internal 
errors only; we discuss instrumental and systematic errors in 
\S\S \ref{instrumental} and \ref{systematics}.

Some of our observations were taken at orbital phases near eclipse,
resulting in blended cross-correlation peaks that were difficult to
separate securely. Indeed, the velocity centroids in these cases can
be adversely affected for at least a couple of reasons. Cross-correlation peaks
in close proximity to one another can suffer from ``peak pulling", 
whereby blending of the two components can cause apparent
shifts in the peak centroids.  In addition, partial blocking of
the stellar disc of one star by the other causes the stellar line
profiles of the blocked star to become asymmetric, narrower, and shifted
with respect to the center-of-mass velocity.
For our orbit solution
analysis we thus discarded measurements obtained within 0.05 phase
of either eclipse, thereby ensuring that the correlation peaks would be
reasonably well separated and that occultation effects would not be
a concern. The only exception to this is the spectrum obtained on
UT 2002 Feb 21, which occurs at the phase where the secondary is almost
completely eclipsed by the primary, so that the secondary spectrum
is practically not present in the observed light.

\begin{figure}[ht]
\epsscale{0.9}
\plotone{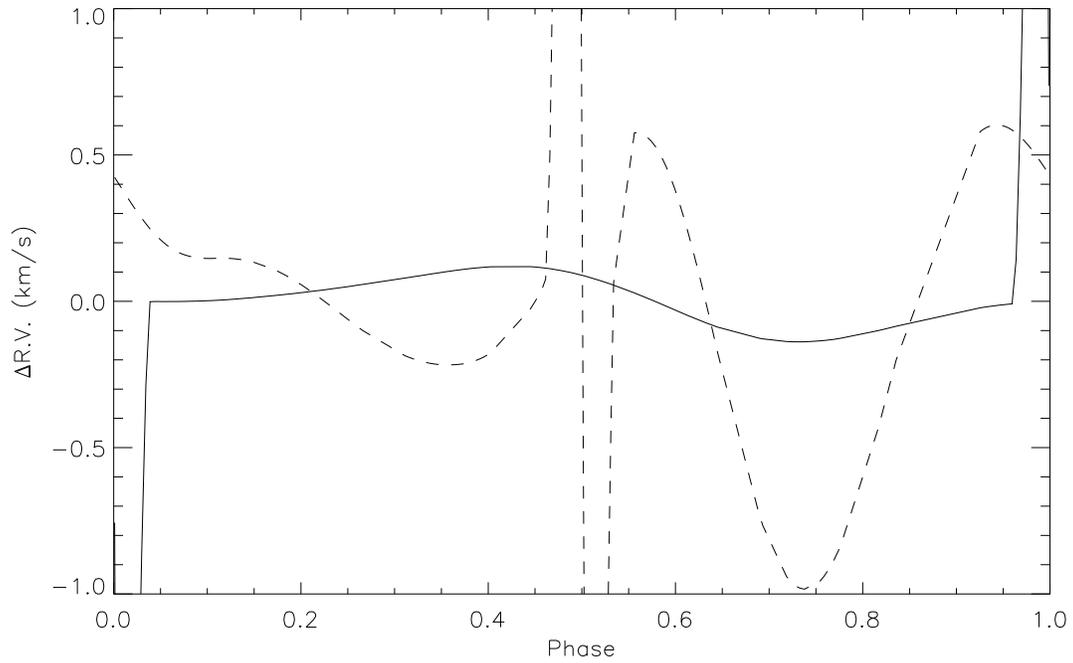}
\caption{Illustration of distortions in observed radial velocities due to
stellar occultation and the presence of surface spots. Distortions in the
primary (secondary) star's radial velocities are shown by the solid (dashed) 
line. Note the strong distortion effects due to occultation at $\pm 0.05$
phase around each eclipse ($\pm 3.7$ km/s and $\pm 4.8$ km/s for the
primary and secondary, respectively).
\label{rv-effects-fig}}
\end{figure}

The presence of spots on the stellar surfaces can also introduce distortions that
affect the observed radial velocities. The importance of this effect for our
observations can be evaluated from the light curve solutions discussed
below, which include the effects of spots. Our model fits
to the light curves obtained in January 2001 show that the effect is
rather small. The theoretical radial-velocity distortions due both to
stellar occultation (the Schlesinger-Rossiter effect; \citet{rossiter,schlesinger}) 
and the presence of surface spots are shown in
Fig.\ \ref{rv-effects-fig}.
For the primary star, the radial velocity distortions due to spots
are undetectable at the precision of our study, having an amplitude
of at most 0.2 km/s, and less at most orbital phases. 
For the secondary star, the distortions are more pronounced, but still have
an amplitude no larger than 1 km/s, and only at certain phases. 
As the light curve data were not obtained simultaneously with the
radial velocity data, we do not correct the
observed velocities for these distortions, noting again that the
effect is in any case probably negligible.

\begin{deluxetable}{rcrrrrrrrr}
\tablewidth{0pt}
\tabletypesize{\scriptsize}
\tablecolumns{10}
\tablecaption{Radial velocities of V1174~Ori\label{vel-table}}
\tablehead{ 
\colhead{\#\tablenotemark{a}} & \colhead{UT Date} & \colhead{HJD\tablenotemark{b}} &
\colhead{Phase} & \colhead{${{{\rm R.V.}_P}\atop{\rm (km/s)}}$} &
\colhead{${{(O-C)_P}\atop{\rm (km/s)}}$} &
\colhead{${{{\rm R.V.}_S}\atop{\rm (km/s)}}$} &
\colhead{${{(O-C)_S}\atop{\rm (km/s)}}$} &
\colhead{N$_P$} & \colhead{N$_S$} 
}
\startdata
2 & 2001 Nov 23 & 2236.84827 & 0.427 & $-9.87 \pm 0.44$ & $-0.14$ & $71.48 \pm 1.56$  & $-2.16$ & 11 & 2 \\
3 & 2001 Dec 10 & 2253.80559 & 0.863 & $83.44 \pm 0.27$ & $0.02$  & \nodata\phantom{.56} & \nodata & 12 & 0 \\
4 & 2001 Dec 18 & 2261.78887 & 0.893 & $72.91 \pm 0.19$ & $0.04$  & $-46.57 \pm 1.23$ & $-5.75$\tablenotemark{c} & 11 & 3 \\
6 & 2001 Dec 22 & 2265.76822 & 0.404 & $-19.90 \pm 0.32$ & $-0.14$ & $88.17 \pm 1.29$ & $0.68$  & 15 & 6 \\
7 & 2001 Dec 23 & 2266.76687 & 0.783 & $101.81 \pm 0.34$ & $0.73$ & $-80.36 \pm 1.29$ & $-0.91$ & 15 & 7 \\
8 & 2002 Jan 26 & 2300.67336 & 0.652 & $88.81 \pm 0.22$ & $-0.57$ & $-62.39 \pm 1.43$ & $0.89$  & 17 & 2 \\
10 & 2002 Jan 28 & 2302.66839 & 0.409 & $-16.27 \pm 0.23$ & $1.31$ & $86.28 \pm 1.91$ & $1.81$   & 20 & 3 \\
11 & 2002 Feb 07 & 2312.64124 & 0.194 & $-48.06 \pm 0.24$ & $-0.87$ & $126.16 \pm 1.29$ & $0.77$ & 17 & 3 \\
12 & 2002 Feb 09 & 2314.63298 & 0.950 & $48.14 \pm 0.30$ & $-0.73$ & $-5.87 \pm 0.98$ & $1.45$   & 17 & 2 \\
14 & 2002 Feb 21 & 2326.61211 & 0.497 & $23.23 \pm 0.29$ & $-0.34$ & \nodata\phantom{.56} & \nodata  & 20 & 0 \\
15 & 2002 Feb 23 & 2328.60823 & 0.254 & $-52.17 \pm 0.27$ & $0.30$ & $131.84 \pm 1.36$ & $-0.85$ & 17 & 8 \\
\enddata
\tablenotetext{a}{See Table \ref{het-table}.}
\tablenotetext{b}{Heliocentric Julian Date (2450000+)}
\tablenotetext{c}{Measurement discarded in preferred solution (see text).}
\end{deluxetable}

\subsubsection{Orbit solution\label{orbit}}
We determined an orbit solution to the radial-velocity measurements in
Table \ref{vel-table} 
by applying a least-squares fit simultaneously to the primary and secondary 
velocities. 
The orbit solution involves seven parameters: the mass ratio, $q$;
the orbital period, $P$; the semi-amplitudes of the primary and secondary 
velocities, $K_A$ and $K_B$; the eccentricity, $e$; the center-of-mass
velocity, $\gamma$; and the time of primary minimum, $T_0$.

With the system ephemeris determined precisely from the eclipse timing
data (see \S \ref{ephemeris}), we chose to fix the values of $P$ 
and $T_0$ to those determined photometrically, and forced the orbit to
be exactly circular (i.e.\ $e \equiv 0$). 
The assumption of circularity is corroborated
by the eclipse timings analysis (\S \ref{ephemeris}), and is
certainly expected for a short-period binary such as this. In any case,
we did try a solution with $e$ as a free parameter, but found a
best-fit value that was statistically insignificantly different from zero.

Using all of the radial velocities resulted in a fit with r.m.s.\
residuals for the primary and secondary of $\sigma_{\rm A} = 0.66$ km/s
and $\sigma_{\rm B} = 2.55$ km/s, respectively. However, the measurement
of the secondary's velocity from the UT 2001 Dec 18 spectrum showed a
residual of $-5.75$ km/s, considerably larger than the other measurements.
Indeed, a close look at the distribution of the secondary residuals
revealed this measurement to be a clear outlier; excluding this 
measurement yielded a fit with $\sigma_{\rm B} = 1.41$ km/s, consistent
with the internal errors of the measurements.

We carefully re-examined (and even reprocessed) the spectrum of
UT 2001 Dec 18 and the cross-correlation functions to see if we could 
discern any reasons for the large discrepancy of this measurement.
The primary's measurement is not obviously
discrepant, and neither is the velocity of the standard star observed
on this night (see \S \ref{systematics}). 
We note however that this spectrum does have
the poorest S/N of all our HET HRS observations (Table \ref{vel-table}). 
Interestingly, our light curve analysis (\S \ref{lightcurves}) suggests that 
spots might be significant near the orbital phase of this measurement
(see Fig.\ \ref{rv-effects-fig}).

\begin{figure}[ht]
\epsscale{0.9}
\plotone{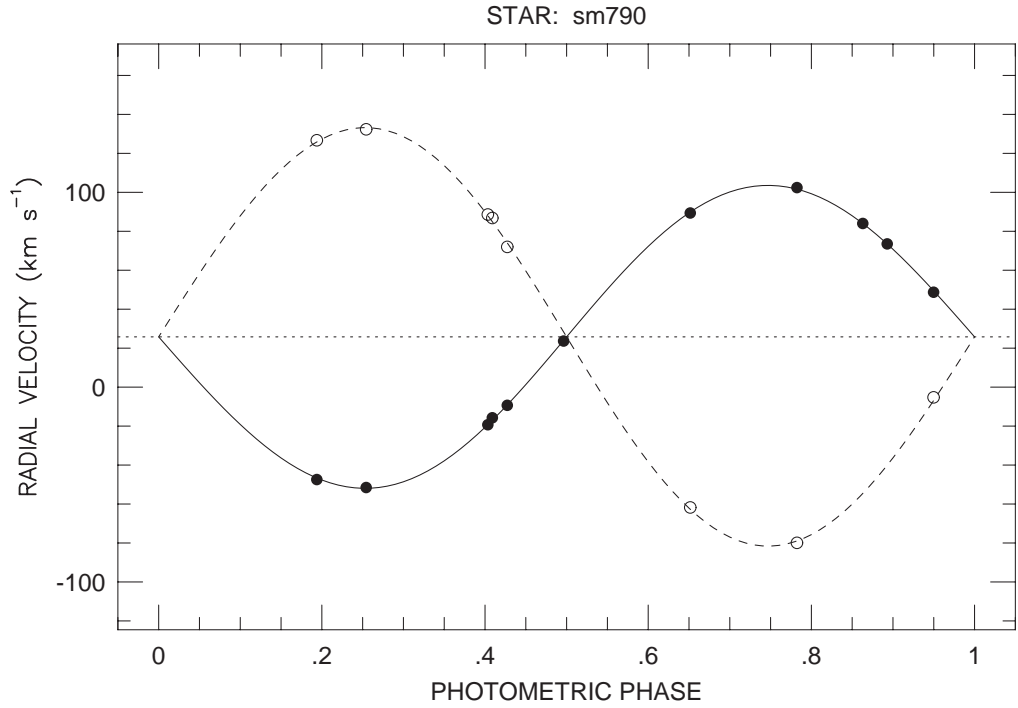}
\caption{Orbit solution for V1174~Ori. Radial velocity measurements
(see Table \ref{vel-table}) for the primary (secondary) star are shown as
filled (open) points. The solutions for the primary and secondary stars are
shown as solid and dashed lines, respectively.
\label{orbit-fig}}
\end{figure}

In Table \ref{orbit-table} we present the orbital parameters 
of \objectname{V1174~Ori} resulting from our best-fit orbit solution, and
display this solution along with the radial-velocity measurements in 
Fig.\ \ref{orbit-fig}.

\begin{deluxetable}{cc}
\tablewidth{0pt}
\tabletypesize{\scriptsize}
\tablecolumns{2}
\tablecaption{Results of V1174~Ori orbit solution\label{orbit-table}}
\tablehead{ \colhead{Parameter} & \colhead{Value} }
\startdata
$P$ (d) & $2.634727 \pm 0.000004$ \\
$\sigma_A$ (km s$^{-1}$) & 0.66 \\
$\sigma_B$ (km s$^{-1}$) & 1.41 \\
\tableline
$q$ & 0.7238 $\pm$ 0.0057 \\
$e$ & 0\tablenotemark{a} \\
$\gamma$ (km s$^{-1}$) & 25.27 $\pm$ 0.19 \\
$K_A$ (km s$^{-1}$) & 77.75 $\pm$ 0.29 \\
$K_B$ (km s$^{-1}$) & 107.41 $\pm$ 0.67 \\
$a \sin i$ (R$_\odot$) & 9.638 $\pm$ 0.041 \\
$M_A \sin^3 i$ (M$_\odot$) & 1.005 $\pm$ 0.015 \\
$M_B \sin^3 i$ (M$_\odot$) & 0.728 $\pm$ 0.008 \\
\enddata
\tablenotetext{a}{Adopted.}
\end{deluxetable}

\subsubsection{Instrumental stability\label{instrumental}}
Our spectroscopic observations allow us to determine an orbit solution
with high precision, that in turn results in very precise ($\sim 1$\%) 
determination of the stellar masses (Table \ref{orbit-table}). As
noted above, the residuals of the radial velocity measurements about
the best-fit orbit solution are $\sigma_A = 0.66$ km s$^{-1}$ and
$\sigma_B = 1.41$ km s$^{-1}$ for the primary and secondary,
respectively. Evidently, the instrumental stability of the HET HRS
is at least as good as 0.65 km s$^{-1}$.

\begin{figure}[ht]
\epsscale{0.85}
\plotone{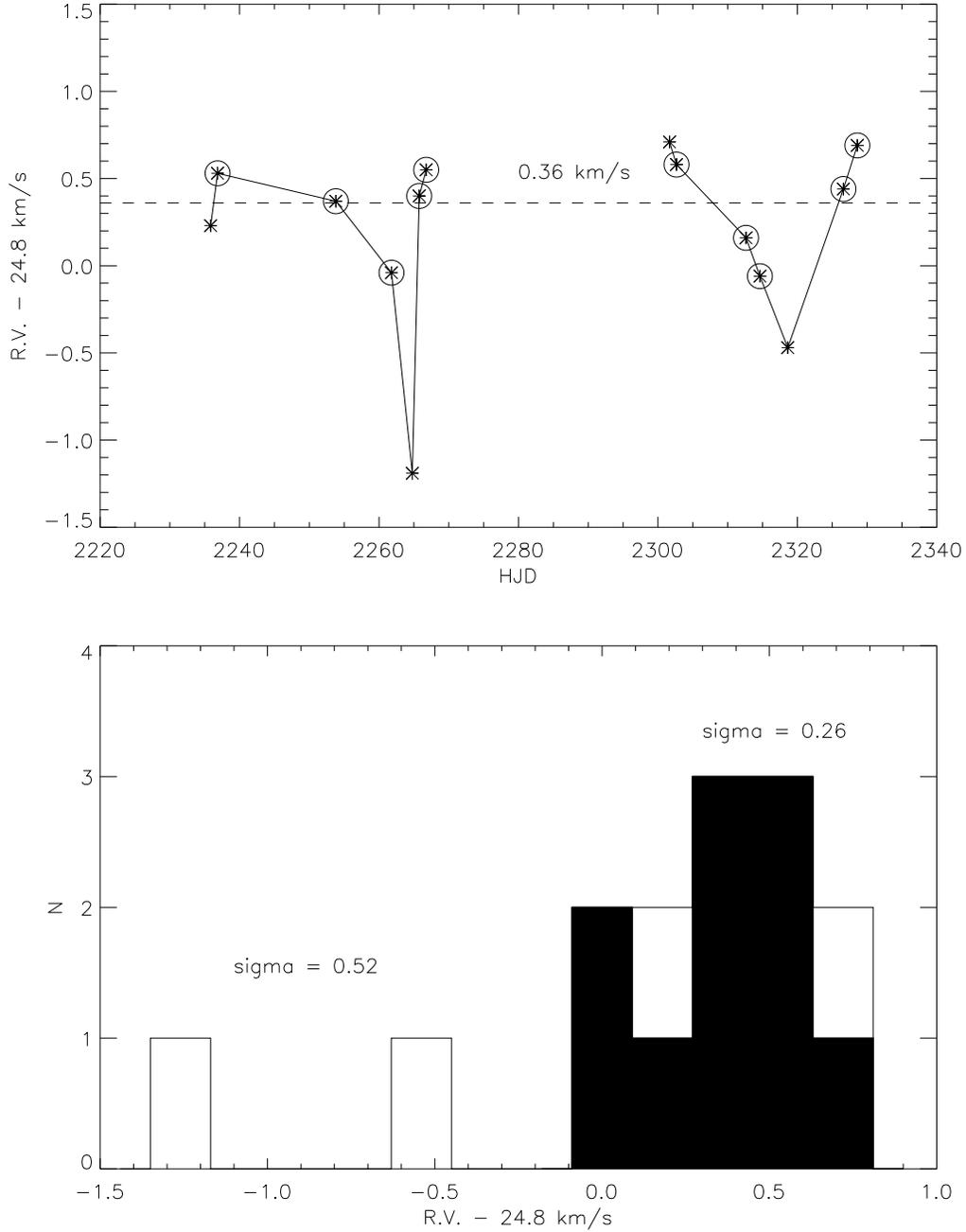}
\caption{\footnotesize Our assessment of mechanical stability of the HET HRS. (a) Shown
as a function of heliocentric Julian date are the radial velocities of the
radial-velocity standard HD~26162 (see Table \ref{vel-table}).
Circled points represent spectra from the same nights that are used in our
orbit solution of V1174~Ori. The horizontal dashed line indicates the mean
of these points, which show a small systematic offset of 0.36 km/s,
likely due to spectral mismatch between HD~26162 and the
standard star (HD~18884) used in our orbit solution of
V1174~Ori. (b) These same radial velocities of HD~26162 are
shown as a histogram (open), with the circled points in (a) represented by
the filled histogram. These latter measurements show an r.m.s.\ scatter of
0.26 km/s.
\label{systematics-fig}}
\end{figure}

To assess the instrumental stability more carefully,
we analyzed our observations of the radial-velocity standard star
\objectname{HD 26162}, which was observed along with \objectname{V1174~Ori}, in the
same way as our \objectname{V1174~Ori} observations. Specifically, we cross-correlated
each \objectname{HD 26162} spectrum against the same radial-velocity
template (\objectname{HD 18884}) that we used to derive the \objectname{V1174~Ori}
velocities. The results are tabulated in Table \ref{systematics-table}
and shown graphically in Fig.\ \ref{systematics-fig}.

The velocity deviations of \objectname{HD 26162} relative to its
published velocity of 24.8 km s$^{-1}$ \citep{udry99} reveal a high level of
stability of the HET HRS system over the 90-day timescale spanned
by our observations. As Fig.\ \ref{systematics-fig} shows, the system
does exhibit low-level, secular drifts, but the dispersion of these
drifts is small. The deviations have an rms of 0.52 km s$^{-1}$
when we consider all of the nights on which \objectname{V1174~Ori} was observed, 
and an rms of 0.26 km s$^{-1}$ when we consider only those nights used in 
our orbit solution\footnote{The lower rms of this subset of the radial-velocity
measurements could be due to chance, given the apparently non-random nature
of the deviations shown in Fig.\ \ref{systematics-fig}.}.

This rms of 0.26 km s$^{-1}$ is
only about 1/2 of the rms residuals in our orbit solution fit to the primary
velocities (0.65 km s$^{-1}$), and about 1/5 of the secondary (1.41 
km s$^{-1}$). Evidently, some other source of measurement uncertainty
dominates the errors in the radial velocity measurements of \objectname{V1174~Ori}.
These could include: the lower S/N of the \objectname{V1174~Ori} spectra
as compared to the \objectname{HD 26162} spectra (Table \ref{het-table}),
particularly for the secondary star; the double-peaked nature of the
\objectname{V1174~Ori} cross-correlation functions; differences in the track length
of the \objectname{V1174~Ori} observations as compared to the \objectname{HD 26162}
observations (1800 sec vs.\ 10 sec); a source of astrophysical
noise (e.g.\ surface activity such as spots, flares, etc.); and/or
the presence of an unseen tertiary companion (see \S \ref{results-fundamental}). 
In any case, instrumental stability does not appear to be the dominant 
source of uncertainty in our radial velocity measurements.

\subsubsection{Systematic errors\label{systematics}}
For this experiment, where the primary goal is to determine {\it accurate}
stellar dimensions for comparison to stellar evolutionary models, it
is critical that we ascertain the extent to which systematic effects may
limit the accuracy of the absolute stellar dimensions that we determine.

The semi-amplitudes of the primary and secondary velocities ($K_A$ and
$K_B$) are the parameters from the spectroscopic orbit solution that 
directly determine the stellar masses, so the masses are directly 
sensitive to systematics in these quantities. 
One possible source of systematic error in these quantities is spectral-type 
mismatch between \objectname{V1174~Ori} and the radial-velocity template.

There is in fact some evidence for a systematic offset of the secondary
velocities relative to the primary. As Fig.\ \ref{systematics-fig}a
shows, the velocities of the radial-velocity standard 
\objectname{HD 26162} show a median velocity in our observations that is
0.36 km/s larger than the published value. We speculate that this offset
is due to the fact that \objectname{HD 26162} has an earlier spectral type
(SpT = K2 III) than the radial-velocity template star (SpT = M2 III).
Our best estimate for the spectral types of the components of \objectname{V1174~Ori}
are K4.5 and M1.5 (see \S \ref{spt}), so a similar systematic shift 
might be expected between the measured velocities for the two components
of \objectname{V1174~Ori}, leading to systematic errors in the radial velocity
semi-amplitudes.

We tried an orbit solution where we allowed for an arbitrary systematic
offset between the primary and secondary velocities to be fit. 
We find a best-fit value of $0.35 \pm 0.55$ km s$^{-1}$ for the offset, 
which is interesting in its similarity to the offset of 0.36 km s$^{-1}$
we find above. 
Including this free parameter and its uncertainty results in slightly
different stellar masses than those listed in Table \ref{orbit-table},
and with slightly larger uncertainties: 
$M_A \sin^3 i = 1.004 \pm 0.016 \; {\rm M}_\odot$ and 
$M_B \sin^3 i = 0.727 \pm 0.009 \; {\rm M}_\odot$.
Thus, while our orbit solution formally allows for an offset of order
1 km s$^{-1}$ at the $2\sigma$ level, it is not statistically significant
and its effect on the final uncertainties on the derived stellar masses
is very small.
For our subsequent analyses and discussion, we adopt the orbit solution
provided in Table \ref{orbit-table}. 

\begin{deluxetable}{cc}
\tablewidth{0pt}
\tabletypesize{\scriptsize}
\tablecolumns{2}
\tablecaption{Radial velocities of HD 26162\label{systematics-table}}
\tablehead{ \colhead{\#\tablenotemark{a}} & 
\colhead{$\displaystyle{\rm Radial~Velocity}\atop{\rm (km/s)}$} }
\startdata
1 & 25.03 $\pm$ 0.08 \\
2 & 25.33 $\pm$ 0.08 \\
3 & 25.17 $\pm$ 0.07 \\
4 & 24.76 $\pm$ 0.07 \\
5 & 23.61 $\pm$ 0.08 \\
6 & 25.20 $\pm$ 0.07 \\
7 & 25.35 $\pm$ 0.07 \\
9 & 25.51 $\pm$ 0.07 \\
10 & 25.38 $\pm$ 0.07 \\
11 & 24.96 $\pm$ 0.08 \\
12 & 24.74 $\pm$ 0.09 \\
13 & 24.33 $\pm$ 0.07 \\
14 & 25.24 $\pm$ 0.07 \\
15 & 25.49 $\pm$ 0.08 \\
\enddata
\tablenotetext{a}{See Table \ref{het-table}.}
\end{deluxetable}

\subsection{Spectral analysis\label{spectral}}
Having determined precise dynamical masses from our spectroscopic
orbit solution, we now require stellar radii and effective temperatures 
for a complete analysis of the \objectname{V1174~Ori} system and for comparison with
PMS stellar evolutionary tracks. Our light curve analysis (\S \ref{lightcurves})
provides precise values for the radii, the inclination, and the {\it ratio\/} of 
effective temperatures. Thus
we need to establish the effective temperature of one of the components
independently.

In this section we conduct a more detailed spectral analysis to determine 
the primary star's effective temperature. To begin, we measure the 
rotational velocities of the two stars and determine their flux ratio. 
This information guides our spectral classification of the primary
by establishing the rotational broadening of the primary spectrum and the
extent to which the secondary spectrum contaminates the lines that we
use to classify the primary. 
The rotational velocity and flux ratio information will also prove useful 
in constraining our light curve analyses below. 
Finally, we measure the strength of
two spectral features of interest in studies of young stars, Li and H$\alpha$,
the former of which serves to establish stellar youth and the latter of
which serves as a proxy for chromospheric activity and/or accretion.

\subsubsection{Rotational velocities\label{vsini}}
We measure the rotational velocities, $v \sin i$, of the two components of
\objectname{V1174~Ori} in two ways. First, the cross-correlation functions we used to
measure radial velocities (\S \ref{radvels}) contain rotational information
in the sense that the peak widths are related to $v \sin i$. Second, we
use the ratio of the widths of the primary and secondary Li lines to
constrain the ratio of $v \sin i$ of the two components.

Because the widths of the cross-correlation peaks reflect the convolution
of the intrinsic stellar $v \sin i$ with both the instrumental response
function and the radial-velocity template spectrum, the peak widths do
not directly yield a measure of $v \sin i$. We calibrated the width
of the cross-correlation peak to $v \sin i$ by artificially rotationally
broadening our narrow-lined (lines unresolved) spectral standards (see
Table \ref{het-table})
and then cross-correlating these broadened spectra against the same
radial-velocity template used with our \objectname{V1174~Ori} spectra. This was done
separately for each spectral order. 
We find that our cross-correlation functions
permit us to discern $v \sin i$ values of $\gtrsim 10$ km/s.

Using this calibration, we measured $v \sin i$ values for the components
of \objectname{V1174~Ori} from each spectrum, averaging together the values from the
multiple orders in a given spectrum. The resulting $v \sin i$ values are
$26.2 \pm 0.6$ km/s and $20.5 \pm 1.6$ km/s for the primary and secondary,
respectively. Here the uncertainties are the standard deviation of the
mean.

We wish to emphasize that the uncertainties quoted above are strictly 
internal. Furthermore, we caution that
we have not used $v \sin i$ standards 
for an external calibration. For our present purposes, what is robust
here is that the rotational velocities for the stars are $\sim$ 20--25 km/s,
and that the primary star rotates more rapidly than the secondary. This
latter point is corroborated by our analysis of the primary and
secondary Li lines below (\S \ref{lithium}).

\subsubsection{Flux ratio\label{fratio}}
We use the temperature-insensitive \ion{Ca}{2} lines at
$\lambda\lambda$6102,6122 to measure the flux ratio between the primary
and secondary components of \objectname{V1174~Ori}. We fit the \objectname{V1174~Ori} spectrum of 
UT 2002 Feb 07 by adding together at various ratios the spectra of our 
K3 and K7 standards\footnote{Our use of the K3 and K7 standards is based
on the preliminary spectral types that we determined for the two components 
of \objectname{V1174~Ori}, but as
the Ca lines used are temperature-insensitive any two standards of
roughly mid-K to early-M spectral type would suffice.} 
(see Table \ref{het-table}) at the appropriate radial velocities and
with the appropriate rotational broadening. We chose the UT 2002 Feb 07
spectrum because on this date the secondary was shifted a full 174 km/s
(3.5\AA\ at this spectral region) redward of the primary. The two 
components are thus well separated, and the $\lambda$6102 line of the
secondary is shifted into a region of the spectrum (6105.5\AA) that is 
free of lines from the primary.

\begin{figure}[ht]
\plotone{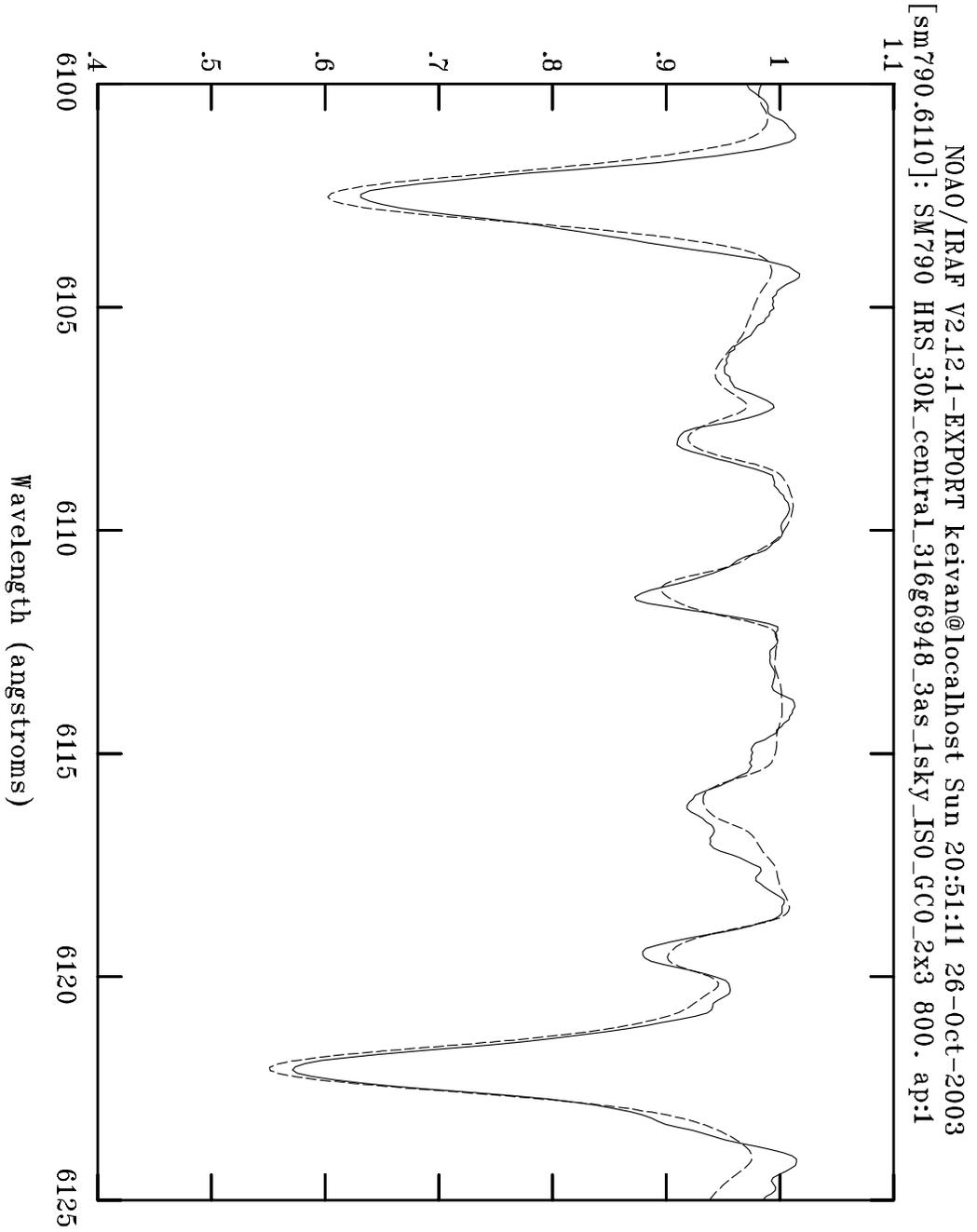}
\caption{Determination of the primary-to-secondary flux ratio for
V1174~Ori. The UT 2002 Feb 07 spectrum (solid line) is compared with a
synthetic spectrum produced by combining the K3 and K7 standards in a
6:1 ratio (dashed line). The standards were first rotationally broadened 
(see \S \ref{vsini}) and shifted to the appropriate radial velocities
(see Table \ref{vel-table}).
\label{fratio-fig}}
\end{figure}

We find a best fit to the \ion{Ca}{2} lines in the \objectname{V1174~Ori} spectrum 
when we combine the two templates in a 6:1 ratio,
(Fig.\ \ref{fratio-fig}), 
although ratios
of 5:1 and 7:1 also fit the observed spectrum reasonably well.

\subsubsection{Spectral classification\label{spt}}
Our high-resolution HET HRS spectra permit spectral classification of
the primary from careful examination of temperature-sensitive line ratios. 
As demonstrated by \citet{basri90} and by \citet{lee}, the ratio of 
\ion{Ni}{1} $\lambda$6108 to \ion{V}{1} $\lambda$6112, the ratio 
of \ion{V}{1} $\lambda$6040 to \ion{Fe}{1} $\lambda$6042, the
ratio of \ion{V}{1} $\lambda$6058 to \ion{Fe}{1} $\lambda$6056, and the
ratio of \ion{Sc}{1} $\lambda$6210 to \ion{Fe}{1} $\lambda$6200
provide a good determination of spectral class from high-resolution spectra. 
These line ratios were also used by
\citet{steffen} in their analysis of the PMS binary
\objectname{NTT 045251+3016}.

\begin{figure}[ht]
\epsscale{0.8}
\plotone{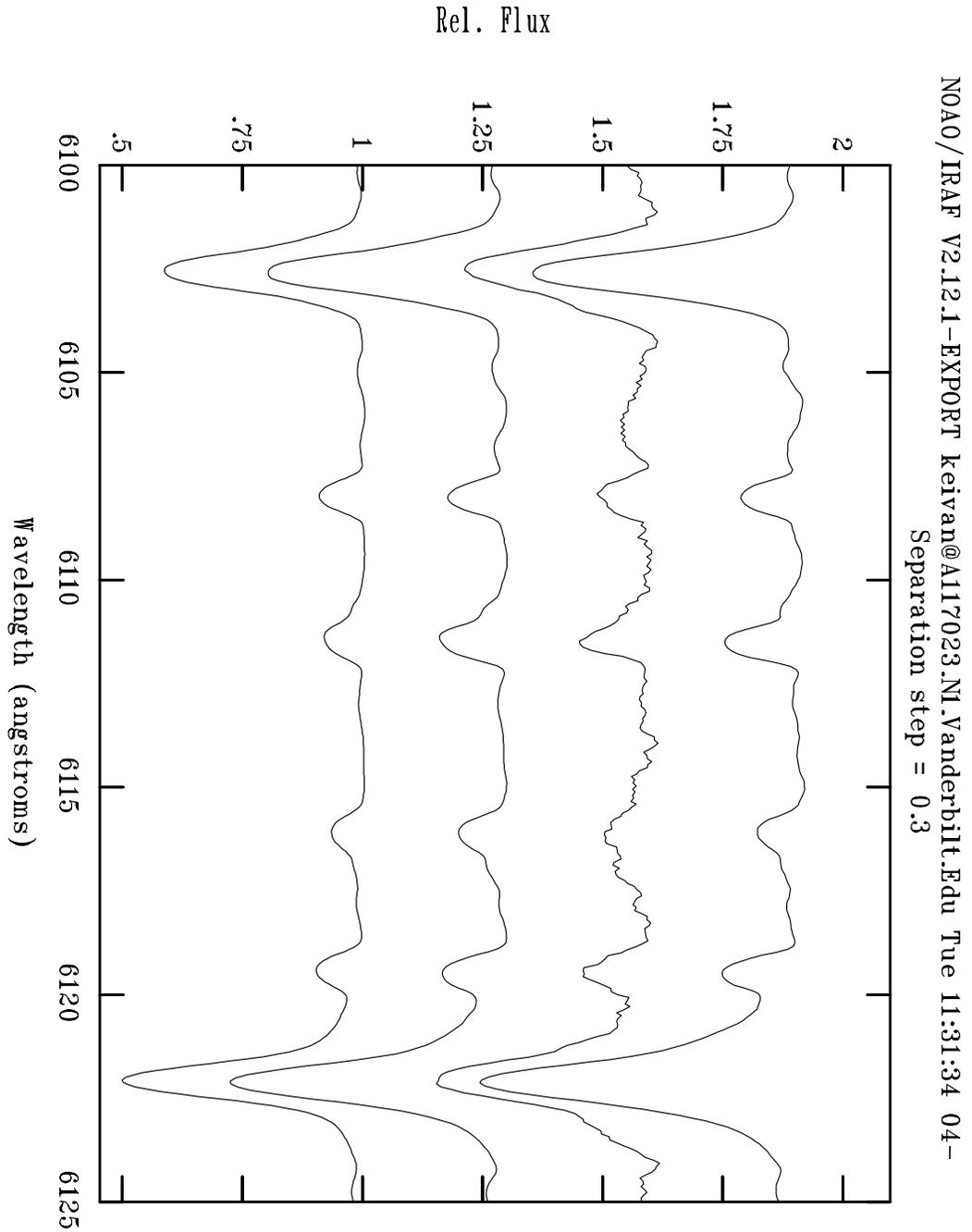}
\caption{Shown is the UT 2002 Feb 07 spectrum of V1174~Ori (second from
top) compared with three spectral standards of type K3 (bottom), K4
(second from bottom), and K5 (top).
The spectral region shown includes the temperature-sensitive lines of 
\ion{Ni}{1} at $\lambda$6108 and \ion{V}{1} at $\lambda$6112, one of
four line pairs that we use to determine the spectral type of the
primary of V1174~Ori. Note that in this spectrum the primary star's
\ion{V}{1} is slightly contaminated by the secondary star's \ion{Ni}{1}
line. Also visible here are the temperature-insensitive lines of \ion{Ca}{2}
at $\lambda\lambda$6102,6122. In this spectrum of V1174~Ori, the secondary
star's $\lambda$6102 \ion{Ca}{2} line is visible as the weak absorption feature 
between the primary star's $\lambda$6102 \ion{Ca}{2} and $\lambda$6108 \ion{Ni}{1}
lines. The $\lambda$6102 \ion{Ca}{2} lines in this spectrum indicate a 
primary-to-secondary flux ratio of $\sim$ 6:1 for V1174~Ori 
(see Fig.\ \ref{fratio-fig}).
\label{spt-fig}}
\end{figure}

We have carefully examined these line ratios in our spectra that
have the secondary sufficiently Doppler-shifted with respect to the primary.
We find a spectral type for the primary of K4--K5, with an uncertainty of 
$\sim$1/2 subclass. As an example,
in Fig.\ \ref{spt-fig} we compare the first line pair above from our HET HRS
spectrum of UT 2002 Feb 07 with four spectral standards of type
K1, K3, K4, and K5. The observed line ratio indicates a K4.5 spectral
type, but after correcting the primary's \ion{V}{1} line for 
contamination by the secondary's \ion{Ni}{1} line (using the flux
ratio determined above), we find a K4 spectral type in this spectrum.

Taking all of these line-pair ratios from all of our observations into 
account, we find a mean spectral type for the primary of K$4.5 \pm 0.1$. 
Due to
the quantized nature of the spectral classification process (0.5
sub-type), this uncertainty is likely unreastically small, so we adopt
a more conservative uncertainty of 0.5 spectral sub-type. 
Using the SpT-$T_{\rm eff}$ conversion of \citet{schmidt-kaler}, this implies an
effective temperature for the primary of $T_{\rm eff} = 4470 \pm 120$ K.

Using this together with the ratio of $T_{\rm eff}$ determined from our 
light curve analysis below (\S \ref{lightcurves}), we infer that
the secondary has an effective temperature of $T_{\rm eff} = 3615$ K,
implying a spectral type of $\sim$M1.5.

\subsubsection{Lithium\label{lithium}}
The $\lambda$6708 line of Li is a commonly used indicator of stellar
youth, as stars rapidly destroy their natal Li content during the
first $\sim$few$\times 10^7$ yr of their evolution. For very young
stars, the Li line at 6708\AA\ can be very strong, with
equivalent widths from $\sim 300$ m\AA\ for early K stars to 
$\sim 750$ m\AA\ for mid-M stars \citep{dolan99}. Indeed,
in our HET HRS spectra, the Li line is one of the few lines from
the secondary that can be unambiguously identified by eye.

We have measured the equivalent widths of the primary and secondary
Li lines in those spectra where the two stars are sufficiently 
Doppler-shifted relative to one another. 
As an example, 
the spectrum of UT 2002 Jan 28 
has the secondary at a radial velocity of $+102.55$ km/s relative to the
primary. The Li equivalent width (EQW) we measure for the primary
star in this spectrum is EQW(Li)$_{\rm A} = 398$ m\AA. 
We measure the Li EQW of the secondary star to be 
EQW(Li)$_{\rm B} = 78$ m\AA\ after correcting for the contribution
of spectral features from the primary. These EQW measures also include
a 5\% correction for the presence of third light, based on our light
curve analysis (\S \ref{third}) which indicates 0\% 
third light at $V$ and 10\% at $I$.

If intrinsically equal, these EQW values would imply a flux ratio of 5.1. However,
adopting spectral types of K4.5 and M1.5 for the two components implies
that the intrinsic EQW of the secondary's Li line is $\sim 25$\%
greater than that of the primary \citep{dolan99}. 
Correcting the secondary's EQW(Li)
accordingly results in a flux ratio of 6.3, consistent with that
derived from our analysis of the temperature-insensitive lines above
(\S \ref{fratio}). Thus from this spectrum we 
infer intrinsic equivalent widths for the two stars of
EQW(Li)$_{\rm A} \approx 465$ m\AA\ and EQW(Li)$_{\rm B} \approx 545$ m\AA.
The other spectra for which we could cleanly separate the primary and
secondary Li features yield values consistent with these. Adopting a
flux ratio of 6:1, we find from the ensemble of our spectra
EQW(Li)$_{\rm A} = 480 \pm 14$ m\AA\ and
EQW(Li)$_{\rm B} = 537 \pm 26$ m\AA.

We can use these EQW measurements to infer Li abundances for the two
stars. From the Li curve-of-growth analysis of \citet{pavlenko}, we find 
$N$(Li)$_A = 3.08 \pm 0.04$ and $N$(Li)$_B = 2.20 \pm 0.06$ assuming LTE, and 
$N$(Li)$_A = 3.00 \pm 0.05$ and $N$(Li)$_B = 1.98 \pm 0.09$ for the non-LTE 
case. Note that these abundance estimates are likely to be slightly 
lower than the actual abundances, as the curve-of-growth values used 
assume $\log g = 4.5$, whereas the components of \objectname{V1174~Ori} have 
$\log g \approx 4.2$ (see \S \ref{results-fundamental}). In addition,
the uncertainties listed reflect only the error in the measurement
of the EQW; including the uncertainty in the stellar \teff\ increases
the abundance uncertainties to 0.2 dex.

Thus, the primary star's Li abundance is consistent with at most
0.2--0.3 dex of depletion (assuming a ``cosmic" Li abundance of 
$\log N$(Li)$=3.3$), whereas the secondary shows $> 1$ dex of
depletion.
As discussed by \citet{dantona-li}
the pattern of PMS Li depletion as a function of stellar mass and age
can provide important constraints on
PMS tracks. We will return to the issue of Li abundances in
\S \ref{luhman}.

Finally, we can also use the primary and secondary Li lines to infer the
relative rotation rates of the two stars. We have already derived 
$v\sin i$ values from the widths of the cross-correlation peaks (\S \ref{vsini}).
Using the ratio of the widths of the Li lines, we can independently 
infer the ratio of $v\sin i$.
We have measured this ratio from our spectra that have the primary and
secondary Li lines sufficiently well separated to allow a clean 
measurement of their FWHM, and find a $v\sin i$ ratio of $1.21 \pm 0.04$,
consistent with the $v\sin i$ values determined from the cross-correlation
analysis.

\subsubsection{$H\alpha$\label{ha}}
Emission of the H$\alpha$ line is also of interest in studies of 
young stars. Indeed, emission stronger than EQW(H$\alpha$) $\approx 10$ \AA\ 
is a defining criterion for classical TTS, which are
believed to be PMS stars actively accreting from circumstellar disks.
Even non-accreting TTS often show some amount of H$\alpha$ in emission,
which is thought to indicate the presence of strong chromospheric
activity.

\begin{figure}[ht]
\epsscale{1.0}
\plotone{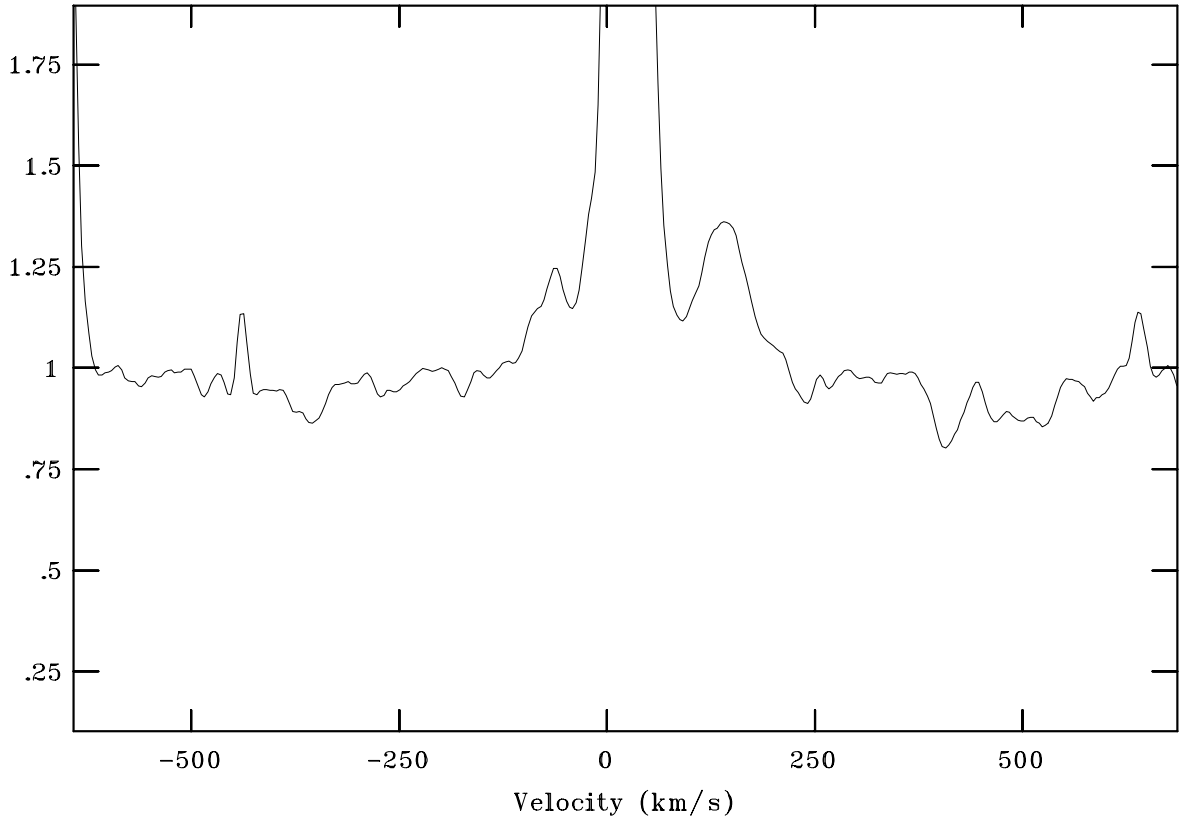}
\caption{Determination of H$\alpha$ equivalent widths for the two components
of V1174~Ori. Shown is the UT 2002 Feb 23 spectrum, plotted vs.\ heliocentric
radial velocity. The strong emission line is nebular. Emission from the primary 
and secondary stars are visible just to the blue and red of the nebular line,
respectively.
\label{ha-eqw-fig}}
\end{figure}

Based on the strength of H$\alpha$ emission, \objectname{V1174~Ori} does not appear 
to be an actively accreting system, with neither component exhibiting
H$\alpha$ emission in excess of a few \AA\ EQW in any of our spectra. 
In Fig.\ \ref{ha-eqw-fig},
we show the spectrum of UT 2002 Feb 23 in the vicinity of H$\alpha$. 
This spectrum has the components of \objectname{V1174~Ori} at their greatest
radial-velocity separation, making it possible to detect their individual
H$\alpha$ emission apart from the very strong nebular emission. The
observed equivalent widths measure EQW(H$\alpha$)$_{\rm A} = -0.4$\AA\
and EQW(H$\alpha$)$_{\rm B} = -4.2$ \AA, after correcting for the
continuum of the companion.

\subsection{Light curve synthesis and modeling\label{lightcurves}}
Having determined an ephemeris for V1174~Ori, and having
conducted a thorough spectral analysis of the system including 
determination of a double-lined orbit solution, we now analyze 
our multi-epoch, multi-band light curves (Table \ref{phottable},
Figs.\ \ref{IVybvlc}--\ref{JHKI94res}) in detail to determine 
the orbital inclination of the system, \teff\ for the secondary, and absolute 
radii for both components. Our analysis procedure makes use of a
Wilson-Devinney (WD) based algorithm that models the observed photometric
variations with synthetic light curves that include the effects of limb
darkening, gravity brightening, mutual illumination and occultation, and spots. 
In addition, 
the code permits the use of either simple Planckian fluxes or detailed
model atmospheres. Along the way, we use 
information provided by our spectroscopic analyses 
($v \sin i$, flux ratio, primary \teff) 
to further constrain the model light curve parameters. 

\subsubsection{Initial considerations}
Although the light curves shown in Figs.\ \ref{IVybvlc},
\ref{IVBI01lc} and \ref{JHKI94lc} are typical for
a system with non-distorted components, apparently well-detached from their
Roche lobes, they also reveal the presence of variations not related to
the eclipses. From careful inspection of the light curves 
it becomes evident that these variations are most likely due to star spots 
that change from one epoch to the next, hindering the use of our photometric 
data from all epochs in a single, integrated analysis. 

Thus, to begin, we used the data from the epoch with the most 
complete set of light curves---those obtained in Jan 2001 in 5 filters
($vbyVI$)---to produce an initial set of light curve solutions. 
These initial light curve solutions were calculated without spots 
by fitting
simultaneously the 5 light curves from Jan 2001 together with the
radial velocity data from Table \ref{vel-table}. 

Our analysis was performed with the WD light-curve synthesis program
\citep{wd71,wd93a,wd93b}. Besides other modifications
\citep{wd95,tycra}, our version of the WD code was implemented with the
possibility of using the SIMPLEX solution method \citep{kallrath87,vieira03}
as an alternative to the traditional least-squares method.
Limb-darkening coefficients for both eclipsing components
were described using the data from \citet{vanhamme93}. All
three limb-darkening laws (linear, linear-log, and linear-square root)
were examined, with the coefficients calculated by
bi-linear interpolations using the current values of $\log g$ and
$T_{\rm eff}$. The linear law gave marginally better solutions and
was therefore adopted. Bolometric albedoes for both components were set 
equal to 0.5 as appropriate for atmospheres in convective equilibrium. 
Gravity-brightening exponents were calculated using the local value
of \teff\ for each point on the stellar surfaces, taking into account
mutual illumination, according to \citet{beta1} and
\citet{beta2}. The radiated flux of both
stars is described using the atmosphere model tables of
\citet{buserkurucz92}, updated relative to the original versions
of the WD code. As the \teff\ of the secondary is lower than
the minimum \teff\ included in the published tables (4000 K), we extrapolated
linearly in a log(theoretical magnitude) vs.\ log(temperature) scale.
Both the radial velocities and the photometric
times of minima indicated no sign of orbital eccentricity, 
so we assumed circular orbits in the light curve analysis.

In performing these initial calculations, we
applied least-squares differential corrections 
between successive iterations to the orbital inclination, the
secondary \teff, the primary luminosity, both stellar surface
gravitational pseudopotentials (stellar radii), the center-of-mass
radial velocity, the mass ratio and an arbitrary phase
shift to be added to the
observations. The luminosity of the secondary was calculated from its
size and \teff. In addition, the limb-darkening coefficients,
normalization magnitudes, surface gravities, and individual velocity
amplitudes were all updated between successive runs to 
correspond to the solution from the previous iteration.

Initially, the WD code settled on a solution in which the secondary star was 
larger than the primary. However, this solution violated the observed luminosity 
ratio (\S \ref{fratio}) as well as the observed rotational velocities if
synchronous rotation is assumed. By forcing the luminosity ratio
constraint, the code found another
solution in which the primary is larger. This solution matches the observed
luminosity ratio and predicts rotational velocities similar to those
observed. This solution thus served as a starting point for introducing
additional parameters into the analysis, such as spots, as we now describe.

\subsubsection{Star spots\label{spots}}
After determining an initial solution as described above, we introduced
spots in the WD calculation in order to model the out-of-eclipse
variations observed.
We started by introducing one cold spot on each component. For each spot
on each star, 
four parameters are added to the WD calculation: the longitude and 
co-latitude of the spot, the spot radius (as seen from
the center of the star), and the spot temperature factor ($< 1$ for cold spots
and $> 1$ for hot spots). 

In this phase of the calculation, we found the SIMPLEX method to
be much more efficient in finding best-fit solutions than the least-squares
method, the latter being very sensitive to small corrections to the spot
parameters, leading to frequent divergence of the solution.
Although the WD algorithm can deal with spots that move in longitude across the 
surface of the star, we assumed that the spots rotate with the same angular 
velocity as the star. After
convergence of each set of spots, we tried many different combinations of
spot number and
configurations, trying to minimize the number of spots on each star.
Note that the WD code treats regions in which multiple spots overlap by 
multiplying their temperature 
factors\footnote{This differs from, e.g., the {\sl Nightfall\/} program,
which uses the average of the temperature factors. See
\url{http://www.lsw.uni-heidelberg.de/users/rwichman/Nightfall.html} .}.

Ultimately, both ``cold'' 
and ``hot''
spots were needed on each star to reproduce
the light curves of Fig.\ \ref{IVybvlc}. 
The cold spots made it
necessary to extrapolate in the atmosphere tables, as mentioned 
above for the secondary star, for some points on the surface of the 
primary star as well.
The calculations with atmosphere tables were compared with calculations
using blackbodies, and we concluded that the extrapolation done
(linearly in a logarithmic scale) was reasonable. Thus, except for the $JHK$
light curves for which \cite{buserkurucz92} did not publish atmosphere calculations,
we thus chose to use atmosphere models instead of simple Planck functions. 

There exist multiple combinations of spots that can reproduce the
observed out-of-eclipse variations reasonably well. 
We believe that the 3 spots on each star used in our final solution 
(Table \ref{spotsall}) are physically realistic. Moreover,
our use of light curves in five different bands makes the spot solution
more robust, as we insisted that one spot configuration reproduce all of
the light curves simultaneously. 
In any case, the goal of including spots
in the light curve modeling is not to determine the properties of the
spots per se, but rather to better constrain the morphology of the
light curves around the eclipses and thereby to minimize the uncertainty
in the parameters of true import: the system inclination, the ratio of
\teff, and the stellar radii.

\subsubsection{Third light\label{third}}
The Jan 2001 light curves proved impossible to fit together 
unless we included $\sim$10\% 
third light in $I$. 
The $V$ light curve required a slightly {\it negative} amount of third light,
which is physically meaningless. We thus forced $L_3(V) \equiv 0$, 
but left this parameter free to be adjusted at the other wavelengths. The
final solution is shown in Figs.\ \ref{IVybvlc}
and \ref{IVybvres} and in Tables \ref{wd5} and \ref{spotsall}.
With the exception of $I$, the significance of the $L_3$ values found for the
light curves
is questionable as they do not follow a Planckian distribution with 
wavelength. They are probably artifacts of the data-reduction process 
(e.g.\ background subtraction in the aperture photometry, scattered light, etc.).

\subsubsection{Final solutions\label{lcfinal}}

\begin{deluxetable}{lclr}
\tablewidth{0pt}
\tabletypesize{\scriptsize}
\tablecolumns{4}
\tablecaption{Final light curve solution for V1174~Ori
\label{wd5}}
\tablehead{ \colhead{parameter} & \colhead{value} & \colhead{parameter} & \colhead{value} }
\startdata
$i$ (\degr)      & 86.974 $\pm$ 0.035 & $r_{\rm A}$ pole & 0.1384 $\pm$ 0.0006 \\
$\Omega_{\rm A}$ & 7.942 $\pm$ 0.026  & \phantom{$r_{\rm A}$} point & 0.1392 $\pm$ 0.0006\\
$\Omega_{\rm B}$ & 7.705 $\pm$ 0.022  & \phantom{$r_{\rm A}$} side & 0.1387 $\pm$ 0.0006\\
$T{\rm A}$ (K) & 4470. (fixed)        & \phantom{$r_{\rm A}$} back & 0.1391 $\pm$ 0.0006\\
$T{\rm B}$ (K) & 3615.0 $\pm$ 9.5     & $r_{\rm B}$ pole & 0.1101 $\pm$ 0.0012\\
$L_3(I)$ (\%) & 9.90 $\pm$ 0.71       & \phantom{$r_{\rm A}$} point & 0.1106 $\pm$ 0.0012\\
$L_3(V)$ (\%) & 0.00 (fixed)          & \phantom{$r_{\rm A}$} side & 0.1103 $\pm$ 0.0012\\
$L_3(y)$ (\%) & 1.73 $\pm$ 0.65       & \phantom{$r_{\rm A}$} back & 0.1106 $\pm$ 0.0012\\
$L_3(b)$ (\%) & 0.03 $\pm$ 0.65       & F$_{\rm out,A}$ & 0.4247 $\pm$ 0.0012 \\
$L_3(v)$ (\%) & 2.47 $\pm$ 0.66       & F$_{\rm out,B}$ & 0.4548 $\pm$ 0.0014 \\
$v_{\rm ratio,A}$& 1.017 $\pm$ 0.026     & \\
$v_{\rm ratio,B}$& 1.001 $\pm$ 0.079     & \\
\enddata
\tablecomments{WD solutions for V1174 Ori, based on the
observations from Jan 2001. Parameters like the limb darkening
coefficients, gravity darkening exponents, and surface gravities were kept
consistent during iterations. The solutions were attained by combining
both the least-squares and the SIMPLEX methods. The errors quoted are the
internal least-squares estimations, which are typically around 3-5 times
larger then those estimated from the vertices of the SIMPLEX hyperpolyhedron
after convergence. The third light contributions are given as percentages
of the eclipsing system total light at the normalization phase (0.25).
The radii resulting from the gravitational effective equipotentials
(in terms of the orbital separation) and the fill-out factors, F$_{\rm out}$
\citep{mochnacki}, are given, which indicate that the eclipsing components
are well inside their Roche lobes. 
The errors quoted on the radii and fill-out factors represent the deviations
caused by the uncertainties in the pseudopotentials and in the mass ratio.
The rotation rates, calculated
from the solutions and the measurements of \S \ref{vsini} are shown,
justifying the assumption of synchronism with the orbit.}
\end{deluxetable}

\begin{deluxetable}{lclcccc}
\tablewidth{0pt}
\tabletypesize{\scriptsize}
\tablecolumns{5}
\tablecaption{Spot parameters used in light curve solutions 
\label{spotsall}}
\tablehead{\colhead{epoch} & \colhead{bands} & \colhead{spot/star} & \colhead{co-latitude (\degr)} & \colhead{longitude (\degr)} & \colhead{radius (\degr)} & \colhead{$T_{\rm factor}$} }
\startdata
Dec/1994&  $I$  &~~1 / A & ~69. $\pm$ 13. & 140.5 $\pm$ 8.6\, & ~7.2 $\pm$ 1.6 & 0.69 $\pm$ 0.24 \\
        &       &~~1 / B & 70.4 $\pm$ 7.3 & ~25.3 $\pm$ 5.5 & ~11. $\pm$ 11. & 1.44 $\pm$ 0.32 \\
2MASS   &  $JHK$&~~1 / A & ~21. $\pm$ 16. & 101.9 $\pm$ 9.6\, & ~27. $\pm$ 5.3 & 0.88 $\pm$ 0.18 \\
        &       &~~1 / B & ~23. $\pm$ 13. & \,336. $\pm$ 21. & ~48. $\pm$ 4.2 & 0.61 $\pm$ 0.14 \\
Jan/2001&$vbyVI$&~~1 / A & 18.2 $\pm$ 7.4 & ~~78. $\pm$ 10. & 15.2 $\pm$ 4.6 & 1.58 $\pm$ 0.74 \\
        &       &~~2 / A & 16.2 $\pm$ 5.7 & 110.8 $\pm$ 6.2\, & 26.0 $\pm$ 4.7 & 0.76 $\pm$ 0.25 \\
        &       &~~3 / A & 13.34 $\pm$ 0.69 & 159.6 $\pm$ 1.3\, & 31.2 $\pm$ 1.5 & 0.508 $\pm$ 0.042 \\
        &       &~~1 / B & 155. $\pm$ 10.\, & ~97.1 $\pm$ 5.5 & 30.2 $\pm$ 5.1 & 0.76 $\pm$ 0.12 \\
        &       &~~2 / B & ~67. $\pm$ 19. & 246. $\pm$ 10. & 17.4 $\pm$ 4.6 &  1.18 $\pm$ 0.11 \\
        &       &~~3 / B & ~76. $\pm$ 17. & 297.4 $\pm$ 5.6\, & 26.0 $\pm$ 5.6 & 0.662 $\pm$ 0.091 \\
Nov/2001&  $I$  &~~1 / A & ~48. $\pm$ 10. & 136.8 $\pm$ 9.6\, & 21.4 $\pm$ 5.2 & 0.74 $\pm$ 0.41 \\
        &       &~~1 / B & ~70. $\pm$ 17. & 134.6 $\pm$ 9.7\, & 12.4 $\pm$ 6.6 & 1.45 $\pm$ 0.91 \\
Nov/2002& $BVI$ &~~1 / A & 22.41 $\pm$ 0.42 &  \,93.3 $\pm$ 7.2 & 31.6 $\pm$ 1.2 & 0.851 $\pm$ 0.057 \\
        &       &~~1 / B & 16.35 $\pm$ 4.1~\, & 306.1 $\pm$ 9.1\, & 56.1 $\pm$ 5.7 & 0.69 $\pm$ 0.11 \\
        &       &~~2 / B & 21.99 $\pm$ 4.2~\, & 348.2 $\pm$ 9.8\, & 44.7 $\pm$ 8.4 & 0.605 $\pm$ 0.095 \\
\enddata
\tablecomments{The characteristics of the star spots
used in the solution of the light curves from Table \ref{phottable} and for
the $JHK$ light curves from 2MASS (one point obtained at March 19th, 1998 and
23 points from February 6th to March 28th, 2000).
The spots are listed for each component in each epoch with increasing
longitudes. As in Table
\ref{wd5}, the errors quoted are those internal to the least squares method.
The standard deviation of the corresponding variables of the SIMPLEX method,
after convergence, are much smaller: of the order of $10^{-3}$ degree for the
angular parameters and of $10^{-5}$ for the temperature factor for both
components. $T_{\rm factor}$ gives the spot temperature in units of \teff.}
\end{deluxetable}

\begin{figure}[ht]
\plotone{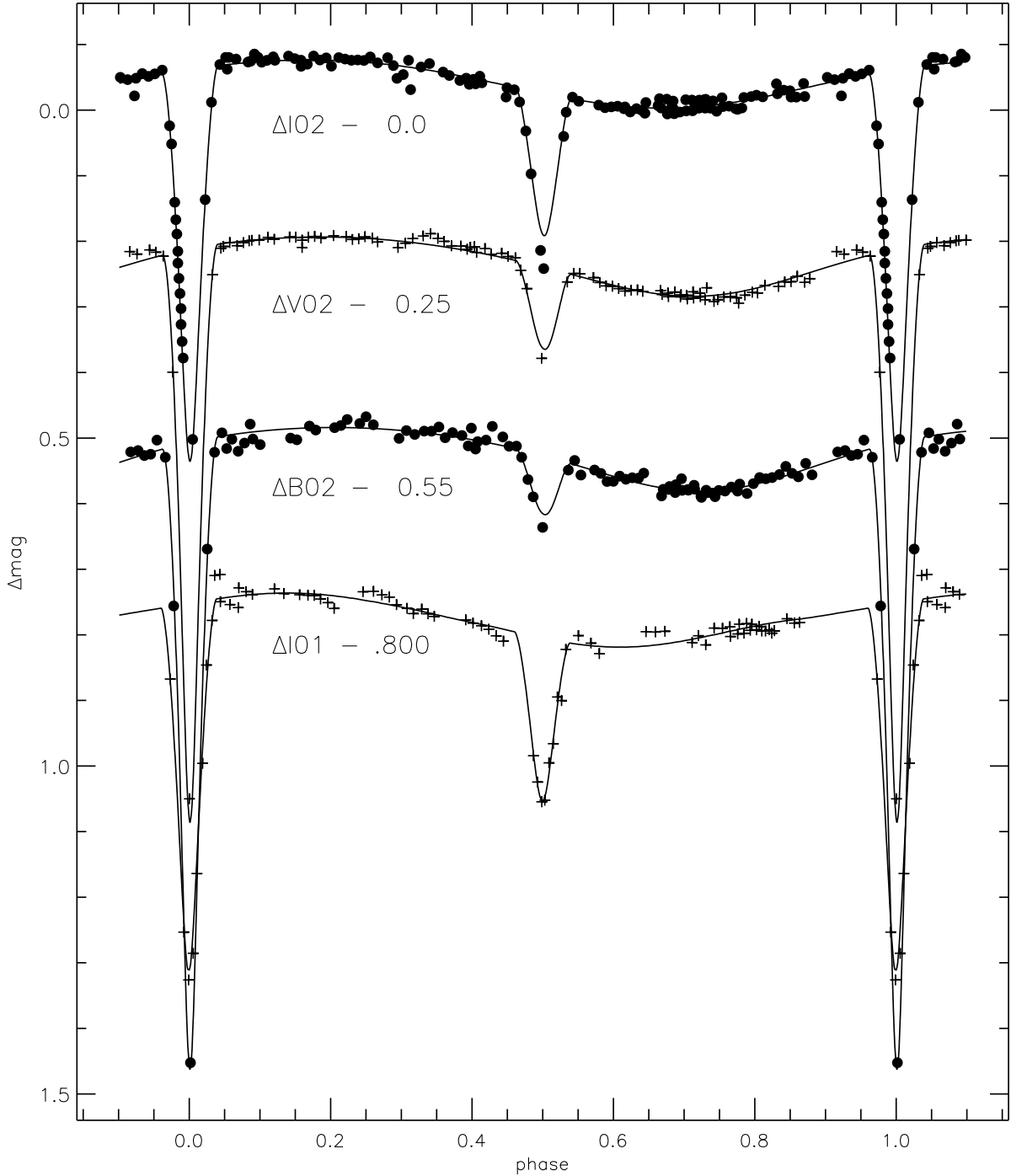}
\caption{The $IVB$ light curves observed in 2002 and the $I$
light curve obtained in 2001 (Table \ref{phottable})
together with the theoretical light curves obtained with the WD
algorithm (Tables \ref{wd5} and \ref{spotsall}). The
light curves are shifted by the amount indicated in the figure for
clarity, but the relative scale is the same for all passbands. Note the
changes in the depth of both eclipses with wavelength 
and the clear change in the form of the light curve outside eclipse from
2001 to 2002.
\label{IVBI01lc}}
\end{figure}

\begin{figure}[ht]
\plotone{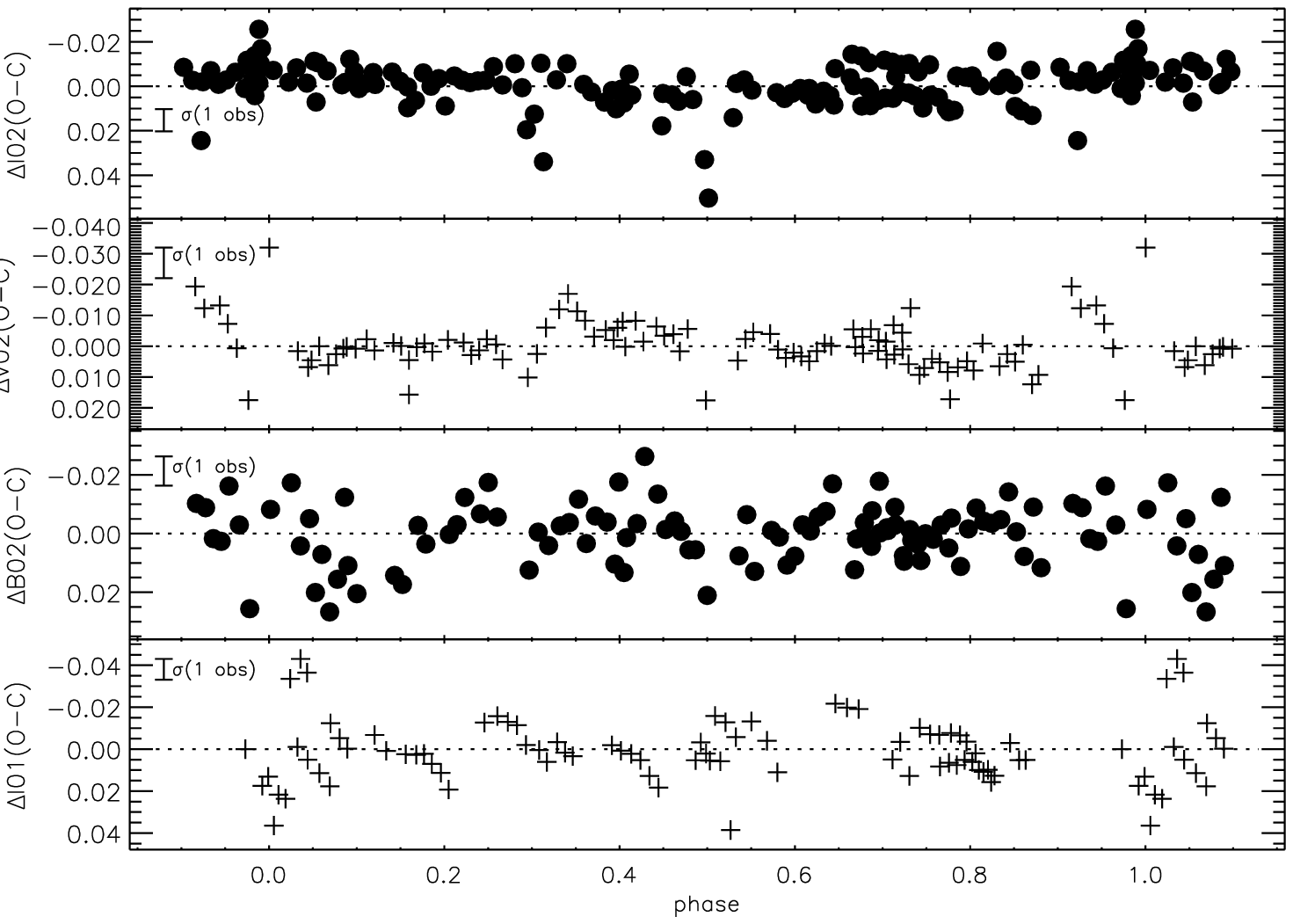}
\caption{The $O-C$ residuals for the solutions of Fig.\ \ref{IVBI01lc} and
Tables \ref{wd5} and \ref{spotsall}. The vertical bar at the
left of each panel is the standard deviation of the residuals.
\label{IVBres}}
\end{figure}

In order to test the validity of the stellar parameters found above from the
light curve solutions
for the Jan 2001 data, we applied the same analysis procedure
to the other sets of light curves (Table \ref{phottable}), now forcing
all the parameters, except those corresponding to the star spots, to remain
fixed. 
The resulting fits for the $I$ light curves obtained in Nov 2001
and the $BVI$ light curves obtained in Nov--Dec 2002 are shown in
Figs.\ \ref{IVBI01lc} and \ref{IVBres}. The characteristics of the
spots used are given in Table \ref{spotsall}.
The amount of third light at the different wavelengths were left as free 
parameters. $L_3(V)$ and $L_3(B)$ tended to become slightly
negative, while $L_3(I)$ was once again $\sim$~10\%. 
We thus used no third light for 
$V$ and $B$, and adopted the value for $L_3(I)$ shown in Table \ref{wd5}.

\begin{figure}[ht]
\plotone{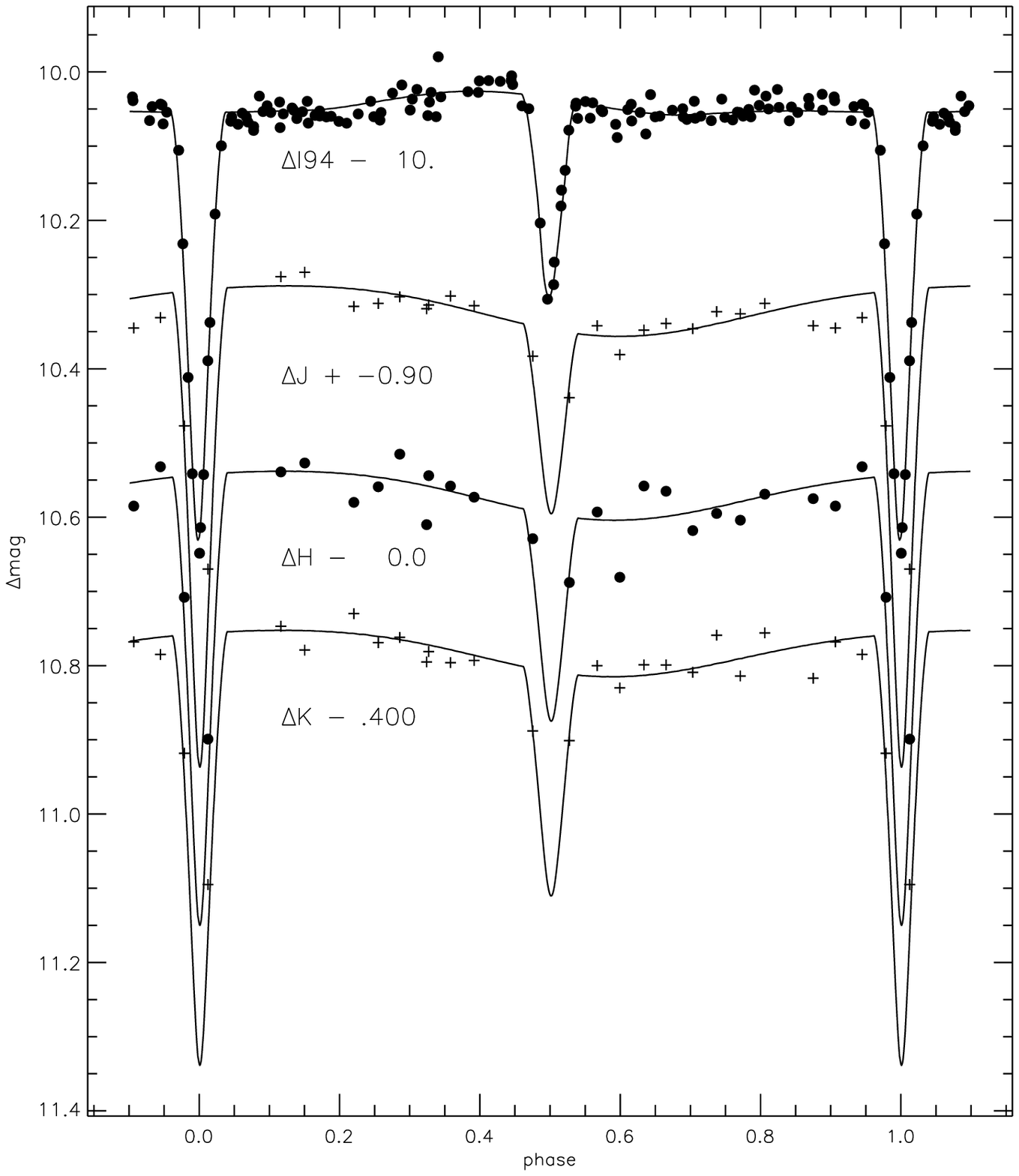}
\caption{The $I$ light curve obtained in Nov 1994 (Table \ref{phottable})
and the $JHK$ light curves obtained by \citet{carpenter01} together with
the theoretical light curves synthesized by the WD algorithm 
(Tables \ref{wd5} and \ref{spotsall}). The
light curves are shifted by the amount indicated in the figure for
clarity, but the relative scale is the same for all passbands. Note the
changes in the depth of both eclipses with wavelength 
and the clear change in the form of the
light curves outside eclipse from 1994 to 2000.
\label{JHKI94lc}}
\end{figure}

\begin{figure}[ht]
\plotone{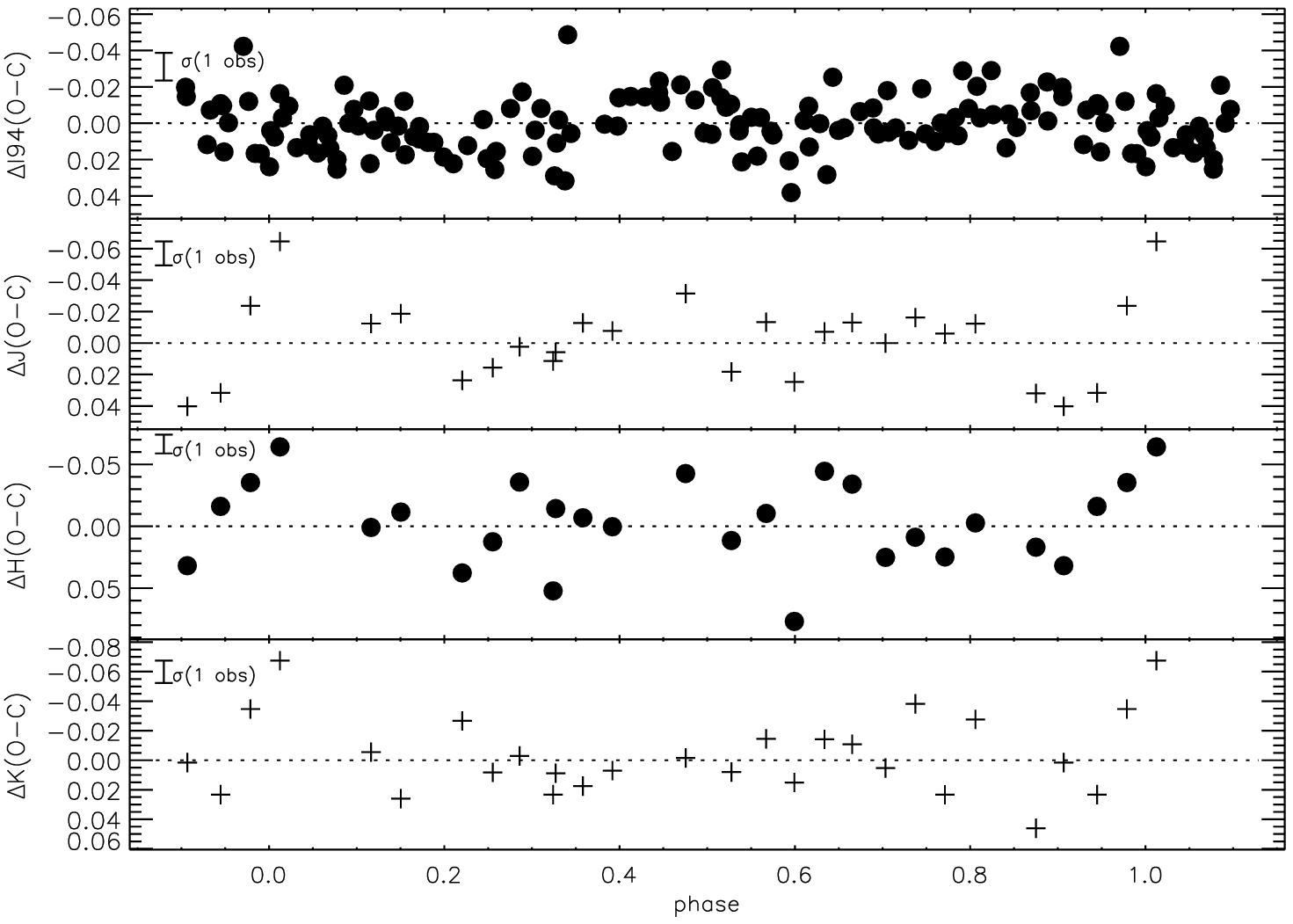}
\caption{The $O-C$ residuals for the solutions of Fig.\ \ref{JHKI94lc} and
Tables \ref{wd5} and \ref{spotsall}. The vertical bar at the
left of each panel is the standard deviation of the residuals.
\label{JHKI94res}}
\end{figure}

The $I$-band discovery light curve obtained in Dec 1994 (Table \ref{phottable})
was also fit
with the solution of Table \ref{wd5} and a different spot configuration. The
amount of third light was again fixed at the value shown in Table \ref{wd5}. Similarly,
the $JHK$ light curves from \citet{carpenter01}, while not providing complete
phase coverage, were also fit with this solution.
No third light was
included in the $JHK$ light curves because the third light parameter
mainly affects the depth of the minima, 
and unfortunately the minima were not
covered by the \citet{carpenter01} observations. 
The solutions are
shown in Figs.\ \ref{JHKI94lc} and \ref{JHKI94res}, and the spots used are
listed in Table \ref{spotsall}.

\begin{figure}[ht]
\plotone{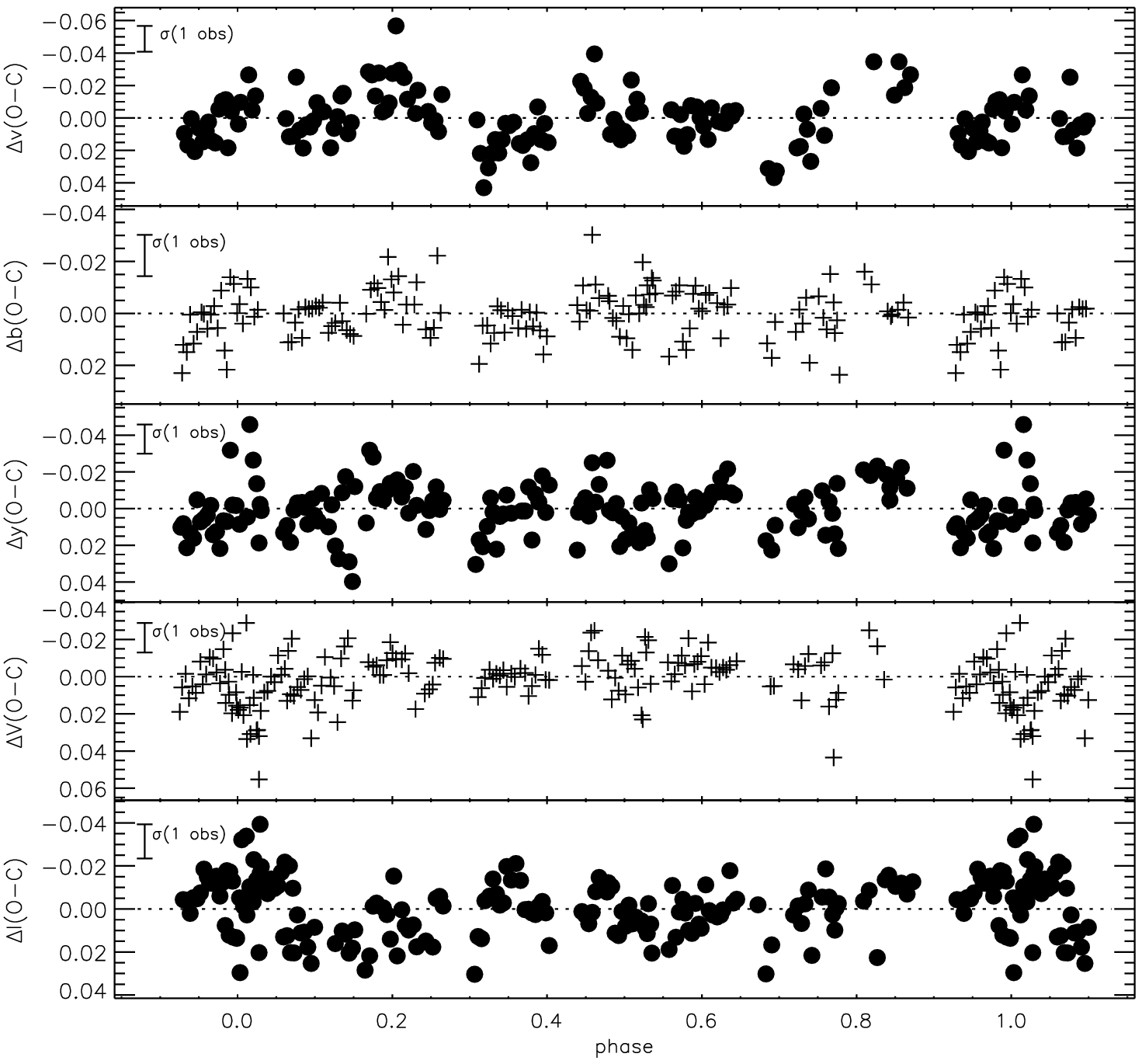}
\caption{The $O-C$ residuals for the solutions of Fig.\ \ref{IVybvlc} and
Tables \ref{wd5} and \ref{spotsall}. The vertical bar at the
left of each panel is the standard deviation of the residuals.
\label{IVybvres}}
\end{figure}

Overall, the quality of the synthetic light curve fits to the data is
excellent. Indeed, all 5 bands from the Jan 2001 observations are
fit extremely well, with only minor systematic effects visible in the
residuals (Fig.\ \ref{IVybvres}). Small systematic residuals do
persist in some of the light curves at other epochs. 
For example, in the Nov--Dec 2002 $I$-band data
there are 2 points that deviate by more than 0.02 mag relative to 
the theoretical light curve near secondary minimum
(Figs.\ \ref{IVBI01lc} and \ref{IVBres}). 
The data points close
to secondary minimum in $V$ and $B$ also show small
deviations, although at only $\sim 0.01$ mag these are not by themselves 
statistically significant. 
A few deviant data points also appear at other phases (e.g.\ phase 0.3 in
$I$). These might be indicative of erratic variations as have been reported in, 
e.g., AK~Sco \citep{aksco} and TY~CrA \citep{tycra}. Indeed, the most discrepant
point near phase 0.5 in the Nov--Dec 2002 $I$-band light curve was
obtained by itself more than two weeks later than the other points near
that phase; perhaps there was small amount of spot evolution or some other 
small change in the system over the course of the Nov--Dec 2002 epoch.

A limitation in our ability to analyze small-scale photometric variability 
over long timescales is that our differential photometric 
observations were not all reduced
in one integrated procedure, but separately for each epoch. 
The \citet{honeycutt} algorithm we employ determines the differential photometric 
variations relative to a statistically defined ensemble of comparison stars,
and this statistical ensemble is not necessarily common to all epochs.
Consequently, the 
photometric calibrations from one epoch to the next are not absolutely tied
to one another.
In both AK~Sco and TY~CrA, 
secular changes in system brightness from one epoch to the next
have been attributed to variable obscuration by dust. Unfortunately,
we cannot determine from our current analysis whether such global 
changes in system brightness
are taking place in \objectname{V1174~Ori} over long timescales.

Even so,
the light curve analyses presented here show clear evidence for changes with
time in both the number
and the characteristics of the spots on both components. The
time coverage of the observations was not adequate to follow the evolution
of the spots in detail, but there are hints in Table \ref{spotsall} that some of the
spots may persist on month timescales, drifting across the surfaces of the stars.
The hot spots apparently come and go more sporadically and, besides 
occurring more frequently on the secondary than on the primary, 
they tend to be smaller than the cold spots.
Finally, although the rotation rates of Table \ref{wd5} indicate
orbital synchronism within the errors, one of the reasons
for spot drifting may be that the stars are not yet completely synchonized. 
The stars may also present differential rotation, as in the
Sun and as recently determined for the rapidly-rotating RS CVn binary
UZ~Librae \citep{uzlib}. 

In the end, while some questions remain as to the origin of the variability
observed in some of our light curves at the finest level of detail, the overall 
match between the synthetic light curves and the data is excellent
(Figs.\ \ref{IVybvlc}--\ref{JHKI94res}), and for the remainder of our
analysis we adopt the solution listed in Table \ref{wd5}.

\section{Results\label{results}}

\subsection{Fundamental stellar parameters\label{results-fundamental}}
Combining our spectroscopic and photometric analyses presented above,
we can establish all of the fundamental stellar properties for both
components of \objectname{V1174~Ori}. We report these parameters in 
Table~\ref{fundamental}.

\begin{deluxetable}{lcc}
\tablewidth{0pt}
\tabletypesize{\scriptsize}
\tablecolumns{3}
\tablecaption{Fundamental stellar properties of V1174~Ori\label{fundamental}}
\tablehead{ \colhead{} & \colhead{Primary} & \colhead{Secondary} }
\startdata
$M$ (M$_\odot$)       & $1.009 \pm 0.015$ & $0.731 \pm 0.008$ \\
$R$ (R$_\odot$)       & $1.339 \pm 0.015$ & $1.065 \pm 0.011$ \\
\teff\ (K)         & $4470$\tablenotemark{a} $\pm 120$ & $3615\tablenotemark{b} \pm 100$\tablenotemark{c} \\
$\log L$ (L$_\odot$)  & $-0.193 \pm 0.048$ & $-0.761 \pm 0.058$ \\
$\log g$              & $4.19 \pm 0.01$    & $4.25 \pm 0.01$ \\
$v_{\rm rot}$ (km/s)  & $26.2 \pm 0.6$     & $20.5 \pm 1.6$ \\
$A_V$                 & \multicolumn{2}{c}{0.32}  \\
Dist.\ (pc)           & \multicolumn{2}{c}{$419 \pm 21$}  \\
\enddata
\tablenotetext{a}{From measured spectral type of K4.5 $\pm 0.5$, and 
SpT--\teff\ conversion of \citet{schmidt-kaler}.}
\tablenotetext{b}{Implies a spectral type of M1.5.}
\tablenotetext{c}{Note that the internal uncertainty on \teff\ for the 
secondary is minuscule. The uncertainty we assign here is inherited from
the uncertainty in the determination of \teff\ for the primary, keeping
the \teff\ ratio fixed at the value determined in \S \ref{lightcurves}.}
\end{deluxetable}

We can in principle determine the extinction, $A_V$, toward the system
by fitting the observed colors (Table \ref{calphot}), assuming main-sequence
intrinsic colors \citep{kenyon-hartmann95} and bolometric corrections
\citep{popper} for the primary and secondary components of \objectname{V1174~Ori}.
There is not a single value of $A_V$ that simultaneously reproduces 
all of the observed colors. 
Formally, however, we find a best-fit value of $A_V=0.32$,
which results in a distance to the
system of 419 pc when we normalize to the observed $V$ magnitude
(see \S \ref{distance} below). The 
colors calculated with this extinction and distance
are compared to the observed values in Table \ref{calphot}. 

The agreement between the observed and calculated colors is best in
$B-V$, where the observed flux is dominated by that of the primary star, and 
in $J-H$ and $H-K$, where the colors become largely degenerate for dwarfs
of mid-K to early-M spectral type. The observed $U-V$ color is bluer than
predicted by 0.11
mag, perhaps the result of a mild UV excess from one or both of the stars,
as is common among chromospherically active PMS stars, and possibly 
consistent with the observed H$\alpha$ emission (see \S \ref{ha})
and hot spots (\S \ref{spots}).

The observed $V-I$ color
is somewhat redder than predicted (by 0.14 mag), roughly consistent with the
level of third light inferred from our light curve modeling 
(\S \ref{lightcurves}) in the $I$ band ($\sim 10\%$). Similarly, the observed $J$
magnitude is 0.30 mag brighter than predicted. 

We can improve slightly the 
fit to the observed $V-I$ color and $J$ magnitude by increasing the $A_V$. 
For example, adopting $A_V = 0.42$ would predict $V-I = 1.52$ and $J = 11.46$.
Of course, this worsens the agreement in the other colors, most notably in
$U-V$, and results in a distance (401 pc) less compatible with that typically
assumed for the ONC ($\approx 470$ pc).

Together, the $V-I$ and $J$
discrepancies suggest the presence of a third, low-mass star in the system
that contributes flux noticeably only at $\lambda \gtrsim 0.9 \mu$m. This
would imply that \objectname{V1174~Ori} is in fact a hierarchical triple system, 
with the
tertiary in a sufficiently wide orbit so as to not manifest itself in the
observed radial velocities (\S \ref{radvels}).
Curiously, TY~CrA \citep{casey}, one of the other
three PMS eclipsing binaries in the literature, is also reported to be
a triple system.

\subsection{Distance to V1174~Ori\label{distance}}
Since we measure the luminosities of the components of 
\objectname{V1174~Ori} directly from \teff\ and $R$, we can infer a
distance to the system assuming a particular $A_V$ derived from the system
colors (see above).
The distance we derive to \objectname{V1174~Ori} of $419 \pm 21$ pc differs 
somewhat from that which has 
been quoted for the ONC in the recent literature. For example, \citet{hill97} 
adopt a distance of 470 pc, and \citet{bally} use a distance of 460 pc.

Various other authors have determined distances to the ONC and to the
various subgroupings of Orion OB1, ranging from 400--500 pc. Based on
photometry of several tens of bright O and B stars,
\citet{warrenhesser78} find a distance of 480 pc to the Trapezium cluster
(Ori OB1d) and 430 pc to the foreground Ori OB1c, while \citet{anthony-twarog},
using the same data, find a distance of 434 pc for Ori OB1d. Using
different photometric indices,
\citet{brown94} find a distance of only 400 pc for Ori OB1c. Perhaps
the most robust constraint is that provided by the BN-KL masers, 
situated just behind the ONC, at 480 pc \citep{genzel81}.

Given precisely determined luminosities for the two eclipsing components 
of \objectname{V1174~Ori}, our distance determination depends primarly upon $A_V$, with
lower values placing the system farther from the Sun. Based on our analysis of
the observed colors, we do not believe that $A_V$ can be much lower than
what we report above. Using the value of $A_V$ reported in Table
\ref{calphot}, the discrepancy between the observed and predicted 
$V-I$ color is already consistent with the amount of third light that we 
find for the $I$ band from our light curve analysis. Decreasing $A_V$ would
increase the discrepancy in $V-I$, which would be incompatible with our
light curve solution.

Once the value of $A_V$ is determined, we compute the distance via the
distance modulus in $V$. Our $V$-band
measurement has a formal uncertainty of 0.03 mag, which translates into
an additional distance uncertainty of 6 pc.
Of course, this uncertainty does not include any possible influence of
spots which, from our analysis above (\S \ref{spots}) has been observed
to be $\sim 0.1$ mag peak-to-peak in $V$. Our calibrated photometry was
obtained in 1995, an epoch at which we unfortunately do not have light
curves that could inform us as to the influence of spots on the calibrated
magnitudes. If, for example, a cool spot
were acting to depress the $V$-band flux by 0.1 mag, then our derived
distance would be underestimated by 20 pc.

Another, if subtler, consideration in our distance determination is the
value assumed for the solar bolometric magnitude, $M_{{\rm bol,}\odot}$.
Lower values of $M_{{\rm bol,}\odot}$ (i.e.\ brighter) result in \objectname{V1174~Ori} 
being farther from the Sun. 
We have adopted the value of $M_{{\rm bol,}\odot} = 4.59$ used by
\citet{caloi}, which produces a
distance roughly consistent with the other distance determinations 
discussed above. However,
the value currently adopted by the IAU, $M_{{\rm bol,}\odot} = 4.75$,
results in a distance to \objectname{V1174~Ori} of 390 pc.

Based on this discussion, we consider it likely that \objectname{V1174~Ori} is a member 
of the older, foreground Ori OB1c at a distance of $\sim 400$ pc,
consistent with the 
slightly older age ($\sim 10$ Myr) that we find below. Indeed, the
position of \objectname{V1174~Ori} approximately 0.3 degrees to the south of 
$\Theta^1$C Ori places it just outside the region defined as Ori OB1d by
\citet{blaauw} and in the region defined as Ori OB1c.

\section{Comparison to theoretical pre--main-sequence evolutionary 
tracks\label{tracks}}

In this section, we use the empirically determined masses, radii, effective
temperatures, and luminosities of the primary and secondary components of
\objectname{V1174~Ori} to test the predictions of PMS stellar evolutionary models. 
We consider in turn four sets of PMS tracks commonly used in the literature:
\citet{bcah98} (\S \ref{bcah98}), 
\citet{ps99} (\S \ref{ps99}), 
\citet{sdf00} (\S \ref{sdf00}), and
\citet{montalban} (more recent version of the \citet{dantona} tracks; \S \ref{dm98}).

These models differ primarily in their choice of atmospheres/opacities,
the equation of state,
and in their treatment of convection. With respect to the former, the
models essentially differ in terms of whether the atmosphere models used
are gray or non-gray, and whether molecular opacity effects are included
at low effective temperatures. 
The differences with respect to convection are somewhat more subtle. 
Except for one set of models that
uses the Full Spectrum Turbulence (FST) formalism of \citet{canuto}, the
models we consider in this paper use a mixing-length theory (MLT) approach,
typically characterized by the mixing parameter, $\alpha$, defined as the
ratio of the convective mixing length to the pressure scale-height, i.e.\
$\alpha \equiv l_{\rm mix}/H_p$. But as discussed by \citet{montalban} and
\citet{dantona-li}, to fully characterize an MLT model requires specification
of a larger set of ingredients than just the single parameter $\alpha$. These
are: the mixing parameter used in the atmosphere calculation ($\alpha_{\rm atm}$);
the mixing parameter used in the calculation of the interior 
($\alpha_{\rm in}$, which can be set equal to $\alpha_{\rm atm}$); 
and the optical depth in the atmosphere at which these two calculations are 
matched up ($\tau_{\rm ph}$). 
Requiring a fit to the present-day Sun generally requires more efficient
convection in the interior than in the atmosphere, that is, 
$\alpha_{\rm in} > \alpha_{\rm atm}$ \citep{bcah98,montalban}. For a given
choice of $\alpha_{\rm atm}$ and $\alpha_{\rm in}$, larger $\tau_{\rm ph}$
produce lower temperatures at the base of the convection zone,
resulting in less Li depletion prior to the main sequence.
The parameters used by the various models considered below are summarized
in Table \ref{models}. All of the tracks considered assume solar metallicity.

\begin{deluxetable}{llcrrrc}
\tablewidth{0pt}
\tabletypesize{\scriptsize}
\tablecolumns{7}
\tablecaption{Parameters of PMS Evolution Models\label{models}}
\tablehead{ \colhead{Authors} & \colhead{Atmosphere} & 
\colhead{Convection} & 
\colhead{$\alpha_{\rm atm}$} &
\colhead{$\alpha_{\rm in}$} & \colhead{$\tau_{\rm ph}$} &
\colhead{Matches Sun?} }
\startdata
MDKH03     & ATLAS9   & FST & \nodata & \nodata & 10  & Y \\
           & ATLAS9   & MLT &  0.5   & 2.3  & 10  &  Y \\
           & NextGen  & MLT & 1.0 & 1.0 & 3\tablenotemark{a}   & N \\
BCAH98     & NextGen  & MLT & 1.0 & 1.0 & 100 &  N \\
           & NextGen  & MLT & 1.0 & 1.9 & 100 &  Y \\
PS99       & gray approx.\ & MLT & \nodata & 1.5 & $\frac{2}{3}$ & N \\
SDF00      & Plez, Eriksson  & MLT & \nodata & 1.6 & 10 & Y \\
\enddata
\tablenotetext{a}{These authors have also computed NextGen-based models
with $\tau_{\rm ph} = 100$, but these are equivalent to the BCAH98 tracks.}
\end{deluxetable}

For each set of tracks, we first consider the positions of the components of
\objectname{V1174~Ori} in the H-R diagram (\teff\ vs.\ $\log L$) and compare the 
stellar masses inferred from the models to those we have measured. We also 
consider whether the theoretical isochrones lie parallel to the observed
positions.  

Comparing the observations to
the models in the plane of the H-R diagram is not necessarily the most robust 
test of the models, however, and certainly does not take full advantage of the
accuracy of the measurements. Placing the stars in the H-R diagram requires
deriving $L$ from $R$ and \teff. While the measurement errors in the
radii are small, the \teff\ values have not been determined directly but
rather converted from the observed spectral type (for the primary)
and then from the \teff\ ratio determined from the light curve analysis 
(for the secondary). Consequently, the uncertainties in the derived 
luminosities are relatively large. 
Thus for each set of tracks we also do the comparison in the mass-radius
plane. In this way, not only do we preserve the accuracy of the empirical 
masses and radii, we also test what is perhaps the most fundamental 
prediction of the models: the mass-radius relationship.

\begin{figure}[ht]
\epsscale{0.8}
\plotone{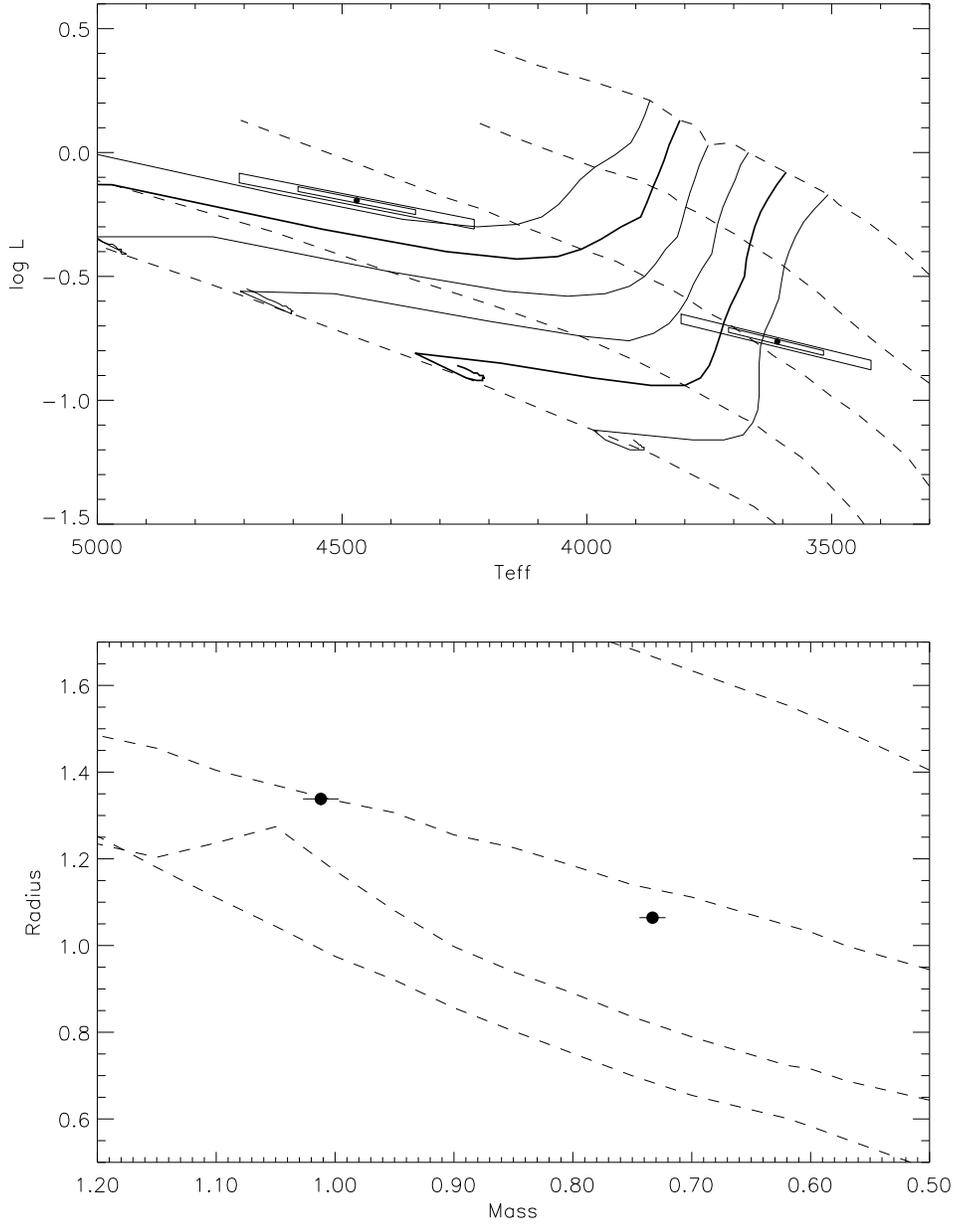}
\caption{\footnotesize BCAH98 tracks with $l_{\rm mix}/H_p = 1.0$.
(a) H-R diagram ($\log L$ vs.\ \teff), with isochrones at
1, 3, 10, 30, and 100 Myr, and mass tracks for 0.6, 0.7, 0.8, 0.9, 1.0, and 1.1
M$_\odot$. Mass tracks at 0.7 and 1.0 M$_\odot$, corresponding approximately
to the masses of V1174~Ori, are thicker than the others. 
Points mark the best-fit values of \teff\ and $\log L$, with
nested boxes indicating confidence regions about those values.
The inner box about the primary corresponds to 0.5 spectral sub-type
uncertainty in \teff\ (see \S \ref{spt}) and $1\sigma$ uncertainty
on the stellar radius. The outer box corresponds to uncertainties twice
as large in each of \teff\ and $R$, and should be taken as the region
of high confidence for the primary. The nested boxes about the
secondary are similar, but the range of \teff\ here is determined by
keeping the secondary-to-primary \teff\ ratio fixed at the value
determined from the light curve analysis (see \S \ref{lightcurves}). Note 
that the positions of the primary and secondary in this figure are
highly correlated; adjusting the position of the primary necessarily
adjusts that of the secondary.
(b) Radius vs.\ mass in solar units. Points with error bars are the
measurements for the two components of V1174~Ori; isochrones at 3, 10, 30,
and 100 Myr are shown as dashed lines.
\label{bcah98-1}}
\end{figure}

\begin{figure}[ht]
\plotone{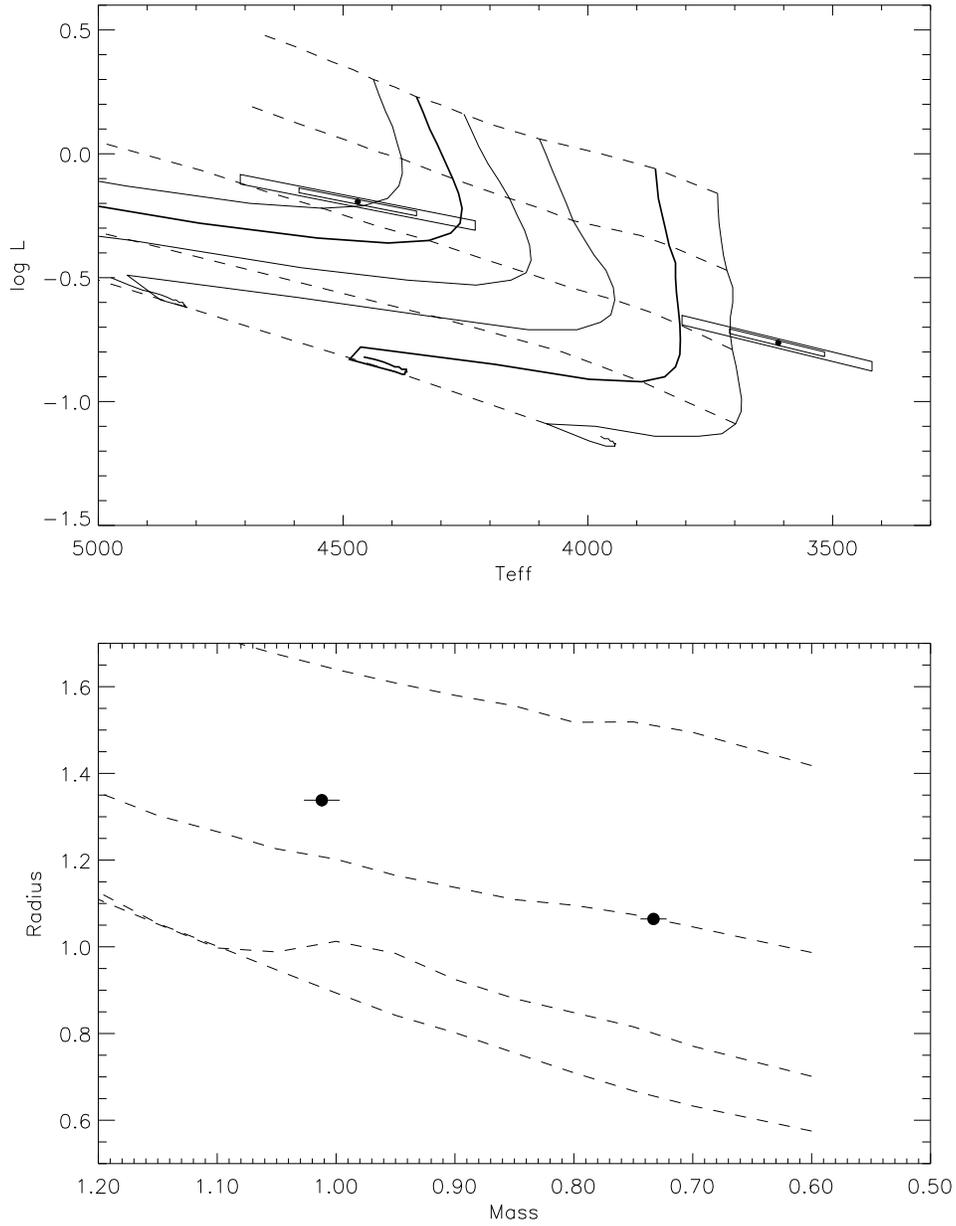}
\caption{BCAH98 tracks with $l_{\rm mix}/H_p = 1.9$.
Symbols and lines are as in Fig.\ \ref{bcah98-1}.
\label{bcah98-1.9}}
\end{figure}

\begin{figure}[ht]
\plotone{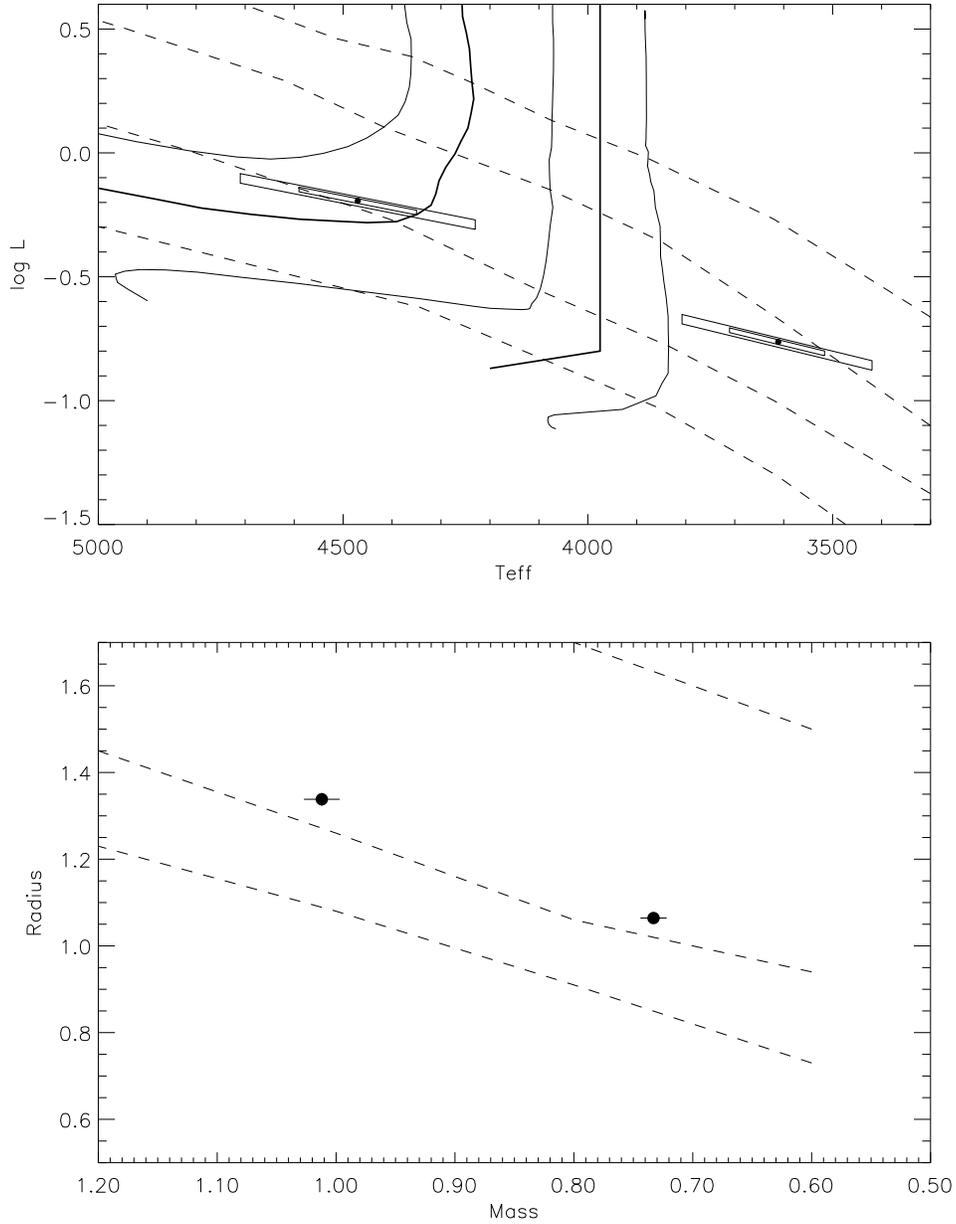}
\caption{PS99 tracks.
Symbols and lines are as in Fig.\ \ref{bcah98-1}, except that
(a) shows mass tracks at 0.6, 0.7, 0.8, 1.0, and 1.2 M$_\odot$
and isochrones at 1, 3, 10, and 30 Myr. Note that the mass track
at 0.7 M$_\odot$ has been visually interpolated between the 
positions of the 0.6 and 0.8 M$_\odot$ tracks. In (b) isochrones are
shown at 3, 10, and 30 Myr.
\label{ps99-1}}
\end{figure}

\begin{figure}[ht]
\plotone{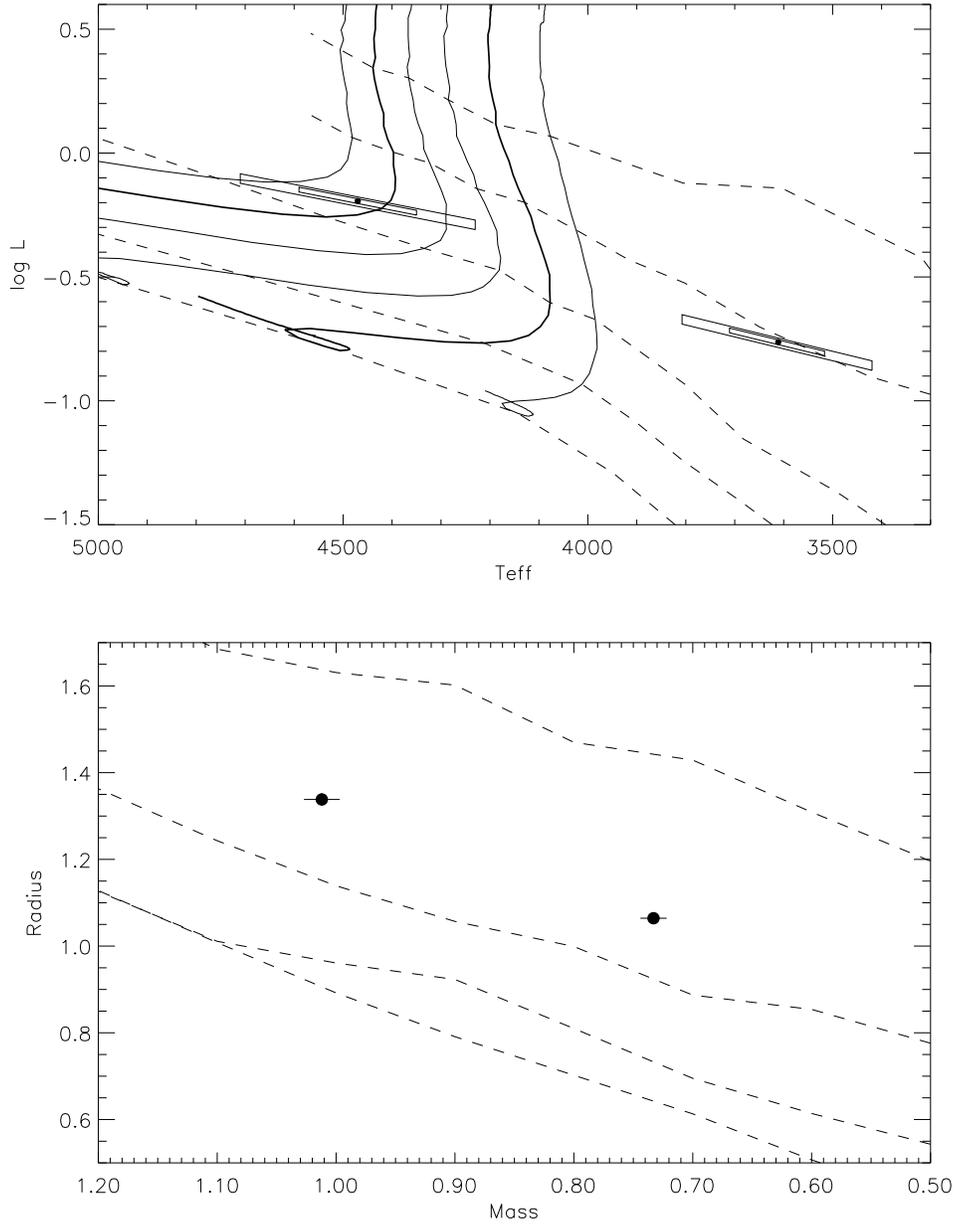}
\caption{SDF00 tracks.
Symbols and lines are as in Fig.\ \ref{bcah98-1}.
\label{sdf00-1}}
\end{figure}

\begin{figure}[ht]
\plotone{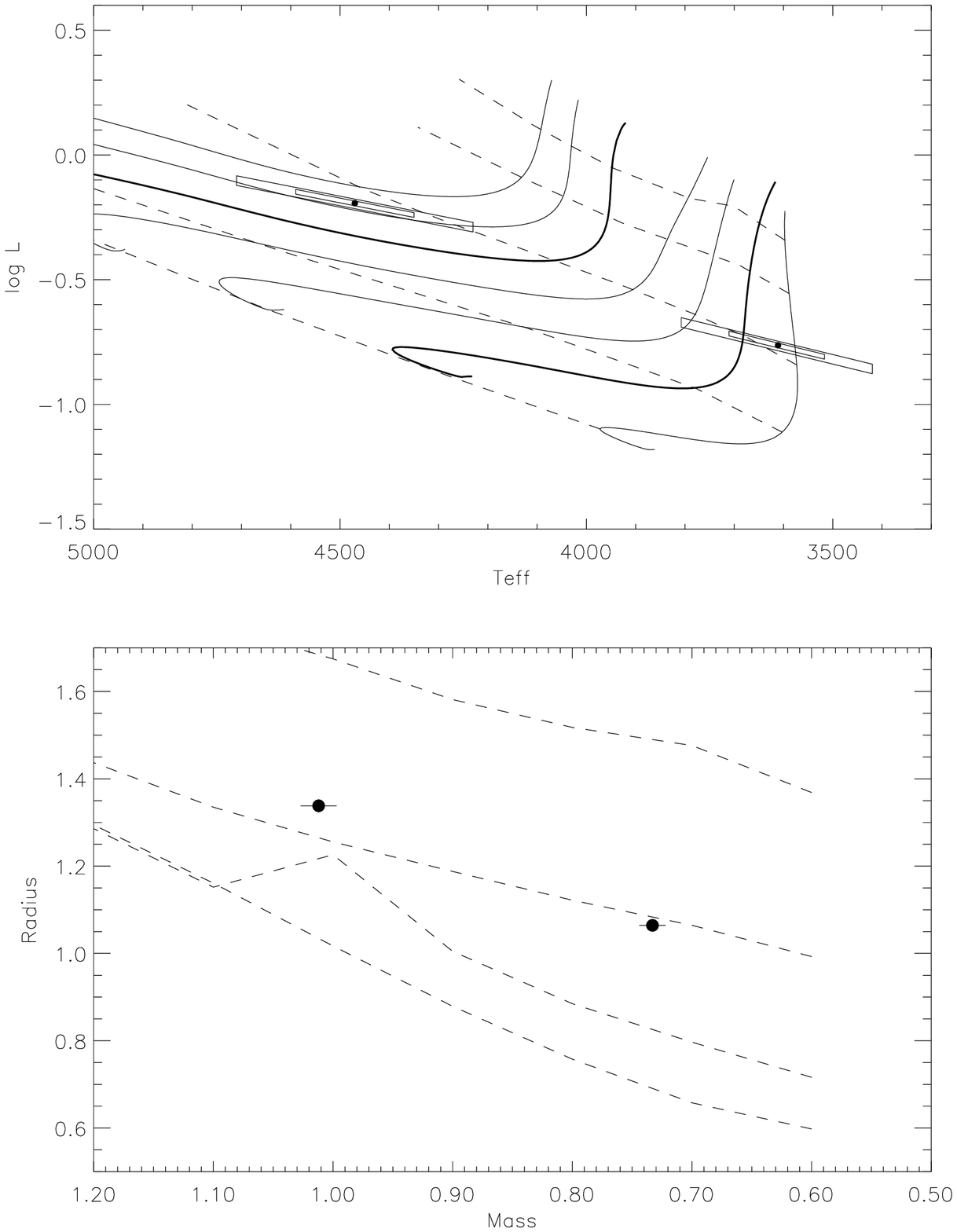}
\caption{\citet{montalban} tracks with MLT convection ($\alpha_{\rm in} = 1.0$,
$\tau_{\rm ph} = 3$) and NextGen atmospheres from \citet{hauschildt}.
Symbols and lines are as in Fig.\ \ref{bcah98-1}. 
\label{dm98-1}}
\end{figure}

\begin{figure}[ht]
\plotone{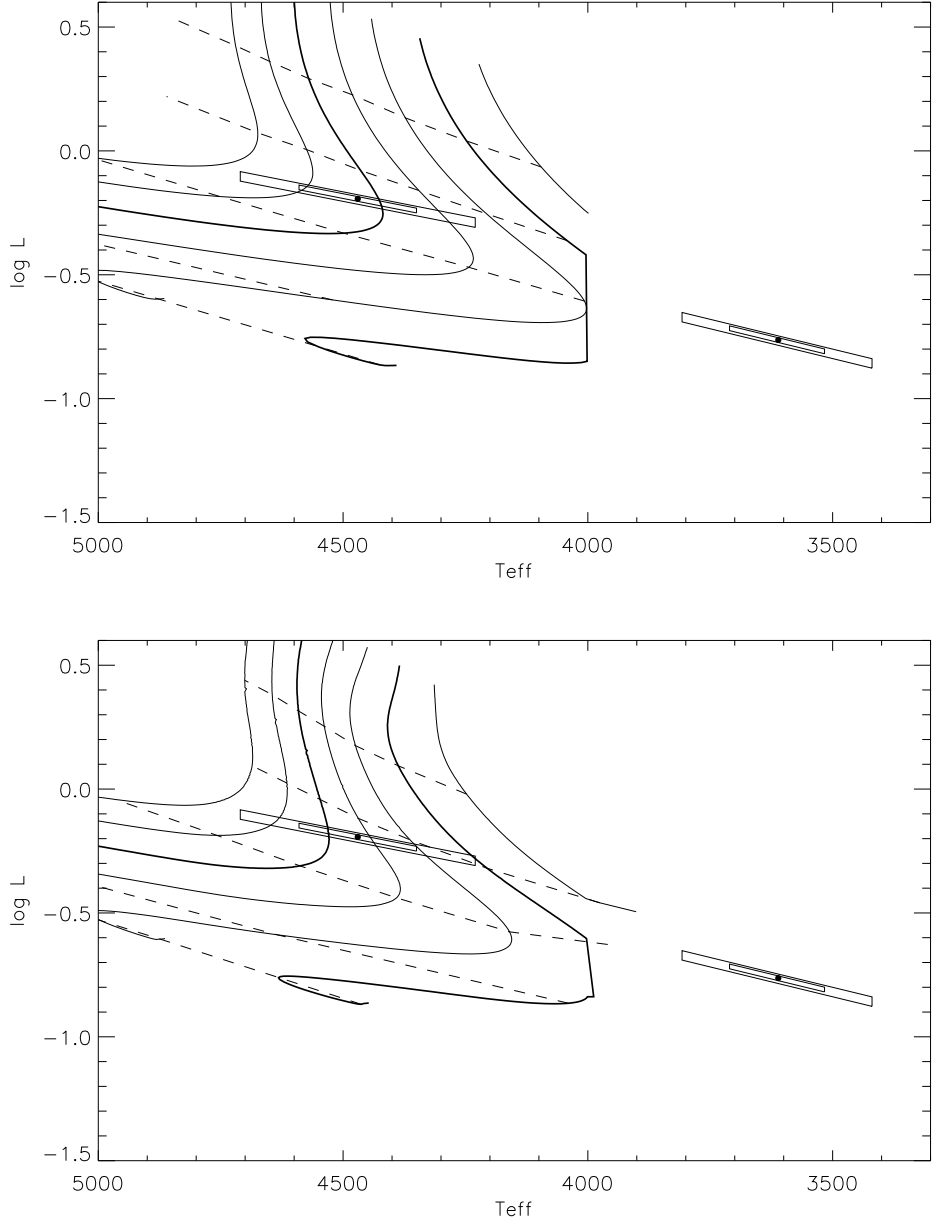}
\caption{\citet{montalban} tracks with (a) MLT convection ($\alpha_{\rm in} = 2.3$)
and (b) FST convection. Both use the ATLAS9 atmospheres from \citet{heiter}.
Symbols and lines are as in Fig.\ \ref{bcah98-1}. Note that the tracks are
truncated below 4000 K as the \citet{heiter} atmospheres do not extend to
cooler temperatures.
\label{dm98-2}}
\end{figure}

One final comment is in order, regarding the placement of the primary
and secondary stars in the H-R diagrams 
shown in Figs.\ \ref{bcah98-1}--\ref{dm98-2}. 
In each H-R diagram we plot two nested boxes around the measured
position of the primary star.
The inner box represents the region of the ($\log L$, \teff) space 
allowed by the combination of the uncertainty in the primary star's 
\teff\ (0.5 spectral sub-type; \S \ref{spt}) and $1\sigma$
uncertainty in its radius ($0.015$ R$_\odot$; Table \ref{fundamental}). 
The outer box corresponds to twice the uncertainty in \teff\ and 
$2\sigma$ uncertainty on the radius. This outer box should thus be
regarded as the region of high confidence for the location of the
primary star in the H-R diagram. 
We plot similar nested boxes about the position of the secondary star, 
but the range of \teff\ here is determined by keeping the 
secondary-to-primary \teff\ ratio fixed at the value determined
from the light curve analysis (\S \ref{lightcurves}), the uncertainty
of which is minuscule compared to the uncertainty in the \teff\
of the primary star. Because the two stars' positions are linked via
the \teff\ ratio, it is important to bear in mind that the positions 
of the two stars in the H-R diagram are very strongly correlated. 
While it may be tempting to adjust the position of each star within its
own bounding box independently, the positions of the two stars {\it relative 
to one another} are in fact rigidly constrained.

\subsection{BCAH98\label{bcah98}}
We begin by considering the PMS tracks of \citet{bcah98}. 
Of the models that we consider here, these are perhaps the
most generally favored in the recent literature
\citep{white,simon,luhman}. The tracks use the non-gray NextGen
atmospheres of \citet{hauschildt} and treat convection using MLT.
These models are shown in Figs.\ \ref{bcah98-1} and \ref{bcah98-1.9}
for two different values of $\alpha_{\rm in}$ (1.0 and 1.9, respectively). 
Note that both models use NextGen atmospheres computed with 
$\alpha_{\rm atm} = 1.0$ (see Table \ref{models}); thus, these models
differ in their convection properties only interior to an optical depth of
$\tau_{\rm ph} = 100$.
The tracks with $\alpha_{\rm in} = 1.9$ are designed to 
match the properties of the present-day Sun.

Neither of these \citet{bcah98} models succeed in simultaneously reproducing
the observed positions of both components of \objectname{V1174~Ori} in the H-R diagram.
The tracks with $\alpha_{\rm in} = 1.0$, favored by \citet{steffen} in their
analysis of the PMS binary NTTS~045251$+$3016, predict a mass for the
\objectname{V1174~Ori} primary of $\gtrsim 1.1$ M$_\odot$. The position of
the primary excludes the 1.01 M$_\odot$ track with $> 3\sigma$ confidence.
Meanwhile, the $0.73\; {\rm M}_\odot$ track is consistent with 
\objectname{V1174~Ori} secondary at the $\sim 1.5\sigma$ level.

The tracks with $\alpha_{\rm in} = 1.9$, favored by the recent analysis of 
\citet{luhman},
are possibly consistent with the position of the $1.01\; {\rm M}_\odot$ 
primary of \objectname{V1174~Ori}, but only by
pushing the primary near the lower extreme of its $2\sigma$ bounding box.
Forcing agreement with the primary in this way necessarily forces the 
secondary to the lower extreme of its $2\sigma$ bounding box as well. 
This would imply a mass for the secondary star of $\sim 0.4\; {\rm M}_\odot$, 
completely inconsistent with its
measured mass of $0.73 \pm 0.01\; {\rm M}_\odot$. The 0.73 M$_\odot$
track is $\sim 400$ K too hot relative to the position that the secondary
would take if the primary were forced into agreement with its mass track.

Of the BCAH98 models considered here, only the $\alpha_{\rm in} = 1.9$ tracks
possess an isochrone in the H-R diagram (at $\sim 7$ Myr) consistent 
with coevality for the components of \objectname{V1174~Ori}. The other tracks 
make the secondary appear over-luminous relative to the primary.

In the mass-radius ($M$-$R$) plane, both sets of tracks imply
an age for the \objectname{V1174~Ori} system of $\sim 10$ Myr, though 
none of the tracks possess an isochrone precisely parallel to the empirical
isochrone defined by the components of \objectname{V1174~Ori}. Interestingly, this is true 
of the $\alpha = 1.9$ tracks as well, despite the fact that these tracks yield
coevality for the two stars in the H-R diagram. Apparently, the empirical
measurements and the theoretical isochrones translate differently between
the $M$-$R$ plane and the H-R diagram, almost certainly the result of
differences in the \teff\ scale. This is most readily seen in the 
$\alpha_{\rm in} = 1.0$ tracks where, in the
$M$-$R$ plane, the secondary is situated below the 10 Myr isochrone and the
primary is above it; in the H-R diagram their positions relative to this
isochrone are reversed.

\subsection{PS99\label{ps99}}
The PMS tracks of \citet{ps99} have been found to show reasonably good agreement 
with PMS
mass measurements in a number of previous analyses \citep{steffen,simon,palla}. 
These models arguably use the most physically realistic initial conditions
of any of the models, that of the stellar birthline \citep{stahler}.
They use a gray atmosphere approximation, and use the \citet{alexander}
and \citet{iglesias} opacities for \teff\ below and above $10^4$ K, 
respectively. Convection is treated with MLT and $\alpha_{\rm in} = 1.5$.
These tracks are compared with the observations
of \objectname{V1174~Ori} in Fig.\ \ref{ps99-1}.

The PS99 models show good agreement with the position of the primary
star in the H-R diagram, with the 1.01 M$_\odot$ track passing within
the inner bounding box of the primary's position. However, the position of the 
secondary star excludes the 0.73 M$_\odot$ track with $>3\sigma$ confidence.
As with the BCAH98 tracks,
the 0.73 M$_\odot$ track is $\sim 400$ K too hot compared to the position
of the \objectname{V1174~Ori} secondary.

The PS99 models moreover yield ages for the two components of \objectname{V1174~Ori}
that are non-coeval. The tracks give ages of 10 Myr and 5 Myr for the
primary and secondary, respectively. Interestingly, this non-coevality
in the H-R diagram is not borne out in the $M$-$R$ plane, where the 
components lie precisely parallel to the theoretical isochrones (9 Myr
in this case). Indeed, of the models considered in this paper, these
models show the best agreement with the slope of the $M$-$R$ relationship as
defined by the components of \objectname{V1174~Ori}.

\subsection{SDF00\label{sdf00}}
Next, we consider the PMS models of \citet{sdf00}. These models,
shown in Fig.\ \ref{sdf00-1},
use the \citet{alexander} and \citet{iglesias} opacities, and the
model atmospheres of \citet{plez} and \citet{kurucz}. They treat convection
with MLT and $\alpha_{\rm in} = 1.6$ to match the present-day Sun.

These tracks behave in a similar fashion as the PS99 tracks. In
particular, the position of the \objectname{V1174~Ori} primary is precisely matched
by the theoretical 1.01 M$_\odot$ track, but the agreement with the
location of the \objectname{V1174~Ori} secondary is extremely poor. Indeed, the
theoretical 0.73 M$_\odot$ track is a full $\sim 500$ K too hot
compared to the observed position of the secondary star. These tracks,
like the others considered above, appear to be too compressed in \teff\
to simultaneously fit the primary and secondary of \objectname{V1174~Ori}.

As with the \citet{ps99} models, these tracks
yield non-coeval ages for the two components of \objectname{V1174~Ori}. Once again,
the inferred ages for the primary and secondary stars differ by a
factor $\sim 2$: in this case, $\sim 7$ Myr and $\sim 3$ Myr for the
primary and secondary, respectively.
But as with the PS99 models, despite non-coevality in the H-R diagram, these
models produce isochrones in superb agreement with the observations in the
$M$-$R$ plane (7 Myr).

\subsection{MDKH03\label{dm98}}
Finally, we consider the new PMS evolutionary tracks of \citet{montalban},
descended from the earlier models of \citet{dantona}.
These authors have produced a variety of models that treat convection via 
MLT or FST, with the non-gray atmospheres from \citet{hauschildt} or 
\citet{heiter}, respectively. For the models using the NextGen 
atmospheres of \citet{hauschildt}, the authors compute families of
models with different values of $\tau_{\rm ph}$. Here we consider three
models as listed in Table \ref{models}.
These models are compared with the measured positions of the 
components of \objectname{V1174~Ori} in Figs.\ \ref{dm98-1}--\ref{dm98-2}.

The MLT models with NextGen atmospheres and $\alpha_{\rm in} = 1.0$ are
most comparable to the BCAH98 tracks in Fig.\ \ref{bcah98-1}. As the
only substantial difference between these models is the depth of the
outer atmosphere ($\tau_{\rm ph} = 3$ for MDKH03 tracks vs.\ 100 for
BCAH98 tracks), these tracks yield very similar results. The only
notable difference appears to be in the rate at which the models contract,
with the $\tau_{\rm ph} = 3$ tracks yielding slightly younger ages, as
seen in the $M$-$R$ plane.

In contrast,
both the MLT and FST models, with the ATLAS9 atmospheres of \citet{heiter}, 
show excellent agreement with the position of the \objectname{V1174~Ori} in
the H-R diagram, with the FST models being $\sim 100$ K hotter.
Unfortunately, these models do not permit a comparison with the
\objectname{V1174~Ori} secondary as the \citet{heiter} atmospheres do not extend below
4000 K. However, the general behavior of these tracks in the H-R diagram
suggests that the 0.73 M$_\odot$ track could intersect the
position of the secondary if the calculation were extended to lower \teff.

Indeed, these ATLAS9-based models differ qualitatively from all of the
other models considered above in terms of the general shape of the tracks.
These models exhibit a dramatic sweep to cooler \teff, and then back to
hotter \teff, as the stars descend their tracks, particularly for
$M < 0.8\; {\rm M}_\odot$. The effect is to spread the mass tracks
over a larger range of \teff\ at ages between 3--30 Myr, possibly 
consistent simultaneously with the primary and secondary of \objectname{V1174~Ori}.

\subsection{Other mass measurements of PMS stars\label{luhman}}
As we have seen,
the empirical mass measurements of \objectname{V1174~Ori} provide important constraints 
on the absolute mass calibration of PMS evolutionary tracks, both in the 
H-R diagram and in the $M$-$R$ plane, between 0.7 M$_\odot$ and 1.0 M$_\odot$.
In this section, we combine our measurements of \objectname{V1174~Ori} with
extant mass measurements of other PMS stars in order to extend our
analysis to a larger range in stellar mass and to compare our
findings among stars in both binary and single systems.

Empirical mass determinations exist in the literature for twelve\footnote{We
exclude from this discussion RS Cha A and B ($M = 1.86\; {\rm M}_\odot$
and $1.82\; {\rm M}_\odot$ respectively), because of their high masses.
We also exclude
BP Tau ($M = 1.32\; {\rm M}_\odot$) because of the large uncertainty
on the mass determination ($\pm 0.2\; {\rm M}_\odot$) and the large 
uncertainty on its distance \citep{simon}.}
other PMS stars---8 in binary systems, 4 single---with masses accurate to
better than $\sim 10\%$ and that are within the range 
covered by the PMS tracks considered above ($\lesssim 1.6\; {\rm M}_\odot$). 
These are:
TY CrA B with $M=1.64\pm0.01\; {\rm M}_\odot$ \citep{casey}; 
AK Sco A and B, both with $M=1.35\pm0.07\; {\rm M}_\odot$ \citep{aksco};
EK Cep B with $M=1.12\pm0.01\; {\rm M}_\odot$ \citep{popper-ty}; 
NTT~045251$+$3016 A and B with $M_A=1.45\pm0.19\; {\rm M}_\odot$ and
$M_B=0.81\pm0.09\; {\rm M}_\odot$ \citep{steffen}; 
RXJ0529.4$+$0041 A and B with $M_A=1.25\pm0.05\; {\rm M}_\odot$ and
$M_B=0.91\pm0.05\; {\rm M}_\odot$ \citep{covino}; 
Lk~Ca~15 with $M=0.97\pm0.03\; {\rm M}_\odot$ \citep{simon}; 
GM~Aur with $M=0.84\pm0.05\; {\rm M}_\odot$ \citep{simon}; 
DL~Tau with $M=0.72\pm0.11\; {\rm M}_\odot$ \citep{simon}; and
DM~Tau with $M=0.55\pm0.03\; {\rm M}_\odot$ \citep{simon}.

\begin{figure}[ht]
\epsscale{0.85}
\plotone{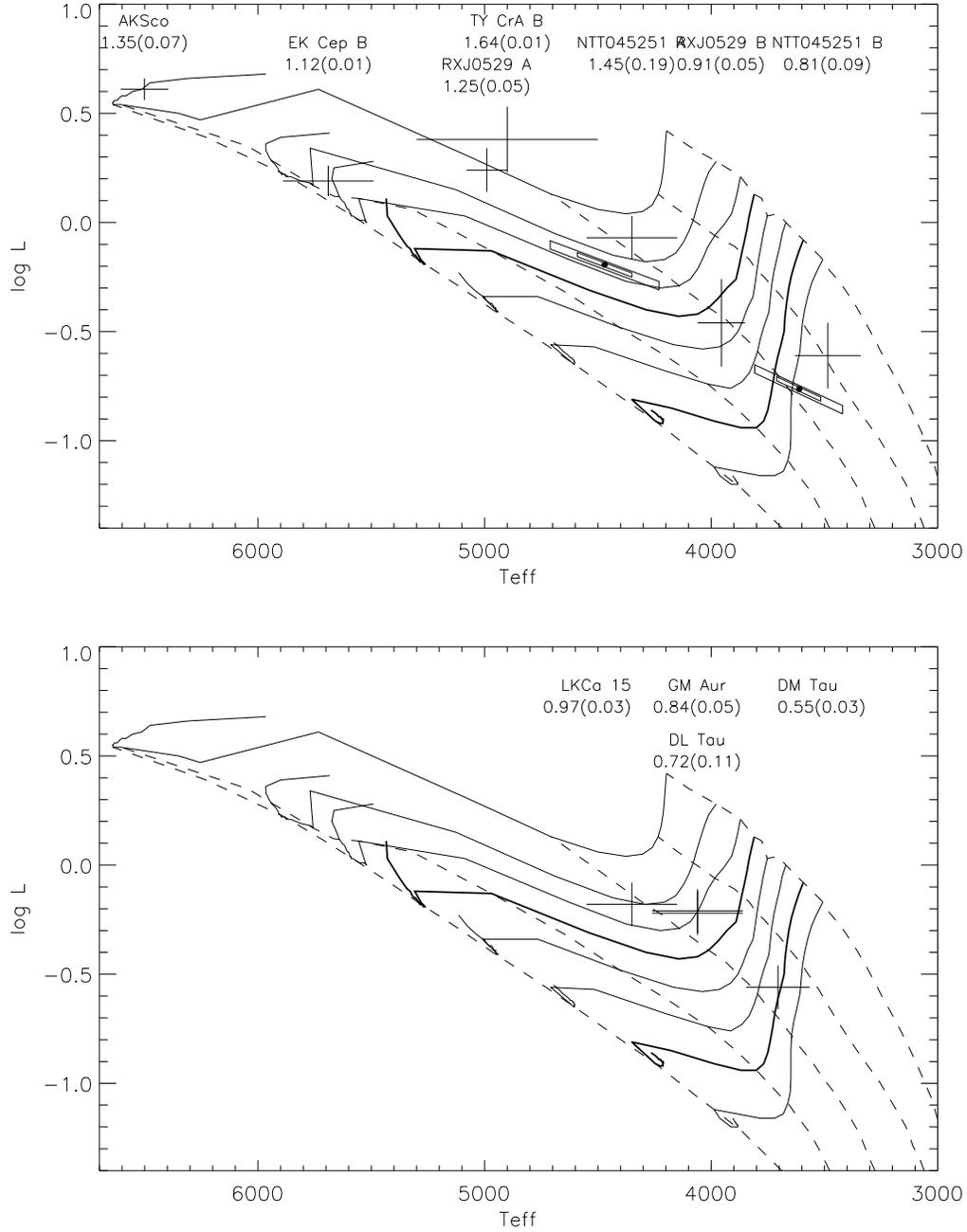}
\caption{Comparison of all PMS stars with accurate empirical mass determinations
and $M \lesssim 1.6$ M$_\odot$ with the tracks of BCAH98, $\alpha_{\rm in}=1.0$. 
(a) Stars in binary systems. 
Masses for each of these are indicated at 
the top of the figure. Mass tracks shown are for 0.6, 0.7, 0.8, 0.9, 1.0,
1.1, 1.2, and 1.4 M$_\odot$. (b) Same as above, but for single stars. 
Note that the uncertainties on the masses of the single stars do not
include distance uncertainties, which can amount to $\sim 15\%$ 
\citep{simon}.
\label{luhman-baraffe10}}
\end{figure}

\begin{figure}[ht]
\epsscale{0.85}
\plotone{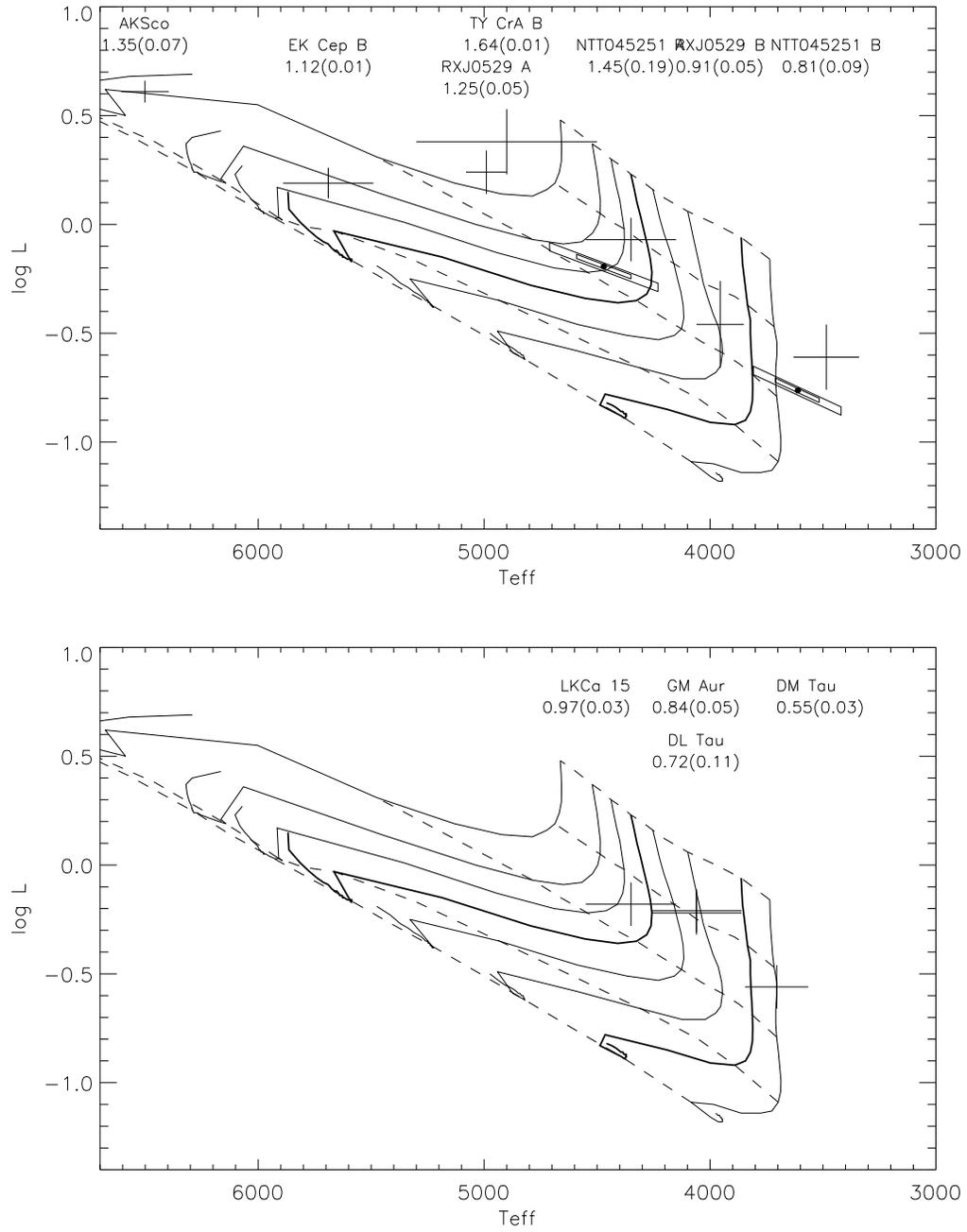}
\caption{Same as Fig.\ \ref{luhman-baraffe10} but for $\alpha_{\rm in} = 1.9$ 
tracks. 
\label{luhman-baraffe19}}
\end{figure}

\begin{figure}[ht]
\epsscale{0.85}
\plotone{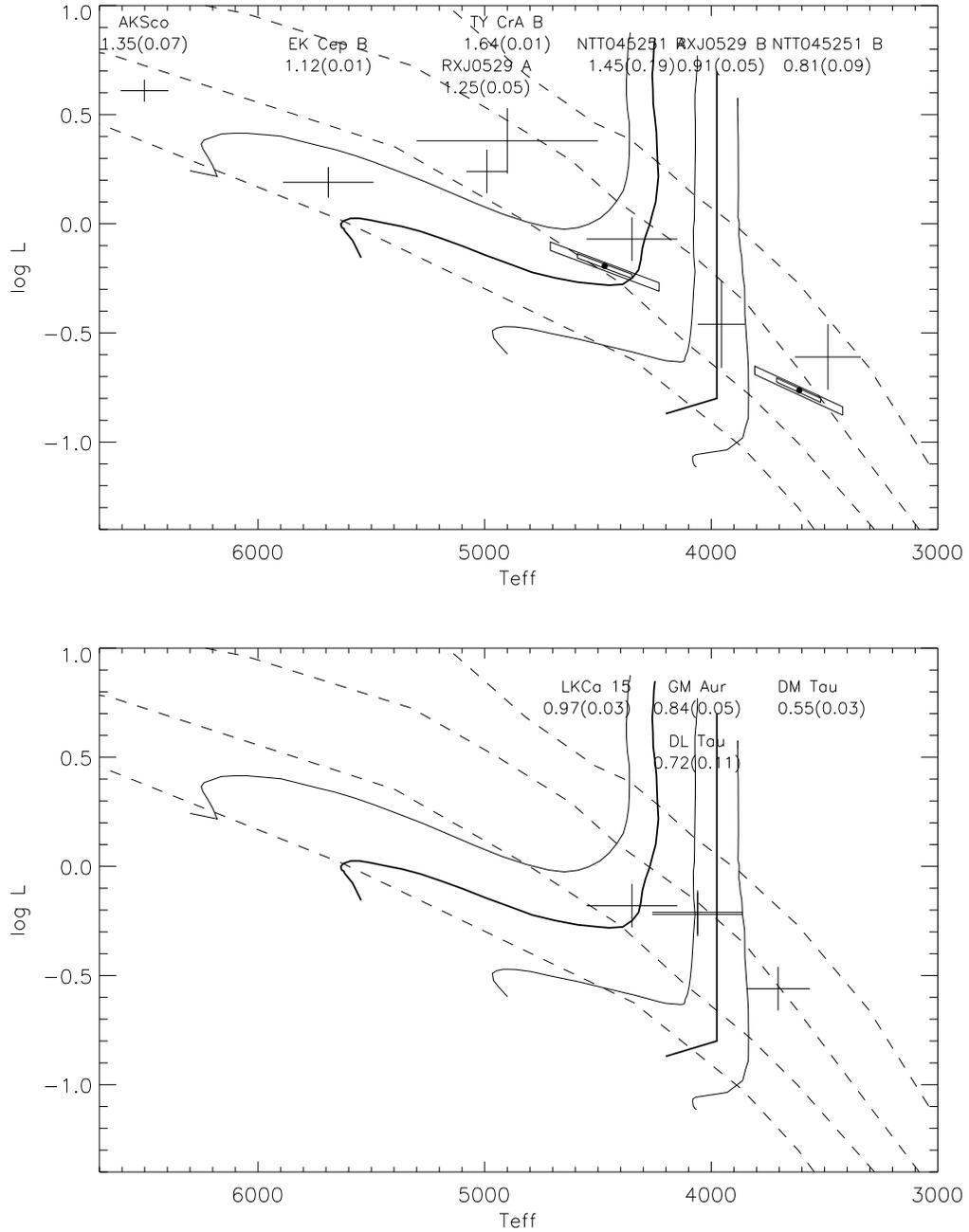}
\caption{Same as Fig.\ \ref{luhman-baraffe10} but for PS99 tracks.
Mass tracks shown are for 0.6, 0.7, 0.8, 1.0, and 1.2 M$_\odot$.
\label{luhman-palla}}
\end{figure}

\begin{figure}[ht]
\epsscale{0.85}
\plotone{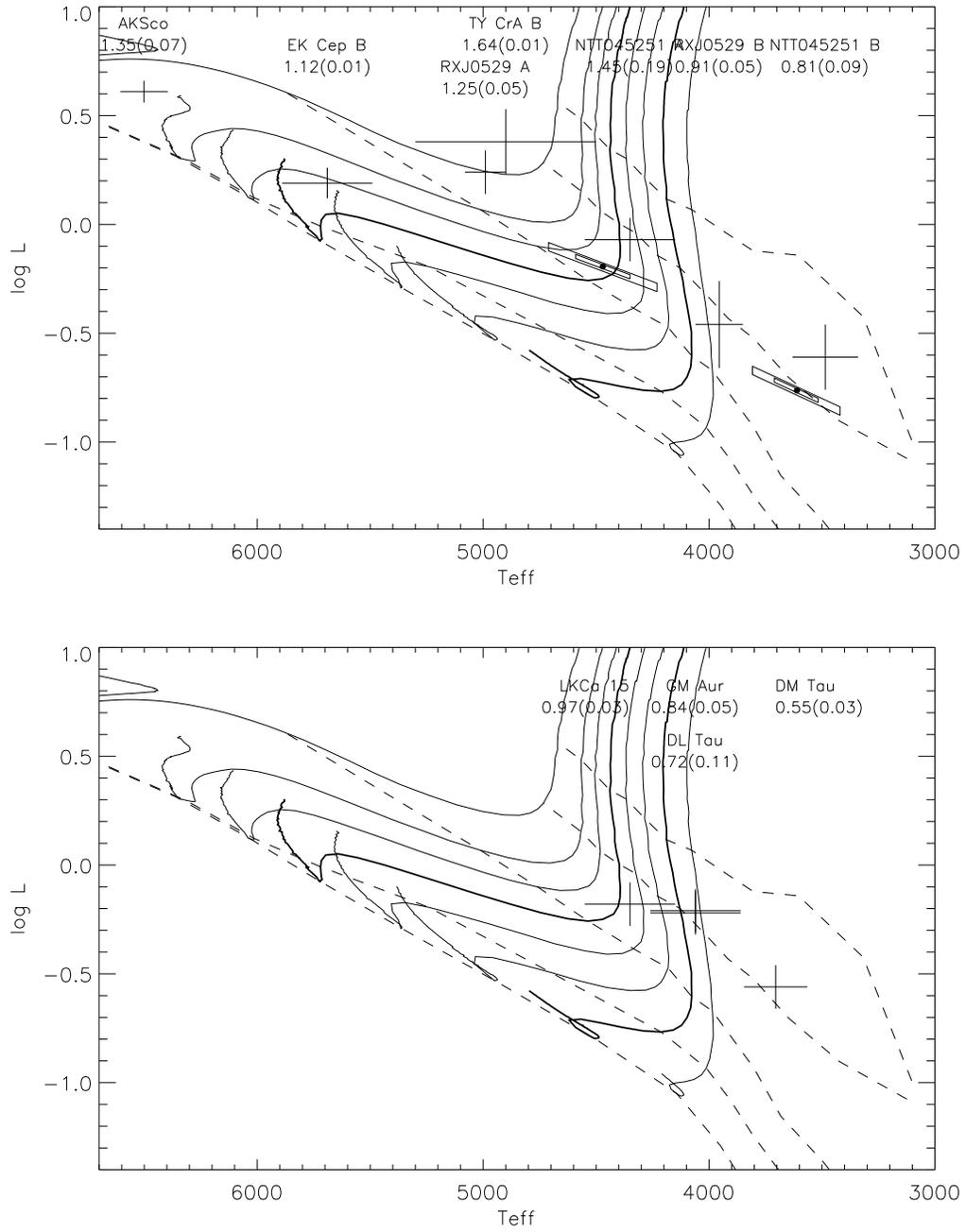}
\caption{Same as Fig.\ \ref{luhman-baraffe10} but for SDF00 tracks.
\label{luhman-siess}}
\end{figure}

\begin{figure}[ht]
\epsscale{0.85}
\plotone{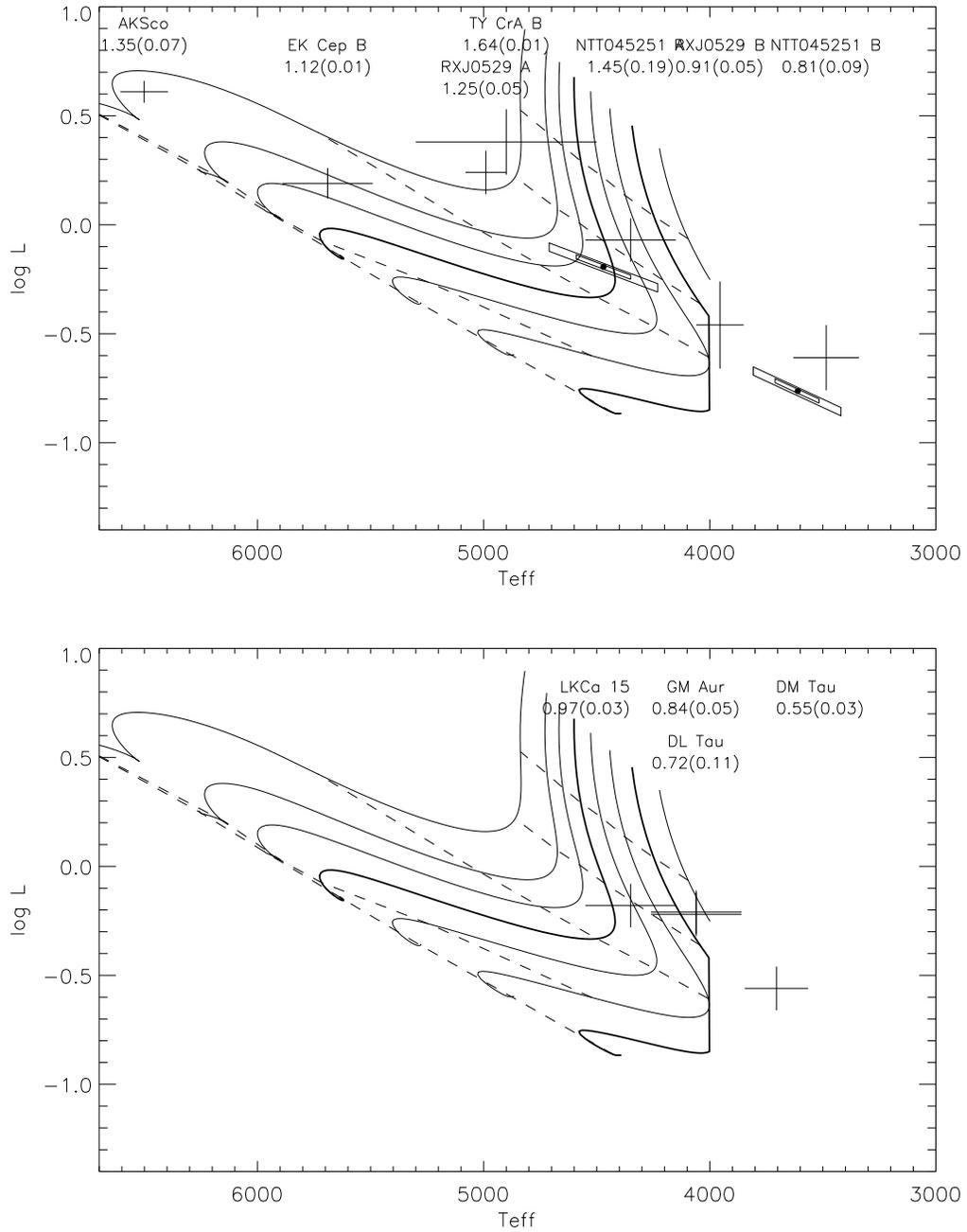}
\caption{Same as Fig.\ \ref{luhman-baraffe10} but for MDKH03 MLT ATLAS9 tracks.
\label{luhman-dm}}
\end{figure}

In Figs.\ \ref{luhman-baraffe10}--\ref{luhman-dm}
we show all of these stars together on H-R diagrams with the
different PMS tracks discussed above. Binary and single stars are
plotted separately in each figure. This ensemble of \teff, luminosities,
and the uncertainties therein
are taken from the compilation of \citet{luhman}, which used the same
temperature scale used in this study. To this
compilation we have added the secondary of TY CrA \citep{casey}, the two
(identical) components of AK Sco \citep{aksco},  and the 
two components of \objectname{V1174~Ori}. Thus the measurements represented in
Figs.\ \ref{luhman-baraffe10}--\ref{luhman-dm} are all based on a consistent
\teff\ scale. 

We note that the listed uncertainties in the masses of the
single stars do not include systematic uncertainties on the distances
to those stars, which can amount to an additional $\sim 15\%$ uncertainty 
on the masses \citep{simon}. We note also that the meanings of the error bars
in $L$ and in \teff\
take on somewhat different meanings for the binary and single stars. For
the single stars and the components of the astrometric binary 
NTT~04251$+$3016, these quantities are measured independently, whereas for
stars in eclipsing binaries these quantities are coupled because $L$ is
derived from \teff\ and the stellar radii. 
Finally, we note that for the binary stars the 
individual masses are coupled to one another by virtue of the measured
mass ratio; adjusting the mass of one component necessarily adjusts
that of the other. These considerations constrain the extent to
which individual objects can be ``fine-tuned" into 
agreement with the models.

We do not find a set of tracks that reproduces the observed locations of 
these stars simultaneously at both high and low masses, and in both 
binary and single systems. 

Consider the BCAH98 tracks with $\alpha_{\rm in} = 1.0$
(Fig.\ \ref{luhman-baraffe10}), 
which have the coolest temperatures of all the tracks we have examined.
These tracks agree with most of the binary stars, so long as the masses of
individual objects are in some cases adjusted within their uncertainties. 
For example, for the tracks to be consistent with the primary of \ntt, 
its mass must be adjusted downward by
$1\sigma$. Of course, doing this requires that the mass of the secondary 
also be adjusted downward by $1\sigma$, implying a true mass of 
0.72 M$_\odot$ for the secondary. Even so,
the tracks miss this star by $\sim 1.5\sigma$. The greatest challenge
to these tracks among the binaries is
the 1.01 M$_\odot$ primary of \objectname{V1174~Ori}, which has the most
accurate mass determination of the lot. This failure to match the
\objectname{V1174~Ori} primary is amplified by the position of LkCa~15; with a very
similar mass of $0.97\pm 0.03\; {\rm M}_\odot$, it is located at almost
precisely the same position in the H-R diagram. And like \objectname{V1174~Ori} A,
LkCa~15 is not matched by these models. 

Indeed, none of the single stars 
are matched by these tracks, the tracks being systematically too cool.
The single stars are better matched by the BCAH98 
$\alpha_{\rm in} = 1.9$ tracks (Fig.\ \ref{luhman-baraffe19}), 
which are $\sim 200$ K warmer,
a conclusion also reached by \citet{luhman}.
However, these tracks have difficulty when compared against the
binaries, especially for those with $M < {\rm M}_\odot$.

The other models show similarly mixed success in reproducing the 
observations. The PS99 tracks (Fig.\ \ref{luhman-palla}) perform 
comparably to the BCAH98 $\alpha_{\rm in} = 1.9$ tracks, matching the
positions of all of the single stars and, with the notable exception of 
\ntt\ A, matching the binaries with $M \ge {\rm M}_\odot$ (including the
primary of \objectname{V1174~Ori}) very well. However, the lower-mass binary 
components---including the secondary of \objectname{V1174~Ori}---are poorly matched
by these tracks. The results for the SDF00 models (Fig.\ \ref{luhman-siess}) 
are very similar as well, though with the warmest and most closely
spaced tracks of those
considered here, the discrepancies among the lowest-mass stars are
particularly severe, even for the single star DM Tau. Finally,
a full evaluation of the MDKH03 MLT ATLAS9 tracks (Fig.\ \ref{luhman-dm}) 
is not possible because they do not extend below 4000 K. However,
based on the qualitative behavior of these tracks it appears possible 
that they could perform well over a broad range of masses and
for both binary and single stars. The lower temperatures of these tracks means 
that they agree with many of the higher-mass binary stars
(though again with \ntt\ A being a clear outlier), and with their
horizontal sweep in \teff---effectively stretching the \teff\ scale for stars
with $M < {\rm M}_\odot$---may be able to simultaneously match the positions 
of the lower-mass binaries and single stars.

\begin{figure}[ht]
\epsscale{1.0}
\plotone{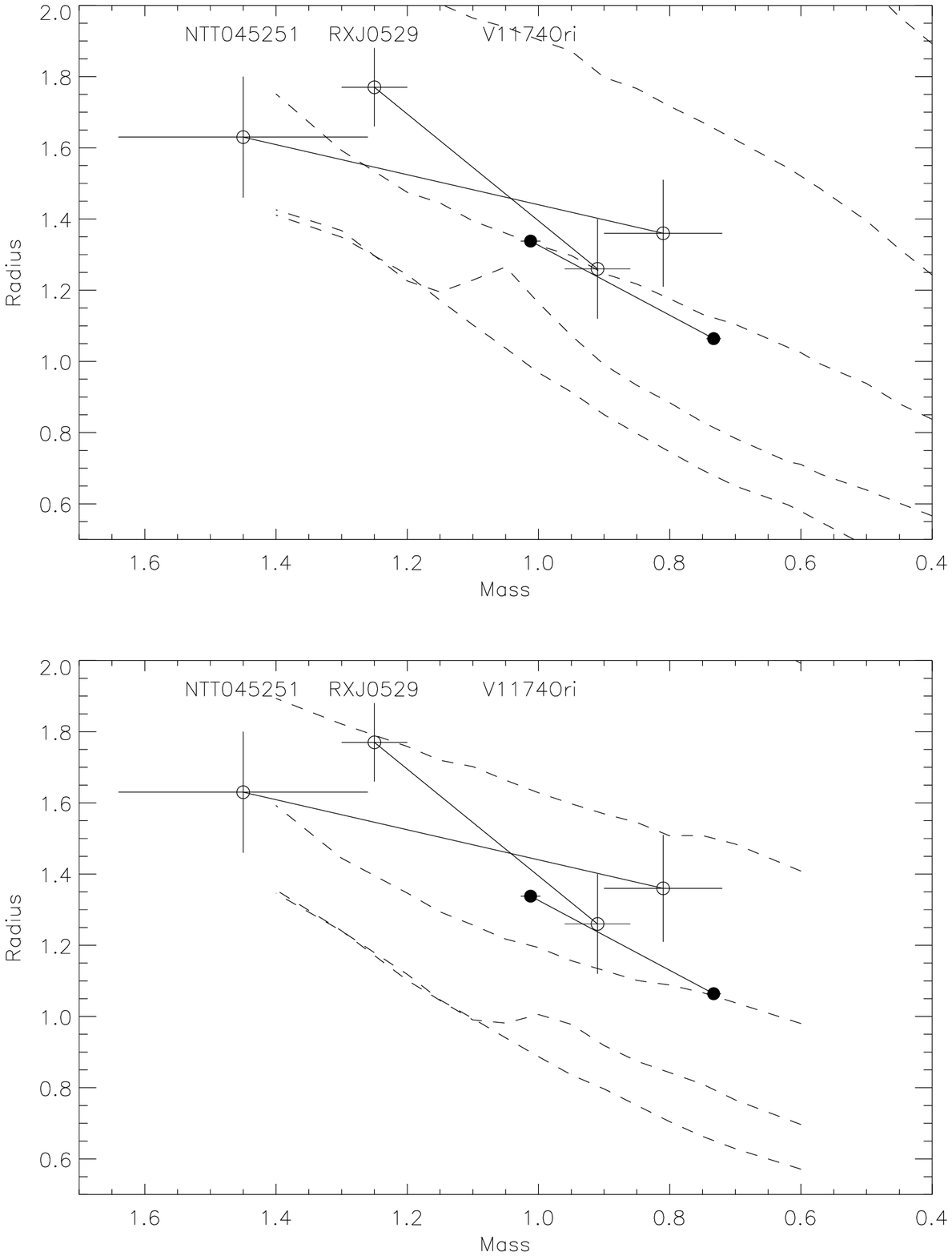}
\caption{Mass-radius relationships (solar units) compared to BCAH98 tracks with 
$\alpha_{\rm in} = 1.0$ (top) and $\alpha_{\rm in} = 1.9$ (bottom).
\label{m-r-relation-1}}
\end{figure}

\begin{figure}[ht]
\epsscale{1.0}
\plotone{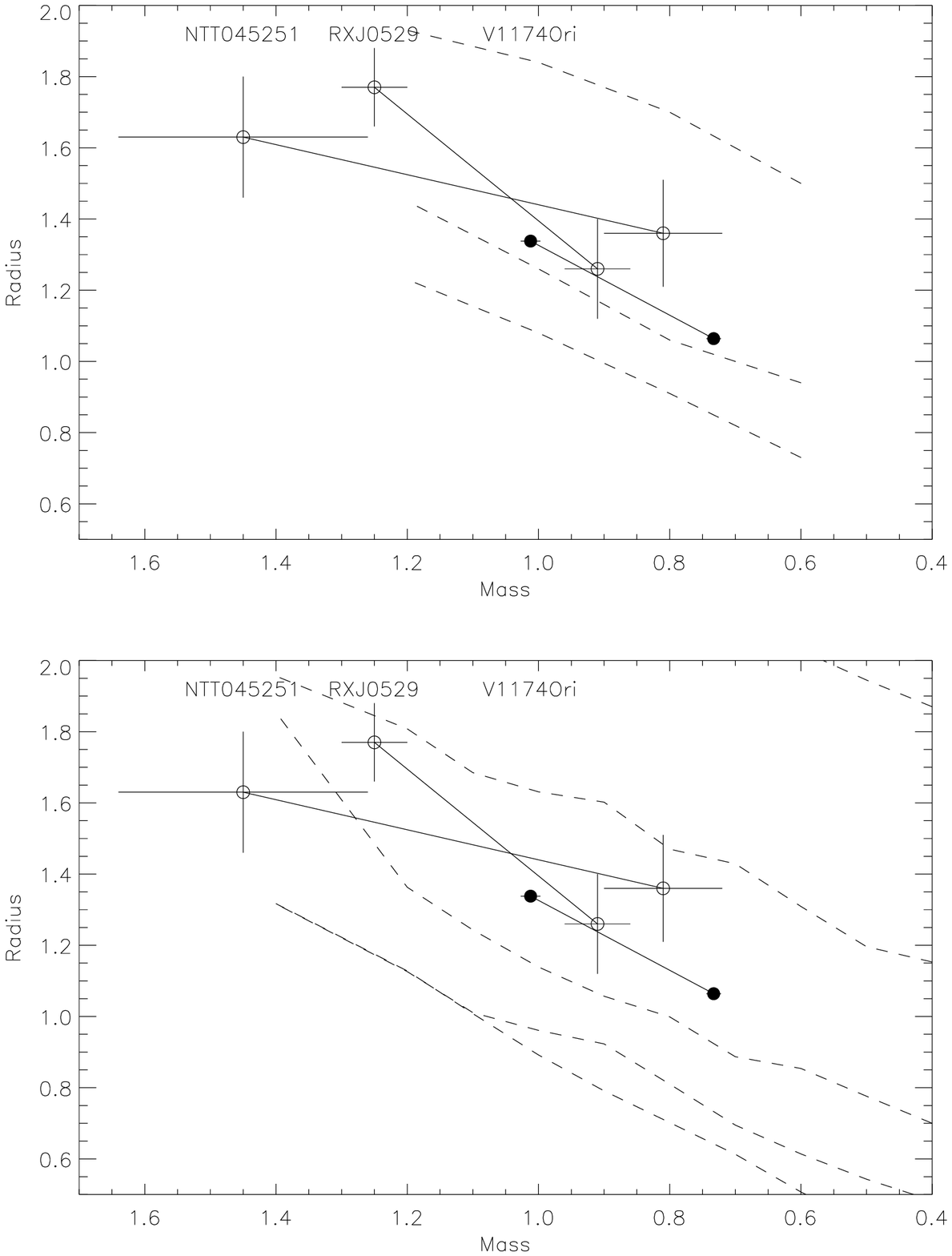}
\caption{Same as Fig.\ \ref{m-r-relation-1} but for PS99 and SDF00 tracks.
\label{m-r-relation-2}}
\end{figure}

For completeness, we show in
Figs.\ \ref{m-r-relation-1}--\ref{m-r-relation-2}
the mass-radius relation for the three PMS binaries for which empirical
mass measurements exist for both components (\objectname{V1174~Ori}, \ntt, and \rxj).
With the relatively large error bars on the components of \ntt\ and
\rxj, we can draw limited conclusions here. All of the
models show qualitative agreement with the general slope of the 
mass-radius relationship. However, the agreement does appear to be best among
the tracks with low values of $\alpha_{\rm in}$, with
the BCAH98 $\alpha_{\rm in} = 1.0$ tracks and the PS99 $\alpha_{\rm in} = 1.5$
tracks providing the best fit to the observations.
The SDF00 $\alpha_{\rm in} = 1.6$ tracks perform reasonably well, whereas
the tracks with large $\alpha_{\rm in}$ (e.g.\ BCAH98 with $\alpha_{\rm in} = 1.9$)
are least consistent with these measurements.

It is valuable and instructive to consider what insights can be gained
from the observations alone, independently of the models. First of all, it
appears clear that the masses of the components of \ntt\ are over-estimated.
The primary star is at a very similar position in the H-R diagram to that of 
the 1.01 M$_\odot$ primary
of \objectname{V1174~Ori}, despite having a measured mass that is considerably larger. 
It does have a somewhat higher luminosity, so it is possible that it does
indeed have a slightly higher mass than does \objectname{V1174~Ori} A. This conclusion
is bolstered by the secondary of \ntt; its position in the H-R diagram,
suggesting a mass slightly less than the 0.73 M$_\odot$ secondary of \objectname{V1174~Ori}, 
belies its measured mass of 0.81 M$_\odot$.
That \ntt\ may be
an outlier may not be surprising given the large uncertainties inherent
to the analysis of this astrometric binary, for which only $\sim 1/4$
of the orbit has been mapped \citep{steffen}. Indeed, simply adjusting
the mass of the \ntt\ primary down by 1--2$\sigma$, while keeping the
mass ratio fixed, would appear to bring the measurements of \ntt\ and
\objectname{V1174~Ori} into closer empirical agreement.

Similar concerns arise with the components of the eclipsing binary \rxj.
Here, the position of the primary star in the H-R diagram coincides closely with 
that of TY CrA B, despite the latter having a precisely determined mass that
is 30\% larger. At the same time, however, the secondary of \rxj\ is located 
beween the two components of \objectname{V1174~Ori}, which is appropriate for its measured mass
intermediate beween those of \objectname{V1174~Ori} A and B. This suggests that
the primary (and perhaps the secondary as well) are simply over-luminous,
which would tend to implicate the stellar radii in this case since,
again, the \teff\ of the secondary appears approximately
correct relative to the components of \objectname{V1174~Ori}. Upon re-examination of the
\citet{covino} results for \rxj, it appears that the uncertainties they
report for the luminosities are underestimated. For example, the fractional 
uncertainty reported by them for the luminosity of \rxj\ A (8\%) is 
smaller than the fractional uncertainty they report (11\%) for the radius 
alone. Including the uncertainty in \teff\ as well gives a more appropriate
uncertainty in the luminosity of $\sim 15\%$.

In some cases, the observations place objects of similar masses at
similar positions in the H-R diagram, bolstering their credibility.
For example, the primary of \objectname{V1174~Ori} and LkCa~15, both with 
$M \approx 1.0\; {\rm M}_\odot$, have very nearly the same \teff\ and $L$.
The same is true for \rxj\ B and GM~Aur, both with 
$M \approx 0.87\; {\rm M}_\odot$. Even though these latter measurements
have somewhat larger uncertainties, the lower luminosity of \rxj\ B
relative to (the presumably younger) GM~Aur in effect provides an empirical 
mass track for $M \approx 0.87\; {\rm M}_\odot$ between $\sim$3--10 Myr.

But there are also cases where the data seem at odds with one another. 
For example, \objectname{V1174~Ori} B and DM~Tau occupy similar positions in the
H-R diagram despite having masses of $0.73 \pm 0.01\; {\rm M}_\odot$ and 
$0.55 \pm 0.03\; {\rm M}_\odot$, respectively. This is worrisome. One way to 
reconcile this apparent contradiction is to have mass tracks, 
like the MLT ATLAS9 tracks of \citet{montalban}, that are double-valued in
\teff\ at low masses, allowing stars with different masses to have similar 
\teff\ at certain times. While these tracks do not presently extend to
the masses of \objectname{V1174~Ori} B and DM~Tau, inspecting these tracks in 
Fig.\ \ref{luhman-dm} one can imagine that these stars' positions
would be well matched by their respective tracks, particularly with
DM~Tau being somewhat more luminous than \objectname{V1174~Ori} B.

Another possible explanation for the discordance between these
measurements is that the uncertainties in the masses for
the single stars are underestimated. Indeed, in contrast to the
distance-independent masses derived for \objectname{V1174~Ori}, the masses determined
for the single stars (from spatially resolved disk rotation profiles;
\citet{simon}) depend directly upon the distance assumed. As noted
by \citet{simon}, the distance uncertainty to any particular star
in Taurus may be as large as $\sim 15\%$, manifesting itself as a
systematic error in the dynamical mass determination.

\begin{figure}[ht]
\epsscale{0.9}
\plotone{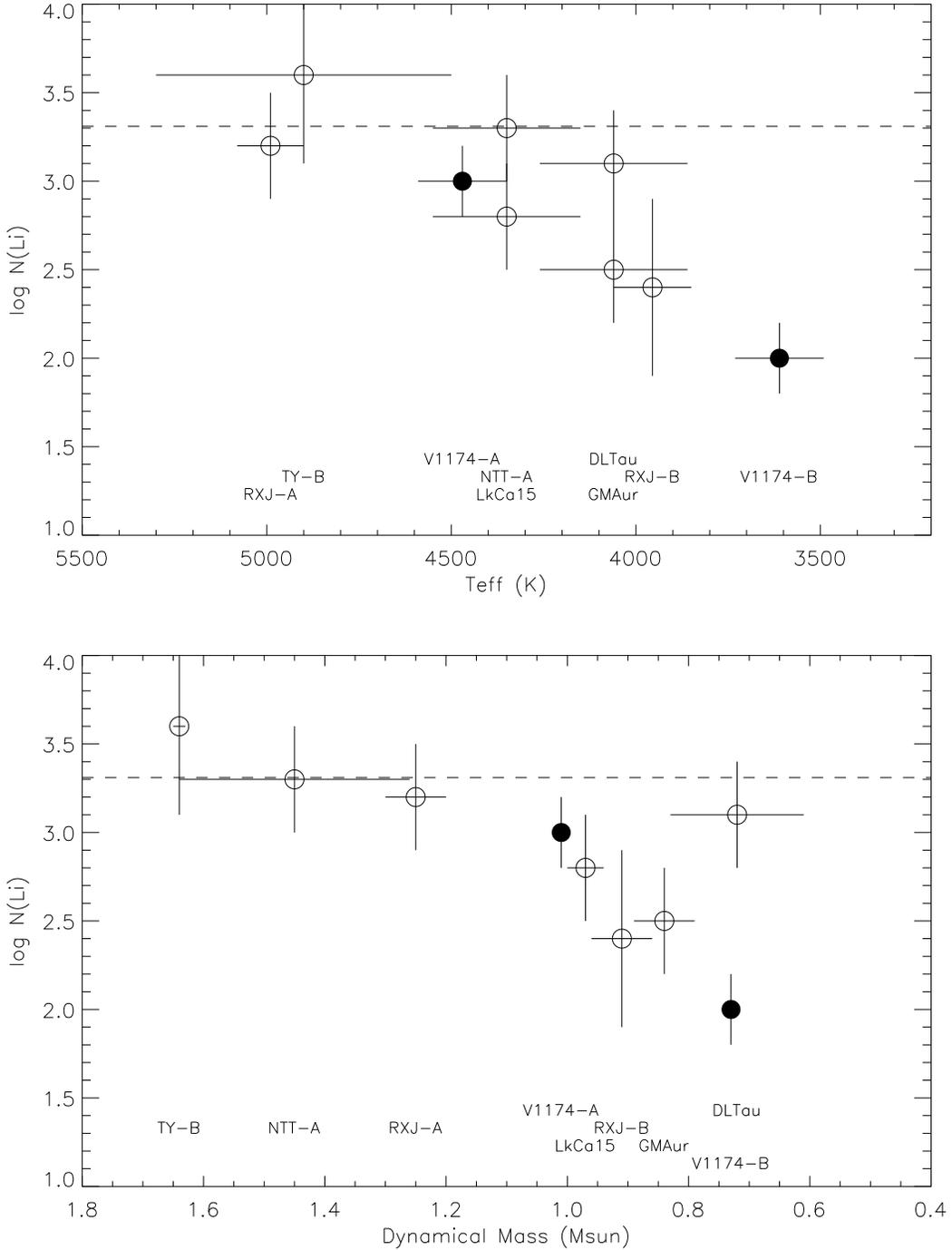}
\caption{\footnotesize Li abundances of PMS stars with empirical mass determinations.
(a) Li abundance plotted as a function of stellar \teff\ (K). The two
components of V1174~Ori are shown as filled circles, other stars are
shown as open circles. (b) Li abundance plotted as function of empirical stellar
mass (solar units). In both plots, the ``cosmic" Li abundance of 3.31 is indicated
by the horizontal line. Abundances are calculated from the observed Li
equivalent widths, using the NLTE curve-of-growth calculations of
\citet{pavlenko}. References for the equivalent width measurements are
as follows: V1174~Ori (this study); \rxj\ \citep{covino}; \ntt\ A
\citep{walter}; TY~CrA B \citep{casey}; LkCa~15 \citep{martin};
GM~Aur \citep{basri}; and DL~Tau \citep{basri}.
\label{fig-li-abund}}
\end{figure}

DL~Tau may serve as a case in point. Its measured mass of 0.72 M$_\odot$
is similar to the mass of 0.73 M$_\odot$ we have measured for the
secondary of \objectname{V1174~Ori}. Yet DL~Tau is 450 K hotter than 
\objectname{V1174~Ori} B,
suggesting a mass more comparable to that of \objectname{V1174~Ori} A. An important
additional clue to the relative masses of these two objects is provided
by their relative Li abundances.
In Fig.\ \ref{fig-li-abund} we show the Li abundances, $\log N$(Li), 
for all of the PMS stars discussed above for which Li measurements exist
in the literature. 
These stars exhibit a clear correlation of $\log N$(Li) with
stellar mass: Stars with $M \gtrsim 1.2\; {\rm M}_\odot$ show
undepleted levels of Li, while less massive stars show a trend of
monotonically decreasing Li abundance with decreasing stellar mass.
DL~Tau is a clear outlier in this trend. Its measured $\log N$(Li), 
even with its large uncertainties, is too high for its mass. In
contrast, \objectname{V1174~Ori} B follows the empirical trend nicely. The
observed Li abundance of DL~Tau implies a lower limit to its mass
of $\gtrsim 0.9\; {\rm M}_\odot$, consistent with its location in the
H-R diagram relative to \objectname{V1174~Ori} A. Whether the masses of 
\objectname{V1174~Ori} B
and DM~Tau can be similarly reconciled is unknown; there is no
published Li abundance for DM~Tau.

The observed pattern of Li depletion with stellar mass provides more
general constraints on PMS evolutionary tracks, particularly with
respect to the efficiency of convection. 
As discussed in detail by \citet{dantona-li}, models with more
efficient convection (i.e.\ larger $\alpha$) result in
dramatically more Li depletion than do models with 
less efficient convection.

Consider a star like \objectname{V1174~Ori} A, having $M = 1.0\; {\rm M}_\odot$ and
situated near the bottom of its Hayashi phase of evolution.
Assuming an initial Li abundance of $\log N$(Li)$ = 3.3$,
a model with $\alpha_{\rm in} = 1.9$ will have $\log N$(Li) $ \approx 2.5$
whereas a model with a more modest $\alpha_{\rm in} = 1.0$ will have
$\log N$(Li) $ \approx 3.2$ \citep{dantona-li}. The primary of \objectname{V1174~Ori} is more
consistent with the low Li depletion predicted by the $\alpha_{\rm in} = 1.0$
model. At lower masses, the depletion predicted for models with
highly efficient convection becomes ever more severe. A star with 
$M = 0.7\; {\rm M}_\odot$ is predicted to have $\log N$(Li) $ \approx -1.5$ 
by the $\alpha_{\rm in} = 1.9$ models, completely inconsistent with
the modest depletion levels seen in Fig.\ \ref{fig-li-abund}, which
agree with the predicted levels of depletion for $\alpha_{\rm in} = 1.0$
models \citep{dantona-li}. Thus the Li depletion measurements strongly
favor models with low $\alpha_{\rm in}$.

At low masses, the Li depletion levels predicted by the models also
become strongly dependent upon the parameter $\tau_{\rm ph}$. 
Larger values of $\tau_{\rm ph}$ result in a larger fraction of the
over-adiabatic region being computed with the low value of $\alpha$
adopted in the model atmosphere. This produces a reduced 
temperature at the base of the convection zone, resulting in lower 
levels of Li depletion. A more detailed analysis of the observed Li 
abundances as a function of stellar mass is beyond the scope of this
paper. But coupled with the precise mass determinations for PMS stars, 
such as we have determined here, the clear pattern of Li depletion 
with stellar mass should in principle provide important additional 
constraints on various key parameters of PMS evolutionary models. 
Moreover, as we have seen here, Li abundances can provide valuable, 
independent mass constraints as well when systematic uncertainties
(e.g.\ distance) may limit the accuracy of the dynamical measurements.

\section{Summary and Conclusions\label{summary}}
We have identified the star \objectname{V1174~Ori} as a previously unknown, 
double-lined, spectroscopic, pre--main-sequence (PMS) eclipsing binary 
system in the Orion star-forming region. 

From high-cadence, multi-epoch, multi-band, time-series
photometry spanning 2900 days, we find the period of the system to be
2.634727 days. From multi-epoch, high-resolution spectroscopy, we
determine a precise, double-lined orbit solution which constrains the
absolute masses of the primary and secondary components to be
1.01 M$_\odot$ and 0.73 M$_\odot$, respectively, accurate to $\sim 1$\%.
Synthetic light-curve modeling and analysis provides absolute stellar
dimensions. We find the stars to have radii of 1.34 R$_\odot$ and
1.06 R$_\odot$, respectively, again with uncertainties of $\sim 1$\%.
These empirical mass and radius measurements are distance-independent.
The secondary is thus the lowest-mass PMS star yet discovered in an
eclipsing binary system, and the first sub-solar PMS star with an
empirically determined mass accurate to better than a few percent.

We determine the distance to the system to be $\sim 400$ pc which,
together with the system's center-of-mass radial velocity, makes 
\objectname{V1174~Ori} a likely member of the Ori OB1c region. It is therefore
likely slightly in the foreground of, and slightly older than, the
Orion Nebula Cluster (ONC) proper.
Given the spread of available estimates for the bolometric 
magnitude of the Sun, we cannot currently
constrain the absolute distance to \objectname{V1174~Ori} more stringently.

Analysis of the system colors suggests the presence of an unseen
tertiary companion in a wide orbit. Follow-up photometry in the 
near-IR is underway, and should reveal substantial third light,
which we predict to be $\sim 30$\% in $J$.

We compare the empirically determined stellar dimensions with a
variety of PMS stellar evolution models in both the H-R diagram
(\teff, $L$ plane) and the mass-radius plane. 
In all cases, the models yield an age for \objectname{V1174~Ori} of $\sim 5$--10
Myr, consistent with membership in the slightly evolved Ori OB1c
region, just in the foreground of the ONC. 

While all of the models predict the slope of the
mass-radius relationship reasonably well, those with less efficient
convection (i.e.\ small convection parameter $\alpha$)
are the most consistent with our measurements, even though larger
values of $\alpha$ are required to match the present-day Sun. 
While some models are consistent with the mass of the \objectname{V1174~Ori}
primary, and others are consistent with the mass of the \objectname{V1174~Ori} secondary,
none of the models correctly predict the masses of both stars
simultaneously. This apears to be the result of the
theoretical mass tracks between 0.7--1.0 M$_\odot$ being
too compressed in \teff\ relative to the components of \objectname{V1174~Ori}.
Recent models by \citet{montalban} employing the ATLAS9
atmospheres of \citet{heiter} seem to show a larger spread of \teff\
in this mass range,
and could possibly fit both components of \objectname{V1174~Ori}, but these
models do not presently extend to masses low enough to test against
the \objectname{V1174~Ori} secondary.

Potential problems with the \teff\ scale are also implicated by a
curious effect in which the theoretical tracks and the empirical
measurements translate differently between the mass-radius plane and
the H-R diagram. Some of the tracks show the components of \objectname{V1174~Ori}
to be coeval in the H-R diagram but non-coeval in the mass-radius
plane, and vice-versa for others.

Analysis of the H-R diagram positions of all PMS stars with empirical
mass measurements yields mixed results for all of the PMS models we
examined, with no one set of models clearly superior to the others.
However, the ensemble of observations appear to be least compatible with the
tracks of \citet{sdf00}, which have the hottest and most compressed \teff\ 
scale of any of the tracks we examined.
Again, the ATLAS9-based models of \citet{montalban} show promise of
being able to reproduce most of the empirical measurements, and we
encourage the extension of these models to lower \teff. 

We have carefully scrutinized the ensemble of empirical PMS mass
measurements for consistency with one another, independent of the models.
Encouragingly, we find instances of superb agreement between independent
mass determinations, such as \objectname{V1174~Ori} A and LkCa~15 (a single star in
Taurus), which have identical
masses and are at identical positions in the H-R diagram.
But, disconcertingly, we also find instances of apparent contradiction,
such as \objectname{V1174~Ori} B and DM~Tau, which are at similar (though not identical)
positions in the H-R diagram yet have very different masses.

These apparently discordant empirical measurements may be reconcilable,
however. PMS evolutionary models that are double-valued in \teff, such
as the ATLAS9-based models of \citet{montalban}, may in fact predict
the H-R diagram positions observed for these stars. 
In some cases, such as the astrometric binary \ntt\ and the eclipsing
binary \rxj, we reconcile the disagreements by adjusting the masses of
the stars within the uncertainties.
In another case---the single star DL~Tau---we find compelling
evidence in the form of an anomalous Li abundance that the reported
mass is incorrect, most likely the result of a systematic error in
the assumed distance.

Indeed, we find that the ensemble of PMS stars with empirical mass
determinations possess a remarkably well-defined pattern of Li depletion as a
function of stellar mass. The mild amount of depletion observed in
the lowest-mass star, \objectname{V1174~Ori} B, strongly constrains the efficiency
of convection that can be acting in the interiors of these stars, and
argues strongly in favor of models with small convection parameters, $\alpha$.
Thus, as argued by \citet{dantona-li}, Li abundance determinations in
PMS stars promise to provide an important observational window into the
interior structure of PMS stars. And, as we have shown in the case of
DL~Tau, Li can provide a valuable means for triangulating against the 
dynamical masses of these stars in the presence of systematic errors.

\acknowledgments
We acknowledge the support of a Hubble Fellowship (KGS) and NSF grant 
AST-0098417 (RDM). We also gratefully acknowledge support from the Brazilian
institutions CNPq, FAPEMIG, and CAPES (LPRV).
We are grateful to the staff of the Hobby Eberly Telescope (HET), particularly
M.\ Shetrone, for generous assistance with the reduction of the HRS
spectra, which were obtained through the first NOAO community program
on the HET HRS, and D.\ Bell for patient support in coordination of
this program which spanned multiple observing seasons. We are also grateful
to NOAO for its commitment to maintaining community access to the small 
telescopes upon which this project relied heavily. Finally, we thank
J.\ Montalb\'{a}n for illuminating discussions on the PMS evolutionary models
and for providing new model calculations in advance of publication.

\clearpage

\clearpage

\clearpage

\clearpage

\clearpage

\clearpage

\clearpage

\clearpage

\clearpage

\clearpage

\clearpage

\clearpage

\clearpage

\clearpage

\clearpage

\clearpage

\clearpage

\clearpage

\clearpage

\clearpage

\clearpage

\clearpage

\clearpage

\clearpage

\clearpage

\clearpage

\clearpage

\clearpage

\clearpage

\end{document}